\documentclass[preprint,3p]{elsarticle}
\usepackage{soul,color}
\soulregister\cite7
\soulregister\ref7
\soulregister\pageref7
\usepackage[utf8]{inputenc}
\usepackage{subcaption}
\usepackage{amsfonts}
\usepackage{amsmath}
\usepackage{stmaryrd}
\usepackage{comment}
\usepackage{dsfont}
\usepackage{subfiles}
\usepackage{fullpage}
\usepackage{tabularx}
\usepackage{hyperref}
\usepackage{amsthm}
\usepackage{placeins}
\usepackage{xcolor}
\usepackage{algorithm}
\usepackage[noend]{algpseudocode}
\usepackage{color, soul}
\usepackage{mathtools}
\usepackage{undertilde}

\DeclareMathOperator*{\argmax}{arg\,max}

\usepackage{siunitx}
\usepackage{amssymb}
\usepackage{amsmath}
\usepackage{amsmath,stackengine}

\usepackage{todonotes}
\usepackage{upgreek}
\usepackage{enumitem}
\usepackage{appendix}
\usepackage{algorithm}
\usepackage{algpseudocode}
\usepackage{enumitem}
\usepackage{stackengine}

\begin{document}
\begin{frontmatter}

\title{A multi-scale probabilistic methodology to predict high-cycle fatigue lifetime for alloys with process-induced pores}

\author[1]{Abhishek Palchoudhary}
\ead{abhishek.palchoudhary@minesparis.psl.eu}
\cortext[cor]{Corresponding author}
\author[1]{Cristian Ovalle}
\author[1]{Vincent Maurel}
\author[1]{Pierre Kerfriden}
\ead{pierre.kerfriden@minesparis.psl.eu}

\address[1]{Mines Paris, PSL University, Centre for Material Sciences (CMAT), CNRS UMR 7633, BP 87 91003 Evry, France}




 

\begin{abstract}
A multi-scale methodology is developed in conjunction with a probabilistic fatigue lifetime model for structures with pores whose exact distribution, i.e. geometries and locations, is unknown. The method takes into account uncertainty in fatigue lifetimes in specimens due to defects at two scales: micro-scale heterogeneity and meso-scale pores. An element-wise probabilistic strain-life model with its criterion modified for taking into account multiaxial loading is developed for taking into account the effect of micro-scale defects on the fatigue lifetime. The effect of meso-scale pores in the structure is taken into account via statistical modelling of the expected pore populations via a finite element method, based on tomographic scans of a small region of porous material used to make the structure. A previously implemented Neuber-type plastic correction algorithm is used for fast full-field approximation of the strain-life criterion around the statistically generated pore fields. The probability of failure of a porous structure is obtained via a weakest link assumption at the level of its constituent finite elements. The fatigue model can be identified via a maximum likelihood estimate on experimental fatigue data of structures containing different types of pore populations. The proposed methodology is tested on a pre-existing high-cycle fatigue data-set of an aluminium alloy with two levels of porosity. 
The model requires lesser data for identification than traditional models that consider porous media as a homogeneous material, as the same base material is considered for the two grades of porous material. Numerical studies on synthetically generated data show that the method is capable of taking into account the statistical size effect in fatigue, as well as demonstrate that fatigue properties of subsurface porous material are lower than that of core porous material, which makes homogenization of the method non-trivial.
\end{abstract}



\end{frontmatter}
\clearpage

\section*{List of Symbols}
\begin{description}[labelwidth=3cm, leftmargin=4cm, style=nextline]
    \item[$R$] Ratio of minimum to maximum stress in a load cycle
    \item[$\sigma_y$] Yield stress of the material
    \item[$\Sigma_a$] Applied stress amplitude
    \item[$\Delta \Sigma$] Applied stress range
    \item[$\Delta \varepsilon^*$] Total strain range in an element for the  stabilized cycle
    \item[$\Delta \varepsilon^{e*}$] Elastic strain range in an element for the stabilized cycle
    \item[$\Delta \varepsilon^{p*}$] Plastic strain range in an element for the stabilized cycle
    \item[$N_{\textrm{R}}^{\textrm{*}}$] Number of cycles to failure of an element
    \item[$V^*$] Volume of an element
    \item[$\utilde{\varepsilon}^*$] Total strain tensor of an element
    \item[$\utilde{\varepsilon}^{p*}$] Plastic strain tensor of an element
    \item[$\utilde{\varepsilon}^{e*}$] Elastic strain tensor of an element
    \item[$\mathcal{T}_c$] Stabilized cycle
    \item[$n^*$] Eigenvector associated to the highest eigenvalue of the stress tensor of an element coming from a separate elastic computation
    \item[$\utilde{\sigma}^{\#*}$] Stress tensor of an element coming from a separate elastic computation
    \item[$A,\alpha,B,\beta,C$] Parameters of the strain-life model
    \item[$\mathcal{W}$] Weibull distribution
    \item[$m, \lambda$] Weibull shape and scale parameters
    \item[$F_{N_{\textrm{R}}^{\textrm{*}}}$] Cumulative distribution function of failure for elements
    \item[$f_{N_{\textrm{R}}^{\textrm{*}}}$] Probability density function of failure for elements
    \item[$F_{N_{\textrm{R}}^{\textrm{s}}}$] Cumulative distribution function of failure for a  specimen
    \item[$f_{N_{\textrm{R}}^{\textrm{s}}}$] Probability density function of failure for the specimen
    \item[$\mathcal{L}$] Likelihood function
    \item[$\mu$] Set of optimized parameters
    \item[$\tilde{\mu}$] Set of intermediary parameters tested by an optimizer
    \item[$\mathcal{H}$] Set of fatigue experiments for geometrically homogeneous specimens
    \item[$\mathcal{I}$] Set of fatigue experiments for a fixed heterogeneous specimen
    \item[$\mathcal{J}$] Set of fatigue experiments for porous specimens
     \item[$\mathcal{K}$] Set of all possible configurations of synthetically generated porous specimens
     \item[$\tilde{K}$] Set of finite synthetically generated porous specimens

\end{description}

\section{Introduction}
\label{intro}

Pores in structures are known to reduce their fatigue lifetimes. For the lifetime prediction of such structures, simplified approaches are available. One approach models the porous material as a homogenised substance with equivalent fatigue properties, eliminating the need to describe individual pores directly \cite{Ezanno2010}. However, the applicability of such models relies on fatigue data from structures made of the specific porous material and cannot be transferred to structures with the same base material but different pore characteristics. Other, more recent approaches rely on the availability of the exact pore distribution in order to predict the fatigue lifetime of the structure, usually obtained via computed tomography (CT) scans \cite{Le2018,Matpadi2024}. However, this may not be suitable for all scenarios, as CT scans of a structure are expensive to obtain and analyse. Moreover, calibration of such a model requires a database of CT scans for a number of specimens tested for failure. Another category of approaches employ statistical methods to consider all possible pore distributions, based on a small representative CT scan of the porous material, when making lifetime predictions of a structure. Some of these approaches do not explicitly model the pore distribution, but use Poisson point processes to generate simplified pore distributions, and make predictions with the help of a Kitagawa Takahashi (KT) diagram that links the fatigue limit to the size of the pores \cite{ElKhoukhi2019, romano2019}. These approaches assume that a single killer pore is responsible for fatigue failure, and do not take into account pore morphology or interactions between pores. Furthermore, obtaining the KT diagram is a lengthy experimental process. Finally, other proposals exist that quantify uncertainty in fatigue lifetime due to statistical variability in meso-scale pores by employing finite element analysis to represent the pores. These approaches typically simplify the representation of pore geometries to spheres or ellipses \cite{Lacourt2019, Shirani2012}, either neglecting inter-pore interactions \cite{Bercelli2021} or restricting the model to two-dimensional structures \cite{Talemi2020,Hou2024}. 

We present a multi-scale method that takes into account uncertainty in fatigue lifetime of structures due to uncertainty in entities at two scales: meso-scale pores and micro-scale heterogeneity. Our approach allows to model the lifetime of structures with varying levels of porosity using a unified fatigue model, thereby reducing the total amount of fatigue data required for model identification as compared to the simplified approaches considering homogenized porous behavior. This is because statistical information on fatigue lifetime is shared between datasets of different porosity grades, as the same base material behaviour around the pores is considered. For the behavior of the base material, a local strain-life model, as proposed by Manson \cite{Manson1953}, is selected. The variability in the base material's lifetime, resulting from micro-scale heterogeneity — including factors like varying grain orientation and the presence of precipitates \cite{Le2018} — is reflected in the strain-life model by making its criterion probabilistic. The physics of this micro-scale heterogeneity are not explicitly modelled. Moreover, the criterion has been carefully selected to address the existence of two fatigue regimes (dependent on the magnitude of loading) that are commonly observed in experiments \cite{astm2015}, and to accommodate multiaxial stress-strain states \cite{Brown1973,Fatemi1988,karolczuk2005} that are likely to be found in the base material around pores. At the meso-scale, the model incorporates the actual three-dimensional geometries of the pores and typical inter-pore spacing by explicitly modelling them using the finite element method. An elasto-plastic material behaviour is adopted to account for the plasticity present in the vicinity of the pores. The probabilistic lifetime of this porous structure is determined by a weakest link hypothesis \cite{Zok2017,Liu2020,Li2022}, on the finite elements that comprise the structure. The entire structure is considered as multiple hot spots may form in porous media due to complex stress fields arising from pore-pore and pore-surface interactions \cite{Matpadi2024}. Due to the presence of ligaments, the location of the critical hot spot is uncertain, making it difficult to apply the concept of critical distances \cite{Taylor1999} in the fatigue criterion. On the other hand, in highly stressed volume (HSV) approaches, the size of the stressed volume is a parameter that requires experimental pre-identification \cite{ElKhoukhi2019, Kuguel1961, Sonsino1997, He2022}.

In the event of uncertainty regarding the meso-scale pore distribution, the overall lifetime of the structure is addressed by introducing statistical variability into the pore distribution within the structure. In other words, a degree of randomness is assumed in the pore distribution due to the unavailability of precise prior information on the pore states in the structure that lead to fatigue failure. The model operates under the assumption that the precise pore distribution responsible for a particular fatigue lifetime data point is unknown. Instead, a computed tomography (CT) scan of a different volume, made from the same porous material as the structure, is obtained to gather statistical information regarding the pores that are likely to be contributing to the observed failure of the structure. This statistical data serves as the input to the proposed lifetime model, which, by considering various potential pore positions and characteristics, computes a lifetime probability for a given structure without the need for precise knowledge of its pore distribution. This approach allows for the uncertainty in lifetime due to all possible pore distributions to be taken into account, including the effect of pore-pore and pore-surface interactions. We also demonstrate the difficulty of bypassing the multi-scale method with a homogenised model; a fatigue model identified for porous material in one type of structure may not always be accurate when applied to other structures. 

Probabilistic lifetime models based on the weakest link approach for non-smooth geometries are identified through iterative optimisation in one of two methods. One approach employs an effective criterion within an equivalent volume to address non-homogeneous stress distributions, thereby reducing the optimisation to a 0D space \cite{munizcalvente2015}. Other authors employ full-field mechanical information in the optimisation process. For instance, the stress is utilised when the loading remains elastic throughout \cite{Lanning2003}, while the strain energy is employed in the event of plasticity \cite{Li2022}. However, the majority of weakest link approaches are concerned with notched structures \cite{Liu2020,Li2022,Lanning2003,Karolczuk2013}. The identification of a lifetime model based on the weakest link approach for a porous structure modelled with the finite element method, where the exact pore distribution is unknown, has yet to be addressed in the literature and will be discussed in this paper.

The criteria of the lifetime model requires access to the full-field elasto-plastic solutions of structures with varying pore distributions subjected to fatigue loading. The presence of local non-linearity in the material model due to localized plasticity around the pores, the need for a high number of degrees of freedom to model the complex geometries of the pores, and the requirement to obtain the stabilised cycle due to a transient hardening phase all make these computations prohibitively costly. Our final innovation consists in accelerating these computations via our previously developed plastic correction algorithm. The algorithm is a Neuber-type plastic correction heuristic coupled with machine learning methods that allows rapid recovery of the tensorial elasto-plastic solution in entire structures for arbitrary (proportional) applied loading \cite{Palchoudhary2024}.

This paper is divided into six sections. Section two presents a data set pertaining to a material with two levels of porosity, which will be used for further developments. Section three presents the lifetime model and the multi-scale methodology for obtaining the lifetime of porous structures. Section four is dedicated to maximum likelihood methods for identifying the parameters of the lifetime model on experimental high-cycle fatigue data sets, including a proposal for dealing with structures with unknown pore distributions. Section five presents the results of calibration for the previously presented high-cycle fatigue data set, while detailing the process of the convergence of the solution and the optimisation behaviour, and also presents an analysis of the fatigue model complexity. Section six is dedicated to numerical investigations of additional natural properties of the model, including its capacity to account for the size effect and pore-surface interaction effects. This section also tests the limits of a naive homogenisation approach with the objective of reducing the cost of utilising the multi-scale approach, which involves explicit pore descriptions. The paper is concluded with an overview of the findings and some elements of discussion.
\section{Experimental data \label{sec:expdata}}
Although the proposed methodology is relatively general by design, the developments are dedicated to the case of a specific porous material for which the following data is available: (i) the number of cycles to failure in the high cycle fatigue (HCF) regime for a relatively large number of traction specimens, and (ii) tomographic data to characterise the distribution and morphology of the pores. This dataset is described below.

\paragraph{Composition, grades and manufacturing process}\mbox{}\\

\noindent An Al-Si7Mg0.3 alloy is considered (i.e. 7\% silicon, 0.3\% magnesium and balance
aluminium in weight \%). Alloy 'B' is produced using the lost foam casting process, followed by a heat treatment, and contains shrinkage and gas pores. The second alloy, designated 'C', is produced using the same lost foam casting process as the alloy 'B', followed by hot isostatic pressing (HIP) at a temperature of $500^{\circ}$C and a pressure of 1000 bars. This results in a virtually porosity-free alloy, which is then subjected to the same heat treatment as the alloy 'B'. The alloy 'C' is therefore referred to as the non-porous alloy. 

The manufacturing process yields slabs of these materials, and specimens for subsequent characterisation (including fatigue tests and CT scans) were obtained from the centre of these slabs, in order to avoid edge effects during casting and thus ensure the accuracy of the microstructure of the materials. Further details on the manufacturing process can be found in the work of other authors \cite{LePhD2016}.

\paragraph{Microstructure and tension-compression fatigue failure mechanisms}\mbox{}\\

\noindent The microstructures and characteristics of pores in these alloys have been previously investigated by other researchers \cite{ElKhoukhi2019, VietDucLe2016, Elkhoukhi2022}. The base material found in the two grades 'B' and 'C', have very similar microstructural characteristics. These include similar values of dendrite and secondary dendrite arm spacing, similar sizes of silicon precipitates, and comparable grain sizes \cite{LePhD2016}.

The porous alloy 'B' has a volumetric fraction of pores of approximately 0.28\%, exhibiting a distribution of varying pore shapes and sizes. The maximum equivalent pore size, defined by the square root of the area observed by optical microscopy scans, reaches up to 300 $\mu m$. In the case of the non-porous alloy 'C', the presence of very small pores (with the maximum effective size $<50$ $\mu m$) has been observed.

In conditions of tension-compression fatigue loading, the porous alloy 'B' is known to fail due to high stress concentration factors in the vicinity of pores. This often occurs as a result of a crack forming from a surface-breaking pore, from the interaction between sub-surface pores and the surface of the specimen, or from the coalescence of cracks originating from several pores \cite{VietDucLe2016, Elkhoukhi2022}. The primary cause of tension-compression fatigue failure in the non-porous alloy 'C' is micro-plastic activity, namely the formation of persistent slip bands from which critical cracks originate \cite{VietDucLe2016}.

\paragraph{Tension-compression fatigue experiments without prior tomography}\mbox{}\\

\noindent A limited data set comprising 32 fatigue experiments for the non-porous alloy and 34 fatigue experiments for the porous alloy is available \cite{VietDucLe2016}. These experiments were conducted in the HCF regime, under symmetric tension-compression conditions with zero mean stress ($R=-1$) at a frequency of 100 Hz. The applied stress range on these specimens ($\Delta \Sigma$) is always within the elastic regime and is defined as the difference between the maximum and minimum applied load in a cycle. The applied stress amplitude is defined as half the applied stress range, expressed as follows: $\Sigma_a = \frac{\Delta \Sigma}{2}$.

\begin{figure*}[!htbp]
    \centering
        \def\svgwidth{.85\textwidth}
        
        \begin{subfigure}[b]{0.90\textwidth}
        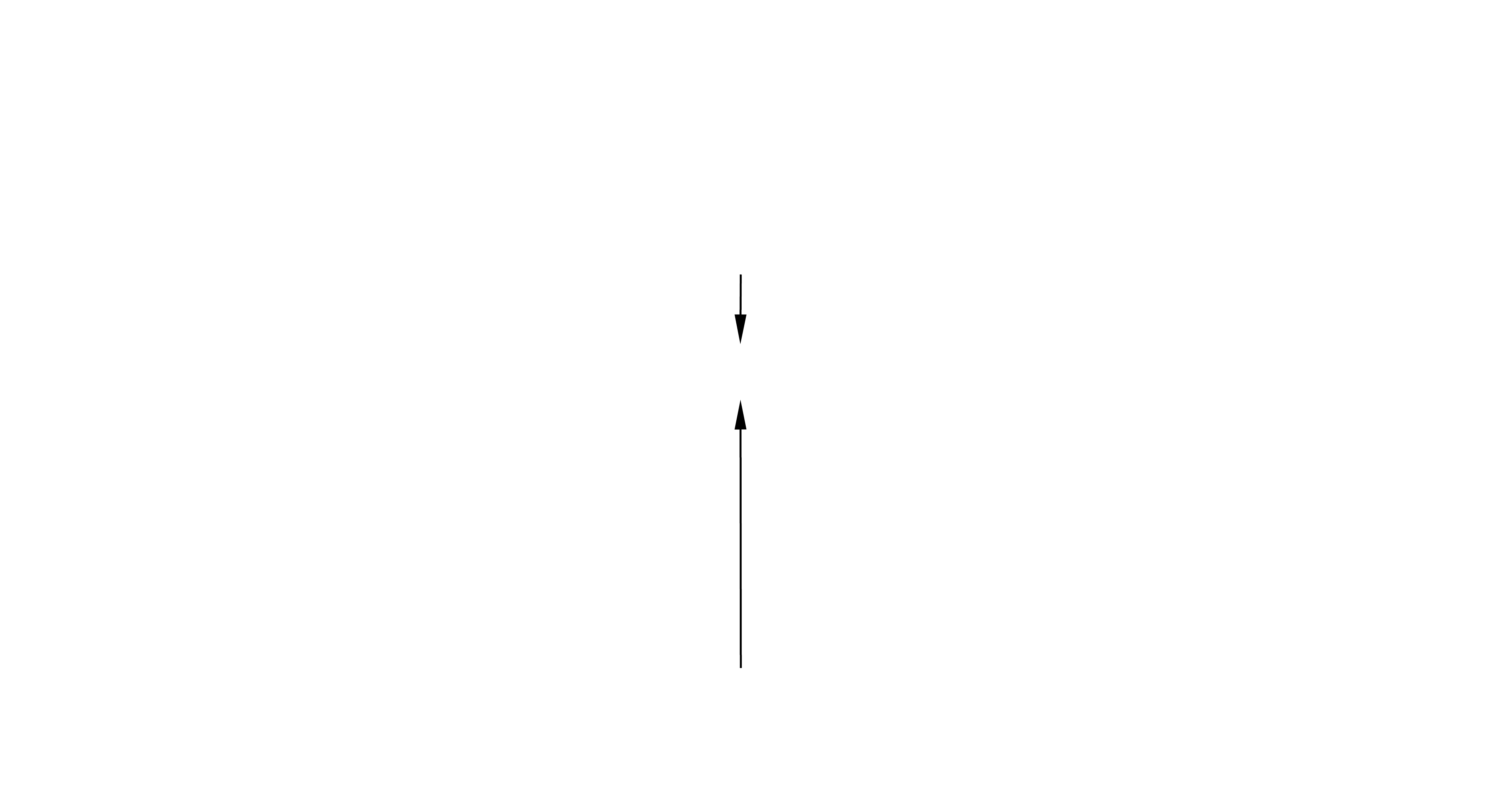
            \caption{}
        \end{subfigure}  
        
        \begin{subfigure}[b]{0.50\textwidth}
         \includegraphics[width=1\textwidth]{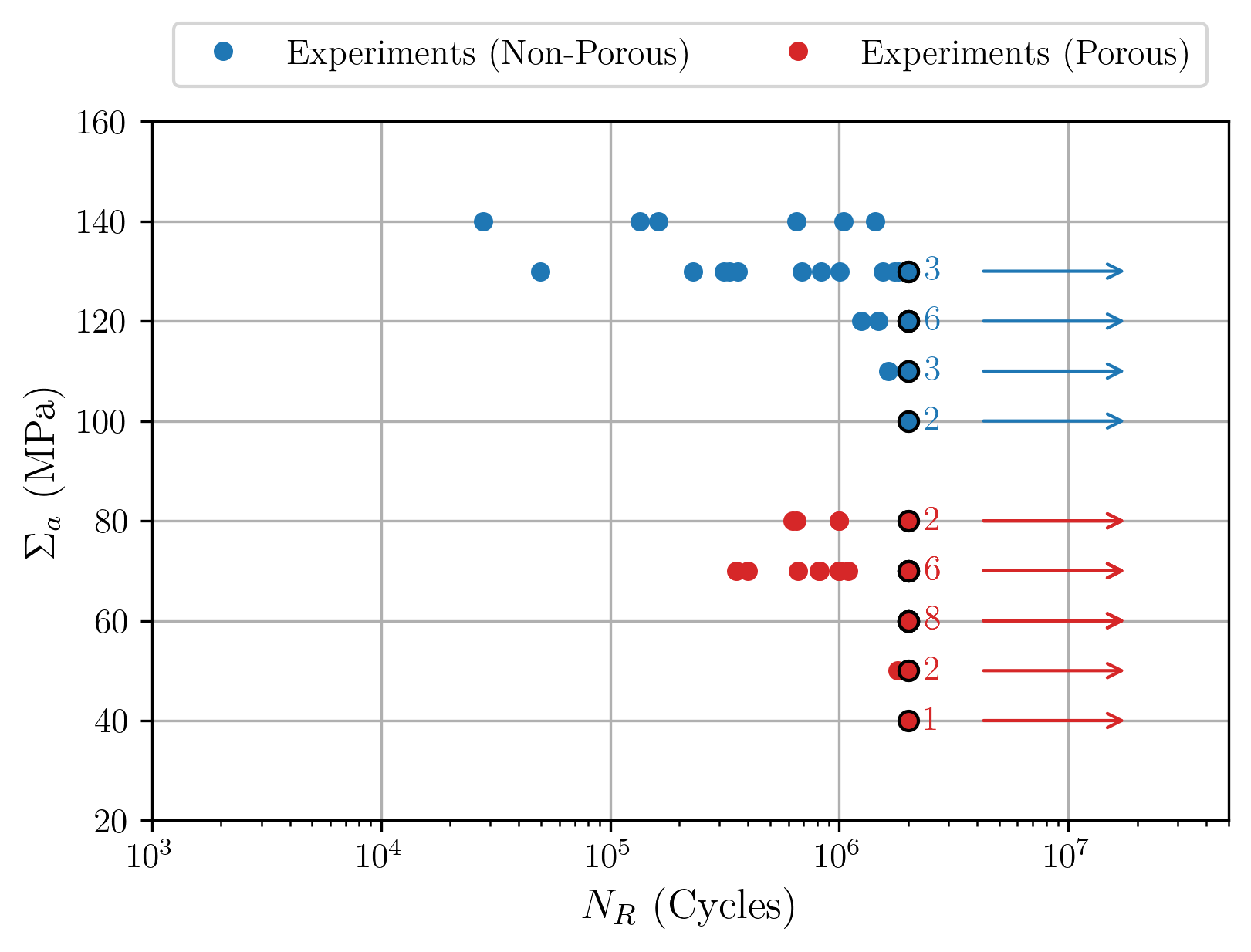}
            \caption{}
        \end{subfigure}
        
        \caption{(a) Geometry of specimens (with values in mm) used for fatigue experiments (from the authors in \cite{VietDucLe2015}) (b) Fatigue experiments reported using these specimens (from the authors in \cite{VietDucLe2015}), with the applied stress amplitude $\Sigma_a$ as a function of the number of cycles to failure $N_R$. The run-outs (at $N_R = 2 \times 10^6$ cycles) are all counted and reported with a number and arrow to indicate possible failure for a larger number of cycles}
        \label{Fig:FatigueExperiments}
\end{figure*}

The geometry of the specimens used is shown in Fig.  \ref{Fig:FatigueExperiments} (a) and the results are recalled in a applied stress amplitude ($\Sigma_a$) - lifetime ($N_R$) plot in Fig. \ref{Fig:FatigueExperiments} (b). The tomographies of the specimens used for fatigue experiments are not available. As the specimens were cut from a bigger slab, the pores in these specimens are volumetric pores created due to the manufacturing process. For all experiments, a maximum limit on the number of loading cycles was set at $2 \times 10^6$ cycles. The specimens which did not break (termed run-outs) were counted and reported at the $N_R = 2 \times 10^6$ mark with an arrow to the right to indicate potential failure for a larger number of cycles (Figure \ref{Fig:FatigueExperiments}(b)).

\paragraph{Computed tomography of a volume with pores}\mbox{}\\

\begin{figure}[!htbp]
    \centering
    \begin{subfigure}[b]{0.9\textwidth}
        \includegraphics[width=\textwidth]{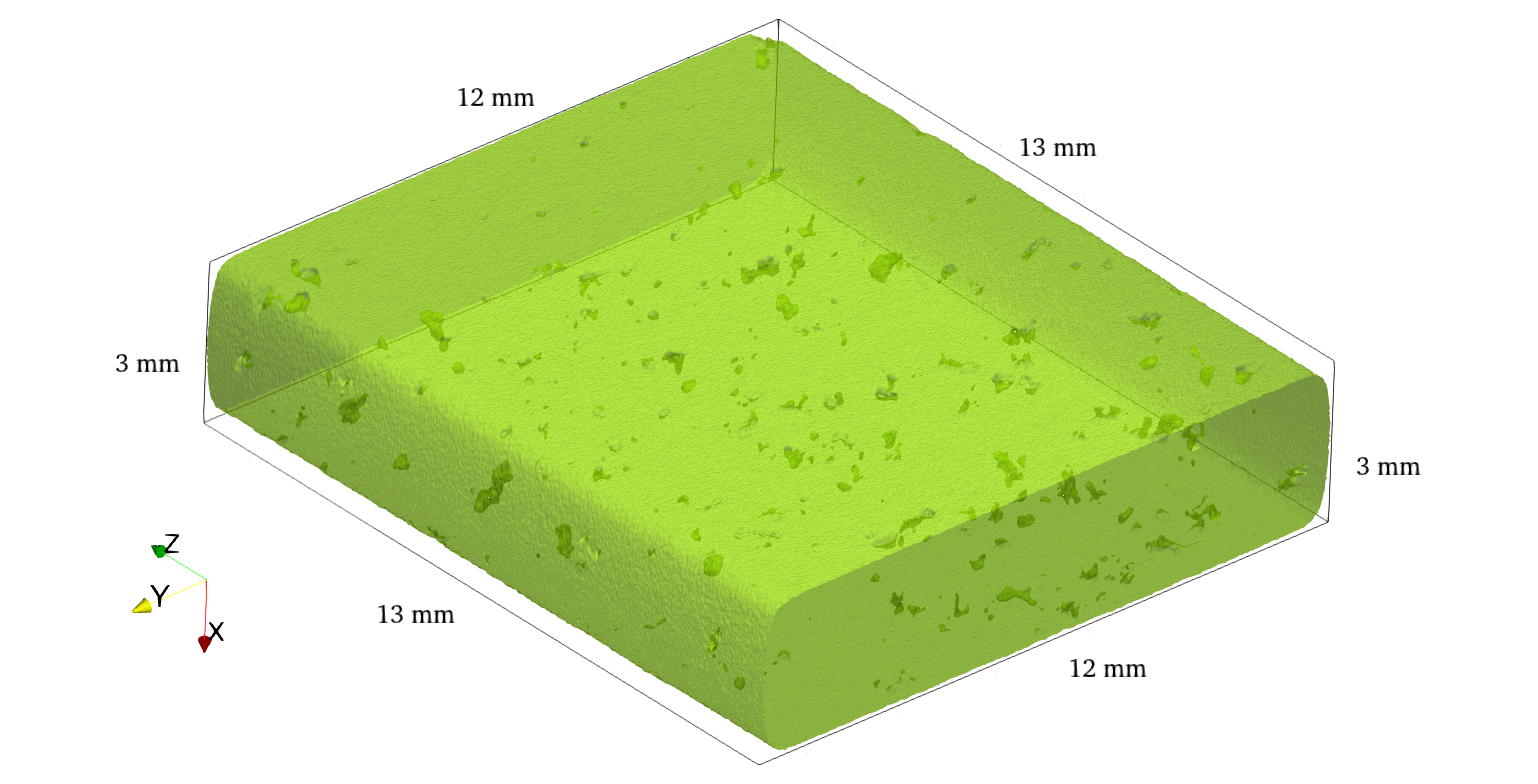}
    \end{subfigure}
    
    \caption{\label{fig:tomo} Tomography of $12\times13\times3 \, \textrm{mm}^{3}$ region of a planar specimen made using the same porous material as the cylindrical specimens used for fatigue experiments} 
\end{figure}

\noindent A computed tomography scan of a $12\times13\times3 \, \textrm{mm}^{3}$ region of a planar specimen, made using the same porous material as the cylindrical specimens used for fatigue experiments, was carried out in order to get information on the pore characteristics likely to have caused fatigue failure. The planar specimen was cut from a larger slab, as was the case with the cylindrical specimens employed in fatigue testing. Thus, the pores in the planar specimen are also volumetric pores resulting from the manufacturing process. The resolution of the tomograph was $\sim 11 \, \mu \textrm{m}$. The tomographic volume of interest was $424\times1211\times1000$ voxels with 16-bit unsigned images. Segmentation of the pores was carried out using a threshold value in grey-level contrast levels, resulting in a binary 3D image of the pores. A surface mesh was generated using Avizo, employing smoothing to remove artefacts. The resulting surface mesh, illustrated in Fig. \ref{fig:tomo} is used in subsequent sections.

\section{Multi-scale probabilistic fatigue lifetime model}\label{sec:TwoScaleModel}
\subsection{Micro and meso scale defects}
        A probabilistic multi-scale approach is proposed for the fatigue lifetime prediction of specimens containing a distribution of pores and microstructural heterogeneity. The first scale, henceforth termed the micro-scale, accounts for the effect of microstructural heterogeneity on the fatigue lifetime. The microstructural heterogeneity is not explicitly modelled; its effect is considered by making the fatigue lifetime model probabilistic \cite{Doudard2004,Ni2004,Pessard2011,Koutiri2013,Li2016}. This approach is adopted as the exact state of the microstructure is not known and would be too costly to model at the scale of the specimens undergoing fatigue loading. The lifetime distribution of an elementary volume subjected to a uniform stress depends on its volume and on the mechanical load to which it is subjected; further details of the construction of this probabilistic model will be explained in the next subsection.

\begin{figure}[htbp]
    \centering
    \def\svgwidth{.95\textwidth}
\begingroup%
  \makeatletter%
  \providecommand\color[2][]{%
    \errmessage{(Inkscape) Color is used for the text in Inkscape, but the package 'color.sty' is not loaded}%
    \renewcommand\color[2][]{}%
  }%
  \providecommand\transparent[1]{%
    \errmessage{(Inkscape) Transparency is used (non-zero) for the text in Inkscape, but the package 'transparent.sty' is not loaded}%
    \renewcommand\transparent[1]{}%
  }%
  \providecommand\rotatebox[2]{#2}%
  \newcommand*\fsize{\dimexpr\f@size pt\relax}%
  \newcommand*\lineheight[1]{\fontsize{\fsize}{#1\fsize}\selectfont}%
  \ifx\svgwidth\undefined%
    \setlength{\unitlength}{924.95690918bp}%
    \ifx\svgscale\undefined%
      \relax%
    \else%
      \setlength{\unitlength}{\unitlength * \real{\svgscale}}%
    \fi%
  \else%
    \setlength{\unitlength}{\svgwidth}%
  \fi%
  \global\let\svgwidth\undefined%
  \global\let\svgscale\undefined%
  \makeatother%
  \begin{picture}(1,0.45425951)%
    \lineheight{1}%
    \setlength\tabcolsep{0pt}%
    \put(0,0){\includegraphics[width=\unitlength,page=1]{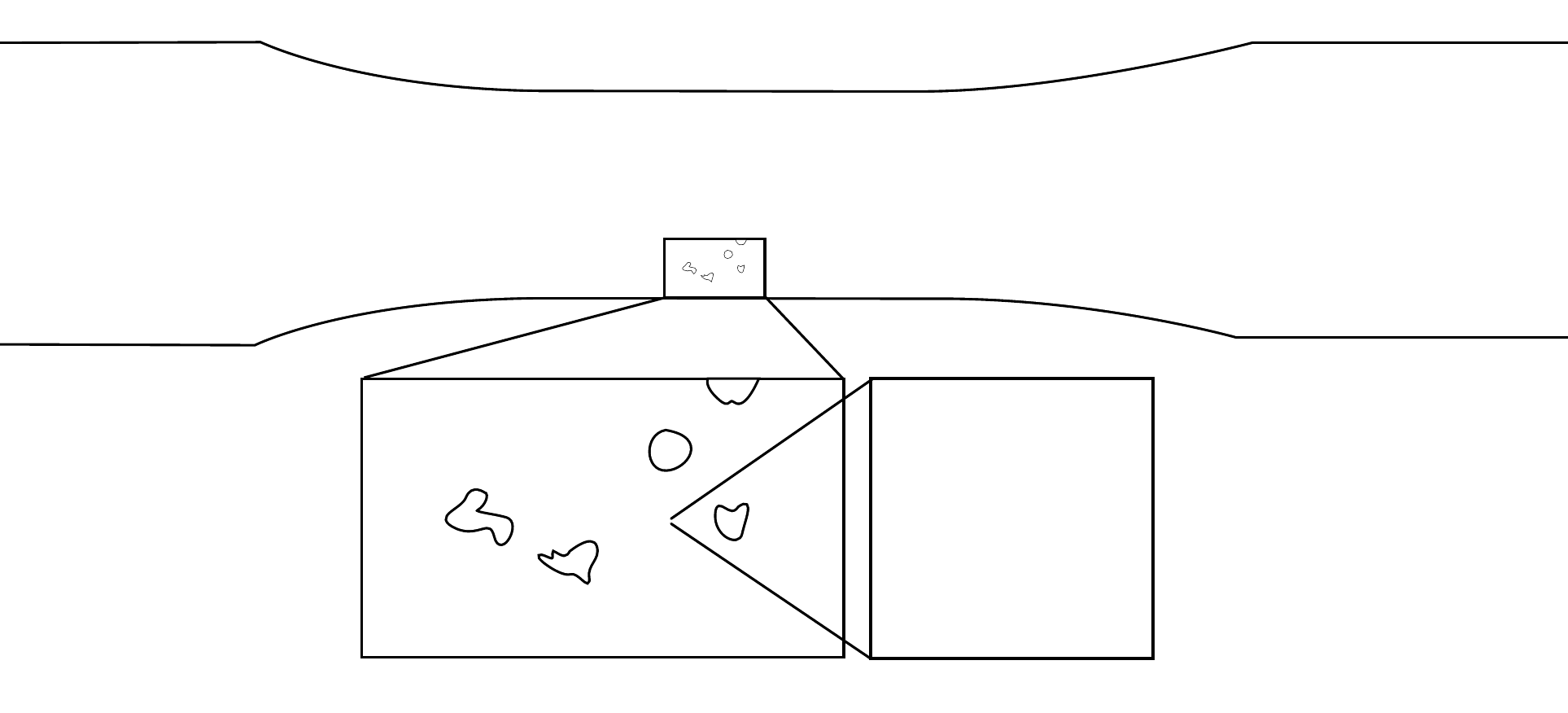}}%
    \put(0.23743443,0.32504132){\color[rgb]{0,0,0}\makebox(0,0)[lt]{\lineheight{1.25}\smash{\begin{tabular}[t]{l}Specimen\end{tabular}}}}%
    \put(0.40420434,0.37124666){\color[rgb]{0,0,0}\makebox(0,0)[lt]{\lineheight{1.25}\smash{\begin{tabular}[t]{l}Subvolume\\of porous\\material\end{tabular}}}}%
    \put(0.24490824,0.1890402){\color[rgb]{0,0,0}\makebox(0,0)[lt]{\lineheight{1.25}\smash{\begin{tabular}[t]{l}Meso-scale: pores\end{tabular}}}}%
    \put(0.24374895,0.04774658){\color[rgb]{0,0,0}\makebox(0,0)[lt]{\lineheight{1.25}\smash{\begin{tabular}[t]{l}Homogenous matrix\end{tabular}}}}%
    \put(0.56731389,0.18888407){\color[rgb]{0,0,0}\makebox(0,0)[lt]{\lineheight{1.25}\smash{\begin{tabular}[t]{l}Micro-scale:\end{tabular}}}}%
    \put(0.56862475,0.1268004){\color[rgb]{0,0,0}\makebox(0,0)[lt]{\lineheight{1.25}\smash{\begin{tabular}[t]{l}Heterogeneity\\in local plastic \\deformation\end{tabular}}}}%
    \put(0,0){\includegraphics[width=\unitlength,page=2]{Schema_ExtHet_general.pdf}}%
  \end{picture}%
\endgroup%

    \caption{\label{fig:Schema_ExtHet_general} Example of a specimen containing a sub-volume of pores, with micro-plasticity activating in the base material between the pores} 
    \end{figure}

\begin{figure}[h!tbp]
    \centering
    \def\svgwidth{.8\textwidth}
    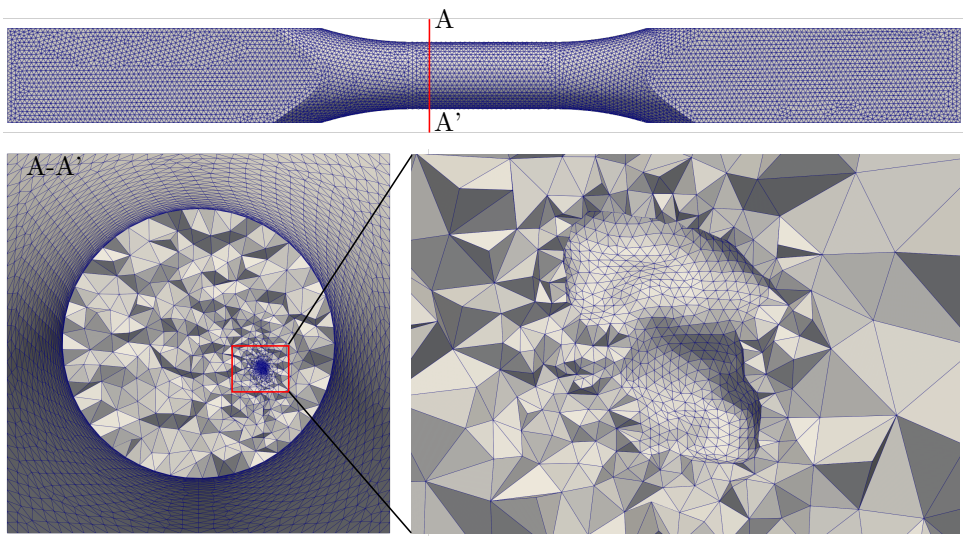
    \caption{\label{fig:FEPores} Example of a specimen containing a sub-volume of pores as a finite element mesh, with a zoom of the mesh around one pore} 
    \end{figure}

        The second scale, defined as the meso-scale, comprises tomography-informed pore distributions that are embedded in the base material containing the microstructural heterogeneity. This is shown schematically in Fig. \ref{fig:Schema_ExtHet_general}. The effects of the pore distribution on the fatigue lifetime are considered by explicitly modelling these pores via a finite element model (see Fig. \ref{fig:FEPores}). The sizes, shapes, and locations of pores with respect to other pores and the surfaces of the specimen all influence the fatigue lifetime. The meso-scale part of the model accounts for uncertainty due to varying pore distributions by synthetically generating them.

        Due to the presence of explicitly represented pores in the fatigue specimen, the mechanical stress in the synthetic specimens is heterogeneous. At macroscopic stress greater than $\sigma_y$, local plasticity is expected to develop around the pores due to stress concentrations. The micro-scale model may be used element-wise to obtain a probability distribution of failure per element.

        The overall life of such a porous specimen is derived from the lifetime probability density of the individual finite elements that constitute it, via the standard weakest link assumption \cite{Zok2017}. This modelling brick will be detailed after the explanation of the micro and meso scales of the model.

\subsection{Probabilistic modelling of the life span at the material point level (micro-scale uncertainty modelling)}
\label{sec:LocalModels}

In this section, a probabilistic element-wise strain life model (with two regimes) is developed. The element-wise model needs to be based on multiaxial criterion, as the local stress state for regions around the pores in specimens depicted in Fig.  \ref{fig:Schema_ExtHet_general} is multiaxial. The fatigue lifetime of an element $* \in \mathcal{E}$ (where $\mathcal{E}$ is the set of finite elements constituting a specimen) is obtained as a probability distribution, given a damage criterion $\frac{\Delta \varepsilon^*}{2}$ and the volume $V^*$ of the element $*$.

\paragraph{Deterministic strain-life approach}\mbox{}\\

\begin{figure}[htbp]
    \centering
    \def\svgwidth{.75\textwidth}
    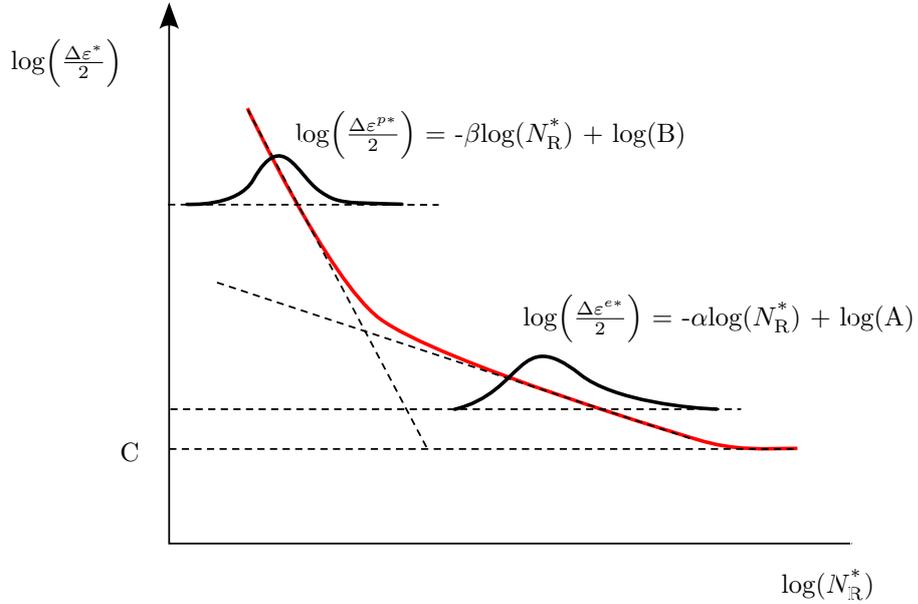
    \caption{\label{fig:Schema_proba} Schematic of the probabilistic strain-life model with two straight lines connected with an elbow in the log($\frac{\Delta \varepsilon^*}{2}$)-log($N_{\textrm{R}}^{\textrm{*}}$) plane for modelling the two regimes (for low and high-cycle fatigue), and $C$ giving the fatigue limit}
    \end{figure}

\noindent The strain life approach assumes that the fatigue lifetime of an element $N_{\textrm{R}}^{\textrm{*}}$ is governed by half the total strain range experienced by the element ($\frac{\Delta \varepsilon^*}{2}$) by a standard two-line model \cite{astm2015}:
\begin{equation}
    \frac{\Delta \varepsilon^*}{2} = g(N_{\textrm{R}}^{\textrm{*}}) = A (N_{\textrm{R}}^{\textrm{*}})^{-\alpha} + B (N_{\textrm{R}}^{\textrm{*}})^{-\beta} + C
\end{equation}

\noindent where $A$ and $\alpha$ correspond to the parameters of the high-cycle fatigue (HCF) line, $B$ and $\beta$ correspond to the parameters of the low-cycle fatigue (LCF) line and $C$ controls the fatigue limit, as illustrated in Fig. \ref{fig:Schema_proba}. As local plasticity may appear depending on the loading in the element, $\Delta \varepsilon^*$ is computed in the stabilized cycle with period $\mathcal{T}_c$, in the following manner \cite{karolczuk2005}:

    \begin{equation}\label{eq:Sigma_0}
        \Delta \varepsilon^* =
        \left[
        \max_{t\in\mathcal{T}_c}({n^*} \cdot \utilde{\varepsilon}^*(t) \cdot {n^*}) 
        - \min_{t\in\mathcal{T}_c}({n^*} \cdot \utilde{\varepsilon}^*(t) \cdot {n^*}) 
        \right]
    \end{equation}
    
    \begin{equation}
        \textrm{where} \quad n^* = \argmax_{n \cdot n = 1} (n \cdot \utilde{\sigma}^{\#*} \cdot n)
    \end{equation}
    
Here, $\utilde{\sigma}^{\#*}$ is element's stress tensor coming from a separate elastic computation and $n^*$ is the eigenvector associated to the highest eigenvalue of $\utilde{\sigma}^{\#*}$. The quadratic form (${n^*} \cdot \utilde{\varepsilon}^*(t) \cdot {n^*}$) of the element's total strain tensor $\utilde{\varepsilon}^*$ is obtained along this critical direction for the full loading sequence. The range of this quantity in the period of the stabilized stress-strain cycle $\mathcal{T}_c$ is computed. This procedure allows to account for the local multiaxiality that may cause a change in the critical direction from element to element. The underlying assumption is that local non-proportionality remains low \cite{Desmorat2002}. 

As $g(N_{\textrm{R}}^{\textrm{*}})$ is monotonic, there exists a function $g^{-1}\left(\frac{\Delta \varepsilon^*}{2}\right)$ defined as:

\begin{equation}
    g^{-1} : \, ]C, +\infty) \to ]0,+\infty)
\end{equation}

\paragraph{Probabilistic strain-life approach}\mbox{}\\

\noindent The fatigue lifetime model will now be made probabilistic. The fatigue lifetime is considered to be stochastic due to randomness in underlying total damage over all cycles that a specimen with volume $V^*$ may undergo before failing. The total damage $\mathcal{G}$ is modelled as a random variable, and the form of the associated probability distribution is considered to be a Weibull distribution that respects weakest link scaling \cite{Zok2017}:
\begin{equation}
    N_{\textrm{R}}^{\textrm{*}} \frac{1}{g^{-1}\left(\frac{\Delta \varepsilon^*}{2}\right)} = \mathcal{G}
\end{equation}
with the cumulative distribution function (CDF) of $\mathcal{G}$ being denoted as $F_{\mathcal{G}}(g)$:
\begin{equation}
    F_{\mathcal{G}}(g) = 1 - \left\{ \exp{\left(-\left\{\frac{g}{g_0}\right\}^m\right)} \right\}^{\frac{V^*}{V_0}}
\end{equation}
where $g_0 = \frac{1}{(\ln{2})^{1/m}}$, which implies that the median of this Weibull distribution is equal to 1 for a volume $V^* = V_0$. Here,  $m$ is the shape parameter of the Weibull distribution. By a change of variables, this leads to the following expression for the fatigue lifetime $N_{\textrm{R}}^{\textrm{*}}$ (with its CDF being denoted as $F_{N_{\textrm{R}}^{\textrm{*}}}(N;\Delta \varepsilon^*,V^*)$):
\begin{equation}\label{N_R^*}
    N_{\textrm{R}}^{\textrm{*}} \sim \mathcal{W}\left(\lambda =  \left\{g^{-1}\left(\frac{\Delta \varepsilon^*}{2}\right)\left\{\frac{1}{\ln{2}}\frac{V_0}{V^*}\right\}^{1/m}\right\},m\right)
\end{equation}
\begin{equation}\label{eq:FNRstar}
    F_{N_{\textrm{R}}^{\textrm{*}}}(N) = 1 - \exp{\left( - \left\{ \frac{N}{\lambda} \right\}^m \right)}
\end{equation}
where $\lambda$ is the scale parameter of the Weibull distribution describing the fatigue lifetime of the element. It is noted that a higher $V^*$ or a higher $\frac{\Delta \varepsilon^*}{2}$ (leading to lower $g^{-1}$) will cause the Weibull scale parameter $\lambda$ to decrease, thus causing the lifetime probability distribution to shift to the left.

The set of parameters introduced, including the strain-life parameters and the Weibull shape parameter, is thus:

\begin{equation}
    \mu = [m, A, B, \alpha, \beta, C]
\end{equation}

Henceforth, the dependence of $F_{N_{\textrm{R}}^{\textrm{*}}}$ on $\mu$ will be introduced as $F_{N_{\textrm{R}}^{\textrm{*}}}(N;\Delta \varepsilon^*,V^*,\mu)$. The probability density function (PDF) of fatigue lifetime of the element $*$ will be denoted as $f_{N_{\textrm{R}}^{\textrm{*}}}(N;\Delta \varepsilon^*,V^*,\mu)$.

\subsection{Statistical modeling of the pore distributions at the meso-scale}
\label{sec:MesoScale}
The meso-scale part of the model takes into account the uncertainty in fatigue lifetime of porous specimens due to varying pore distributions by synthetically generating them. In the context of our research, the precise positions and morphology of pores within the specimens for which we aim to predict lifespan are unknown and not sought, resulting in inherent uncertainty. If the positions and morphologies of the pores were known, the model could be identified directly from the exact tomographies of the tested specimens rather than from synthetic ones. This approach would then align with fatigue studies on deterministic geometries, such as notches \cite{Li2022,Lanning2003}. The distinct feature of our work lies in addressing the randomness of pore positions rather than deterministic pore distribution configurations, highlighting the unique challenge of random pore distribution in predicting material fatigue.

The synthetically generated porous specimens have the same geometry as the specimens used for fatigue testing (Fig. \ref{Fig:FatigueExperiments}(a)), but with the pores inspired by the computed tomography (Fig. \ref{fig:tomo}). These synthetically generated porous specimens thus resemble the traction specimens used for fatigue testing. The resulting meshes have linear elements with a piece-wise constant approximation of the stresses.

\paragraph{Synthetic pore field generator from CT scans}\mbox{}\\

\begin{figure}[htbp]
    \centering
    \includegraphics[width=\textwidth]{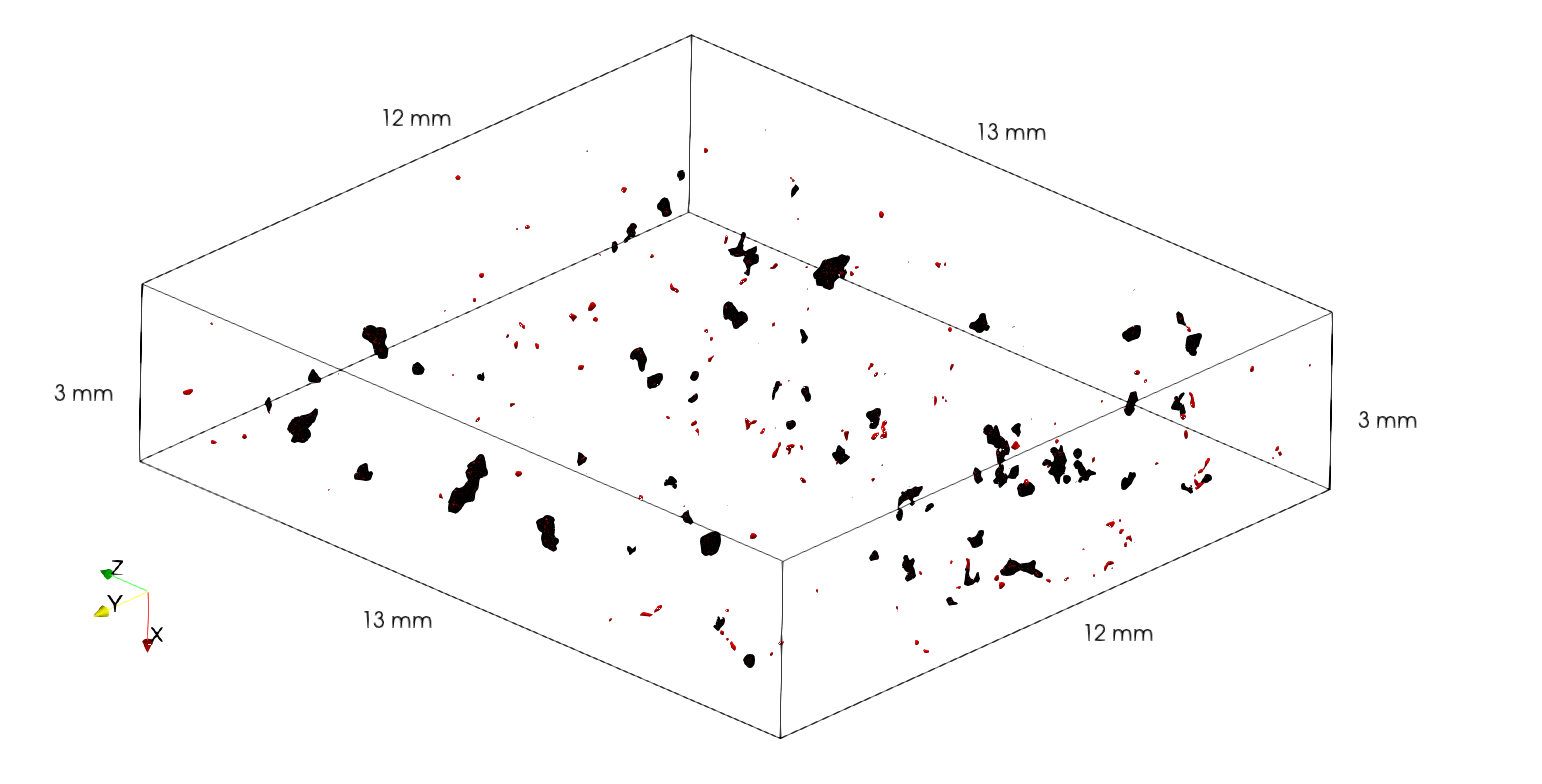}
    
    \caption{\label{Fig:PoresGen1_tomo_sizefilter} Segmented defects, with accepted defects (black) and rejected defects (red) based on their size (acceptance for effective radius of a sphere of same volume as the defect $\geq$ $50$ $\mu$m)} 
\end{figure}

\noindent The tomographic scan of the porous specimen used for getting information on pores was shown in Fig. \ref{fig:tomo}. A volume threshold is applied on this scan to exclude small pores. A pore is accepted if the radius of a sphere with the same volume as the pore is greater than $50$ $\mu m$. This is justified by other authors, who show that in presence of large sized porosity, smaller pores have a negligible effect in fatigue life and that in the high cycle fatigue regime, cracks systematically originate from large pores \cite{Kitagawa1976, VietDucLe2015, ElKhoukhi2019}. This size filter step is illustrated in Fig. \ref{Fig:PoresGen1_tomo_sizefilter}. 

\begin{figure}[htbp]
    \centering
    \begin{subfigure}[b]{0.48\textwidth}
        \includegraphics[width=\textwidth]{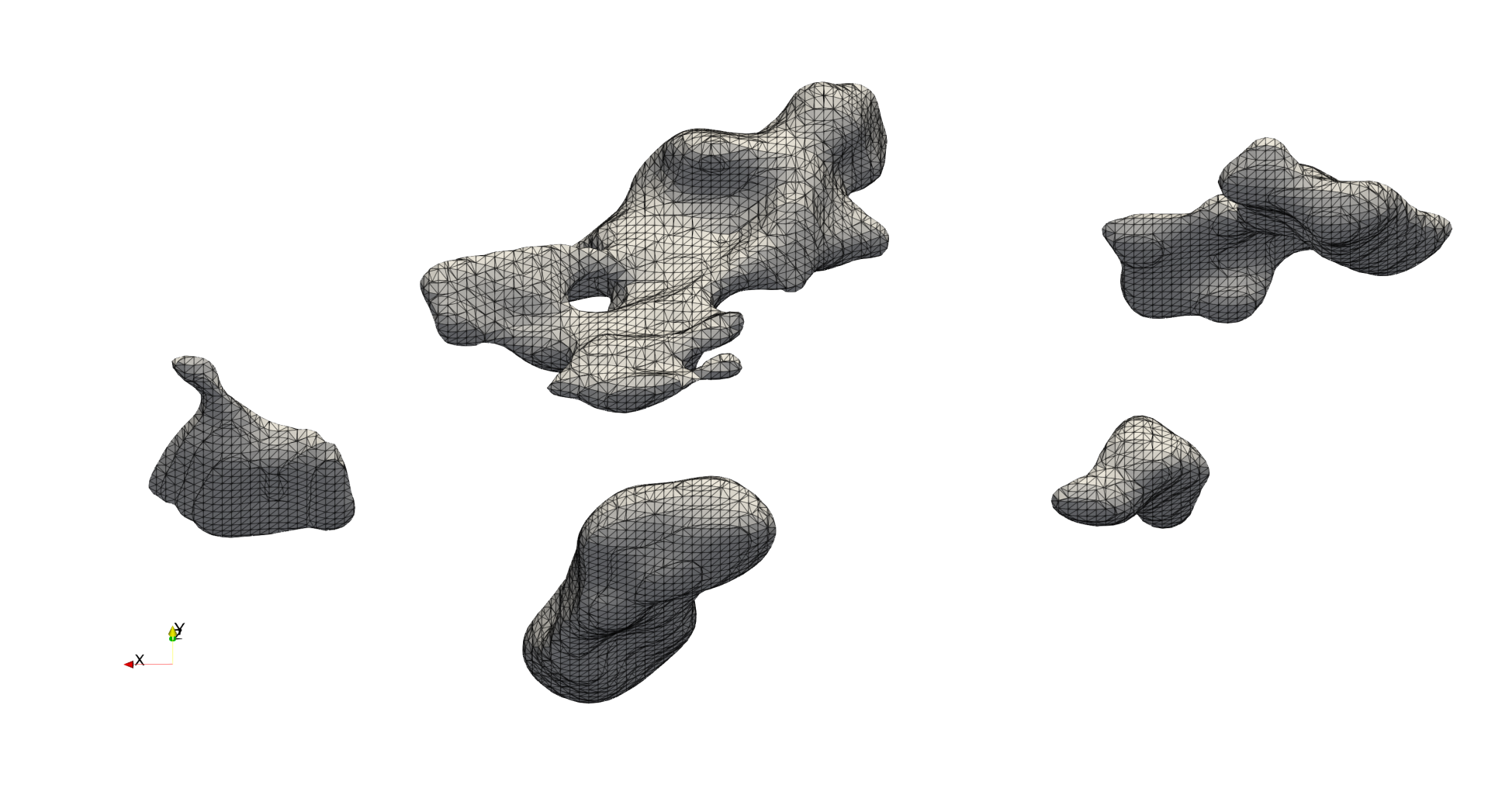}
        \caption{Before smoothing}
    \end{subfigure}\hfill
    \begin{subfigure}[b]{0.48\textwidth}
        \includegraphics[width=\textwidth]{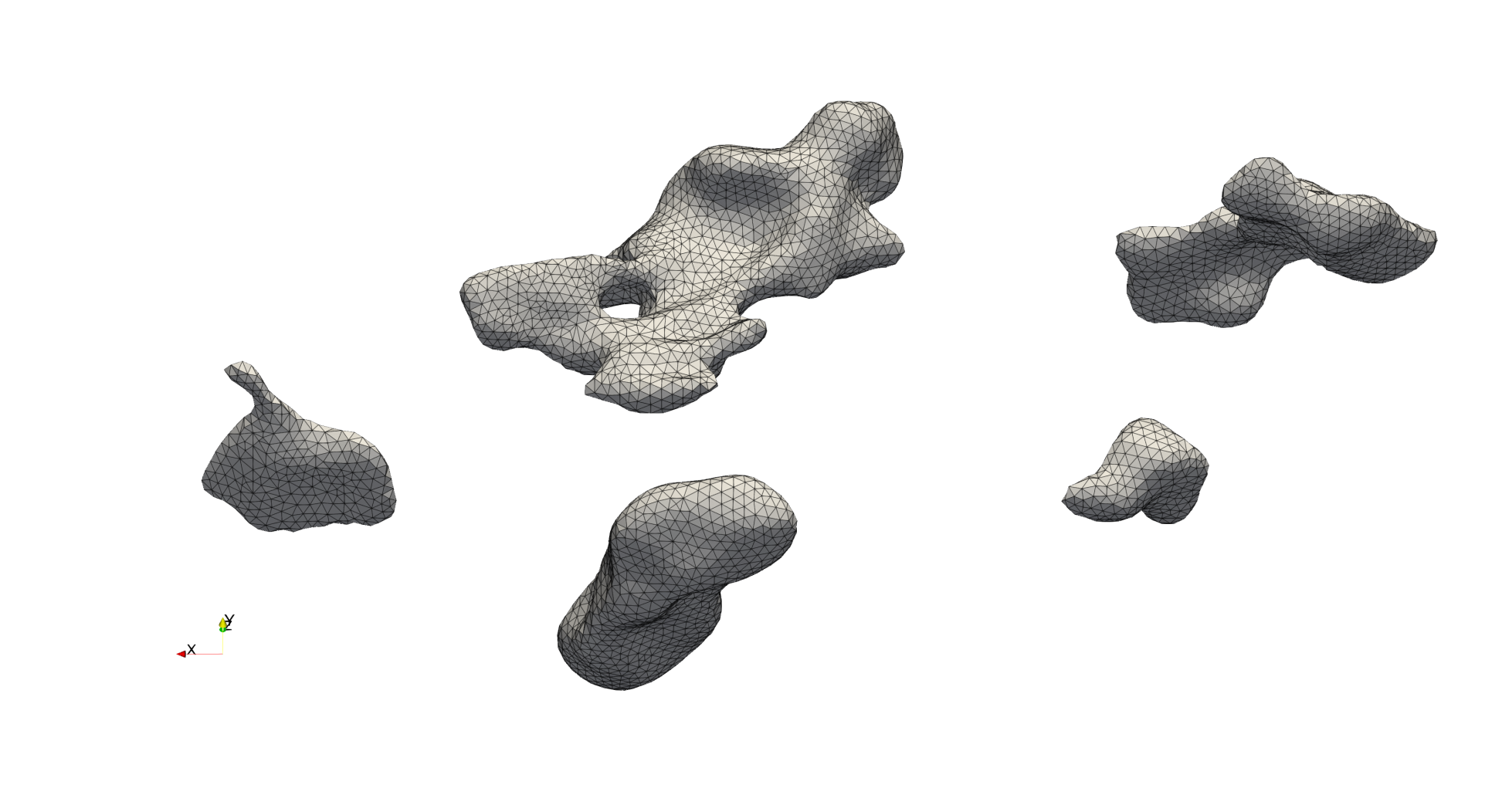}
        \caption{After smoothing}
    \end{subfigure}
    
    \caption{\label{Fig:PoresGen2_remeshporesurfaces} Curvature based surface re-meshing of the defect surfaces} 
\end{figure}

Surface meshes of the remaining pores after the size-filtering step were obtained using smoothing based on a criterion on sharp angles between surfaces of the pores - the original surfaces from the tomography and the remeshed surfaces are shown in Fig. \ref{Fig:PoresGen2_remeshporesurfaces}. This smoothing is done to facilitate finite element meshing.

\begin{figure}[htbp]
    \centering
    \begin{subfigure}[b]{0.37\textwidth}
        \includegraphics[width=\textwidth]{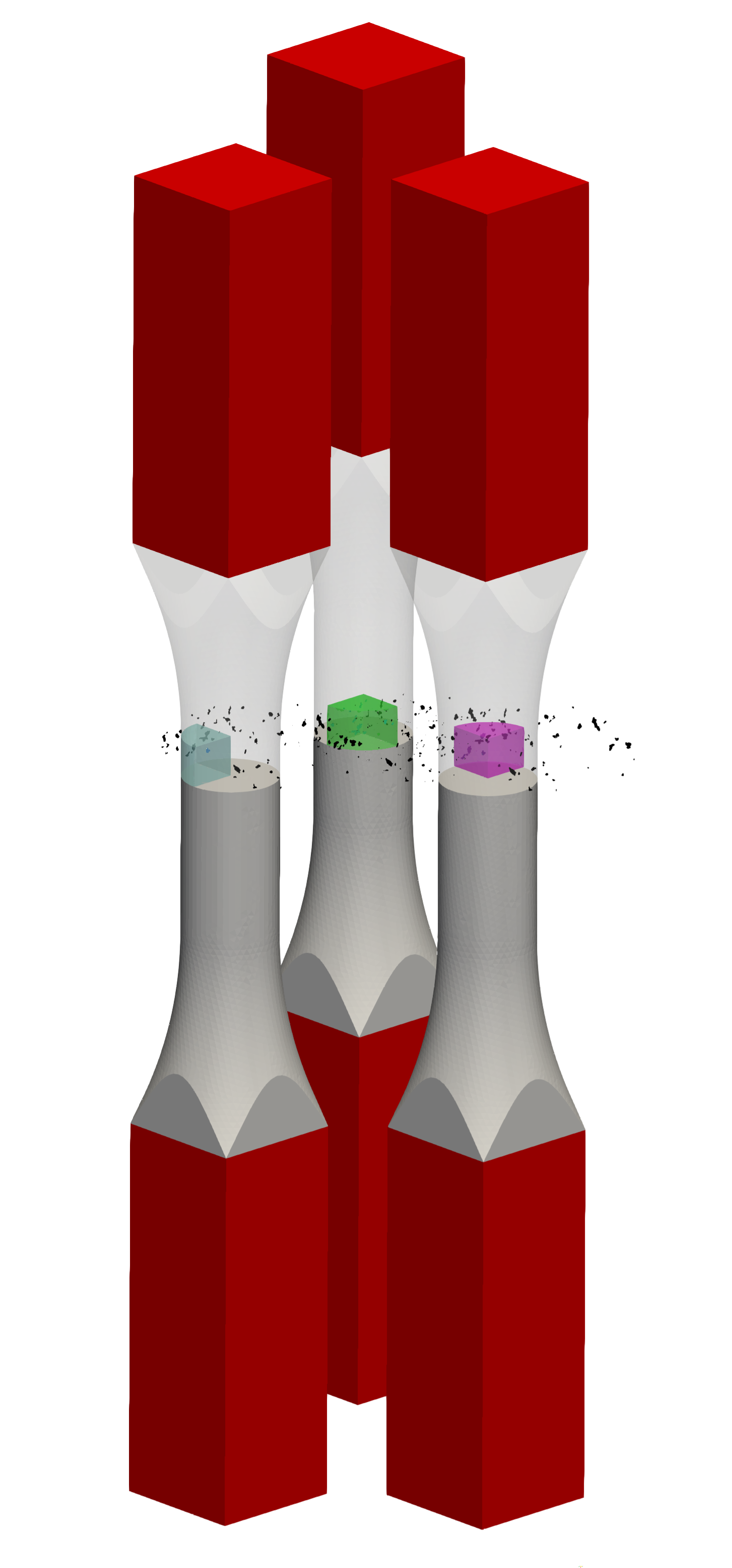}
        \caption{}
    \end{subfigure}\hfill
    \begin{subfigure}[b]{0.61\textwidth}
        \includegraphics[width=\textwidth]{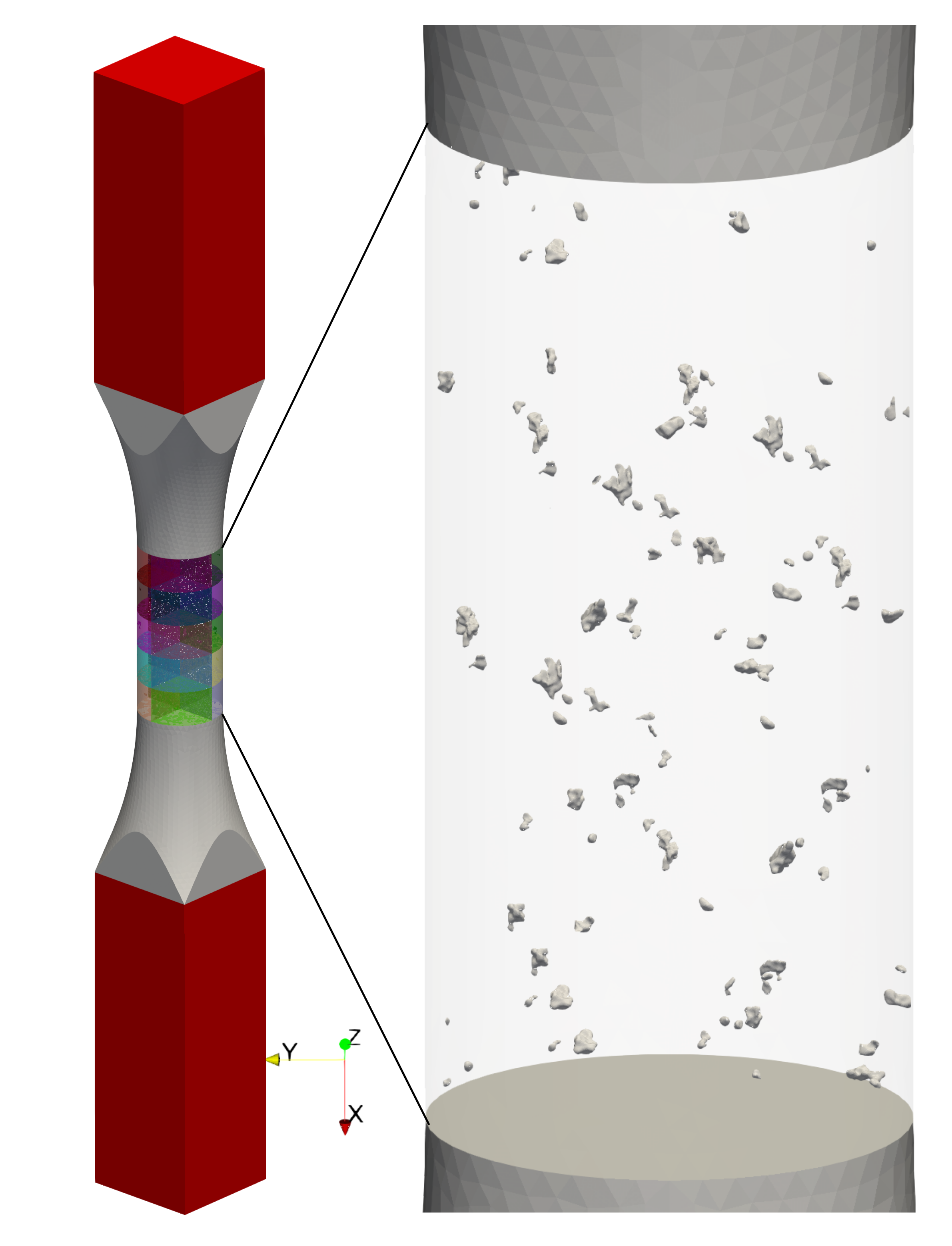}
        \caption{}
    \end{subfigure}
    
    \caption{\label{Fig:PoresGen3_randomcutsassemble} (a) Random intersection of the defects with the geometric CAD representation of the fatigue specimens, with the defects being meshed into the specimen if it falls within the colored sub-volumes (process repeated 100 times) (b) Assembly of randomly selected porous sub-volumes to obtain a synthetically generated porous specimen with defects in its gauge section} 
\end{figure}

A set of synthetically generated porous specimens (with varying pore distributions) was created by using an algorithm based on this set of size-filtered and smoothed pores. 
The volume of the gauge section of the cylindrical fatigue specimens, being around $593$ $\textrm{mm}^{3}$, contains a very large number of pores (and therefore elements). A splitting of the computational domain into sub-volumes is therefore carried out to avoid memory limitations. The following steps were done to synthetically generate porous specimens:

    \begin{description}
       \item[1.] A CAD geometry of the cylindrical specimen used for fatigue experiments was created using GMSH \cite{gmsh2020}.
       \item[2.] The gauge section of this CAD geometry was randomly superposed with the representative volume of pores, and the set of pores intersecting with the geometry were retained for meshing.
       \item[3.] A meshing operation is carried out in the base material between the surfaces of the specimens and the pores, with adequate refinement of the mesh close to the pores, using GMSH. Thus, only one region or sub-volume of the gauge section contains pores, henceforth referred to as a 'porous sub-volume'.
       \item[4.] The previous two steps were carried out 100 times, to get 100 such specimens each with a different region of pores (illustrated in Fig. \ref{Fig:PoresGen3_randomcutsassemble} (a)).
       \item[5.] Elements in the 'porous subvolumes' of these 100 specimens are extracted. A random combination of porous sub-volumes that equal the volume of the gauge section of the cylindrical specimen makes up one synthetically generated porous specimen (illustrated in Fig. \ref{Fig:PoresGen3_randomcutsassemble} (b)).
    \end{description}
    
    %
    

\paragraph{Elasto-plastic computations}\mbox{}\\

\noindent The multi-scale probabilistic fatigue model developed requires the computation of the local criterion ($\Delta \varepsilon^*$) for all the finite elements comprising the porous specimens due to applied stress amplitude on the specimen ($\Sigma_a$) for computation of the specimen's fatigue lifetime probability distribution. As the presence of pores in the specimen causes high stress concentration factors leading to local yielding, we consider a Chaboche-type plastic material behavior, consisting of non-linear isotropic and kinematic hardening. The equations are recalled in table \ref{tab:MaterialModel}. 


\begin{table}[h!tbp]
    \begin{center}
    \caption{Set of equations used for the mechanical behavior - with $\protect\utilde{\sigma}$ being the stress tensor, $\protect\utilde{\sigma}_d$ being the deviatoric stress tensor, $\mathcal{C}$ the stiffness tensor, $\protect\utilde{\varepsilon}$ the total strain, $\protect\utilde{\varepsilon}^{e}$ and $\protect\utilde{\varepsilon}^{p}$ the elastic and plastic strain tensors respectively, $\mathcal{J}$ the von Mises yield function, $p$ the cumulative plastic strain, and $\protect\utilde{\varepsilon}_d^p$ the deviatoric plastic strain \cite{Chaboche1989}}
    \label{tab:MaterialModel}

    \begin{tabular}{ll}
    Elasticity & $\utilde{\sigma} = \mathcal{C}:(\utilde{\varepsilon} - \utilde{\varepsilon}^p)$\\
    &\\
    \hline
    &\\
    Strain partitioning & $\utilde{\varepsilon}={\utilde{\varepsilon}}^{e}+\utilde{\varepsilon}^{p}$\\
    &\\
    \hline
    &\\
    Yield function & $f_y\textrm(\utilde{\sigma};\utilde{X},p) = \mathcal{J} \left( {\utilde{\sigma}} - {\utilde{X}} \right) - \sigma_y - R(p)$\\
    &\\
    \hline
    &\\
    Evolution of yield function & ${f_y} \dot{p} = 0 \quad \textrm{and} \quad {f_y}{\leq 0}$ \\
    &\\
    \hline
    &\\
    Isotropic Hardening & $R(p)=Q\big(1-\exp{(-b p)}\big)$\\
    &\\
    \hline
    &\\
    Kinematic Hardening & $\dot{\utilde{X}} = \frac{2}{3}C\dot{\utilde{\varepsilon}}^p - D\utilde{X}\dot{p}$\\
    &\\
    \hline
    &\\
    Flow rule & $\dot{\utilde{\varepsilon}}^p = \dot{p} \left(  \dfrac{3}{2} \dfrac{\utilde{\sigma}_d -\utilde{X}\;}{\;\mathcal{J}(\utilde{\sigma}_d - \utilde{X})} \right)$\\
    &\\
    \hline
    &\\
    Cumulative plastic strain  & $\dot{p} = \sqrt{\frac{2}{3}\dot{\utilde{\varepsilon}}_d^p:\dot{\utilde{\varepsilon}}_d^p}$\\
    \end{tabular}
    
    \end{center}
\end{table}%

However, time integration of non-linear material behavior is time-consuming and resource heavy, especially for large meshes. Therefore, a plastic corrector is used, which allows rapid approximation/post-processing of the full-field elasto-plastic response of specimens for any proportional loading sequence when given elastic FEA results at one timestep \cite{Palchoudhary2024}. An example of the approximation quality of the criterion with respect to a reference computation is shown in \hyperref[appendixB]{Appendix B} on a subvolume of pores.

The following steps highlight the process:
    
    \begin{description}
    \item[1.] Elasto-static computations are performed on the sub-volumes of the synthetically generated specimens with the FEniCS \cite{fenics2015} software.
    \item[2.] The approximate elasto-plastic solution is reconstructed in a space-time domain, for several nominal loading levels, by a plastic corrector that approximates elasto-plasticity for any form of proportional applied loading via post-processing of elasto-static finite element results \cite{Palchoudhary2024}
    \item[3.] The local total strain amplitude ($\Delta \varepsilon^*$) is extracted for the stabilized cycle for all the elements comprising the gauge section of the generated specimens and for all the nominal loading levels.
    \end{description}






\subsection{Probabilistic lifetime model at the structure level : weakest link assumption}\label{sec:structurelevel}
A structure with a heterogeneous stress distribution, for example, a structure with pores that are explicitly meshed, or a structure with a notch, will have different values of the criterion ($\Delta\varepsilon^*$) in different elements of the structure for a given applied stress amplitude $\Sigma_a$. We formally define a function $\Delta\varepsilon^*(\Sigma_a) : \Sigma_a \mapsto \Delta \varepsilon^*$ that maps the load applied to the structure as boundary condition to the fatigue criterion in element $*$. This function is computed by FEA. The probabilistic strain-life model developed in section \ref{sec:LocalModels} can subsequently be applied to all the elements of the FE mesh. The CDF of fatigue lifetime of a given element in the geometrically heterogeneous case is thus obtained as $F_{N_{\textrm{R}}^{\textrm{*}}}(N;\Delta \varepsilon^*(\Sigma_a),V^*,\mu)$, as given in equation \eqref{eq:FNRstar}.

A weakest link hypothesis is used to obtain the fatigue lifetime density of the full structure. In this model, the failure of the structure is given by the failure of the first element of the FE mesh. The weakest link hypothesis assumes that the failure of the elements are statistically independent events.
The probability of the survival of a structure  $s$ (modelled by the random variable $N_{\textrm{R}}^{\textrm{s}}$) is given by:
    \begin{equation}
        \textrm{Prob}(N_{\textrm{R}}^{\textrm{s}} \geq N) = \prod_{* \in \mathcal{E}} \textrm{Prob}(N_{\textrm{R}}^{\textrm{*}} \geq N)
    \end{equation}
where $\mathcal{E}$ is the set of elements comprising the structure. The analytical form of the probability distribution of failure of the structure is the same as that of a 2-parameter Weibull distribution with Weibull scale parameter denoted as $\lambda^s$ and given by (the proof is given in \hyperref[appendixC]{Appendix C}):
\begin{equation}\label{N_R^s}
    N_{\textrm{R}}^{\textrm{s}} \sim \mathcal{W} \left( \lambda^s = \frac{1}{\left\{\sum\limits_{*\in \mathcal{E}}  \frac{1}{\lambda^{m}}\right\}^{1/m}}, m \right)
\end{equation}
The CDF of fatigue lifetime of the structure is:
\begin{equation}\label{eq:FNRstruct}
    F_{N_{\textrm{R}}^{\textrm{s}}}(N) = 1 - \exp{\left( - \left\{ \frac{N}{\lambda^s} \right\}^m \right)}
\end{equation}

The computation of $\lambda^s$ is dependent on $\Sigma_a$ as the computation of $\lambda$ needs the function $\Delta\varepsilon^*(\Sigma_a)$. It is also dependent on the parameters of the fatigue model. Henceforth, the cumulative distribution function (CDF) and the probability density function (PDF) of the specimen's fatigue lifetime are denoted as $F_{N_{\textrm{R}}^{\textrm{s}}}(N;\Sigma_a,\mu)$ and $f_{N_{\textrm{R}}^{\textrm{s}}}(N; \Sigma_a,\mu)$, respectively.
 
\section{Identification of the probabilistic lifetime model from experimental data by maximum likelihood estimation\label{sec:Identification}}
Identification of the probabilistic lifetime model can be done on experimental fatigue lifetime data. In the following we will expose how this can be done in the case of (i) specimens with homogeneous stress distribution in the gauge section (ii) structures with heterogeneous stress distributions and (iii) structures with heterogeneous stress distributions that arise due to pores whose exact geometries and locations are unknown.

\subsection{Specimens with homogeneous stress distributions}
\label{sec:ParamsIdentification_Homogenous}
We recall from section \ref{sec:structurelevel} that $f_{N_{\textrm{R}}^{\textrm{s}}}$ is the PDF of the fatigue lifetime of a structure with a heterogeneous stress distribution. For specimens with a homogeneous stress distribution in their gauge section, i.e. without stress concentrations, the distribution of fatigue lifetime may be obtained without FEA.

The standard log-likelihood function is defined as \cite{Pollak2009, Lee2023}:
    \begin{equation}
        \ln \mathcal{L}(\mu) = \sum_{i\in \mathcal{H}} \ln f_{N_{\textrm{R}}^{\textrm{s}}}(N^{i}; \Sigma_a^{i}, \mu)
    \end{equation}
    
where $\mathcal{H}$ is the index set of all the specimens with uniform stress distribution tested for fatigue failure, $N^i$ is the number of cycles to failure for an applied stress amplitude $\Sigma_a^{i}$, and $\mu$ is the set of parameters of the fatigue model. A Nelder-Mead algorithm is used for optimisation:

    \begin{equation}
        \mu =
        \argmax_{\tilde{\mu}} {\left( \sum_{i\in \mathcal{H}} \ln f_{N_{\textrm{R}}^{\textrm{s}}}(N^{i};\Sigma_a^{i},\tilde{\mu})  \right)}
    \end{equation}

Furthermore, experimental high-cycle fatigue data-sets usually include some data points corresponding to specimens that did not break before reaching the maximum number of cycles allowed $N_{\textrm{max}}$. The latter, termed "run-outs", have a finite probability of failing for some number of cycles greater than $N_{\textrm{max}}$. This probability is expressed as:

\begin{equation}
    P(N>N_\textrm{max}) = 1 - F_{N_{\textrm{R}}^{\textrm{s}}}(N_{\textrm{max}}; \Sigma_a, \mu)
\end{equation}

\subsection{Structures with heterogeneous stress distributions}
\label{sec:ParamsIdentification_HeterogenousDeterministic}
We remind the reader of the PDF of the fatigue lifetime of structures with a heterogeneous stress distribution $f_{N_{\textrm{R}}^{\textrm{s}}}$ from section \ref{sec:structurelevel}. 
Structures with multiple distinct geometries can be taken into account here, provided that the exact geometries are known. This could, for example, be structures with different notch geometries, or porous specimens whose exact pore distribution is known for each of the specimens tested. In this case, $f^i_{N_{\textrm{R}}^{\textrm{s}}}$ for each distinct geometry $i$ with fatigue lifetime $N^i$ at load $\Sigma_a^i$ is needed for identification of the model.
For identification of the parameters of such a model, previous studies have used Bayesian calibration \cite{Liu2020} or manual adjustment of parameters \cite{Karolczuk2013}, especially on notched geometries. We use the maximum likelihood method:

    \begin{equation}
        \mu =
        \argmax_{\tilde{\mu}} {\left( \sum_{i\in \mathcal{I}} \ln f^i_{N_{\textrm{R}}^{\textrm{s}}}(N^{i}; \tilde{\mu}, \Sigma_a^{i})  \right)}
    \end{equation}


where $\mathcal{I}$ is the set of all the structures with heterogeneous stress distributions tested for fatigue failure, $N^i$ is the number of cycles to failure at an applied stress amplitude $\Sigma_a^{i}$ and $\tilde{\mu}$ is the set of parameters of the fatigue model. A Nelder-Mead process is used for optimisation, outputting parameters $\mu$ of the lifetime model that are the most probable given the experimental fatigue results, by maximizing the likelihood function.

To illustrate the computational complexity involved, we note that the computing the Weibull scale parameter of the structure, $\lambda^s$, requires summing the Weibull scale parameters, $\lambda$, of all elements in the FE mesh (equation \eqref{N_R^s}). This result is then used to compute $f_{N_{\textrm{R}}^{\textrm{s}}}$. The scale parameters of the elements in turn require the computation of the inverse function $g^{-1}\left(\frac{\Delta \varepsilon^*}{2}\right)$ for each element (equation \eqref{N_R^*}). The criterion $\Delta \varepsilon^*$ is obtained for all the elements by one elasto-static finite element computation combined with the Neuber-type element-wise plastic correction algorithm, and does not need to be re-computed during the optimisation process. As the inverse function $g^{-1}\left(\frac{\Delta \varepsilon^*}{2}\right)$ depends on some of the parameters of the fatigue model ($A$, $B$, $\alpha$ and $\beta$), it needs to be re-computed for all the elements at every iteration of the optimisation process.

\subsection{Structures with pores whose exact distribution is unknown}
\label{sec:ParamsIdentification_Heterogenous}
We now treat the case of structures with a fixed macroscopic geometry, that contains pores whose exact geometries and locations are unknown. In this case, no observations of the pore distribution in these structures were realized before fatigue testing, which makes the methodology presented in section \ref{sec:ParamsIdentification_HeterogenousDeterministic}, with deterministic geometries, inapplicable. The purpose is to identify a lifetime distribution on fatigue data of structures accounting for all possible pore geometries and localization.
This is done by taking an expectation over all the possible configurations of pore distributions in the structures. For example, if $\mathcal{J}$ is the set of all the porous specimens of a given macroscopic geometry tested for fatigue failure, and $\mathcal{K}$ is the set of all possible synthetically generated configurations of porous specimens, the parameters of the lifetime distribution are obtained using the following optimisation statement:

    \begin{equation}
        \mu =
        \argmax_{\tilde{\mu}} {\left( \sum_{i\in \mathcal{J}} \ln \left( \mathop{\mathbb{E}}_{k\in \mathcal{K}} f^k_{N_{\textrm{R}}^{\textrm{s}}}(N^{i};\Sigma_a^{i},\tilde{\mu})  \right)\right)}
    \end{equation}

    Where $N^i$ is the number of cycles to failure at an applied stress amplitude $\Sigma_a^{i}$. The expectation is taken over all realisations of the specimen with random meso-scale pore distributions. A Monte Carlo method is used to approximate this expectation over a finite number of synthetically generated porous specimens (denoted as the set $\tilde{K}$):

    \begin{equation}\label{eq:approxexpectation}
        \mu =
        \argmax_{\tilde{\mu}}  {\left( \sum_{i\in \mathcal{J}} \ln \left(  \frac{1}{n_\text{k}}\sum_{k\in \tilde{K}} f^k_{N_{\textrm{R}}^{\textrm{s}}}(N^{i};\Sigma_a^{i},\tilde{\mu})  \right)\right)}
    \end{equation}
    
    The computational cost of this optimisation is broken down here. Elasto-static finite element computations are required to be performed once for $n_\text{k}$ synthetically generated porous specimens. The approximation of the criterion $\Delta \varepsilon^*$ in all the elements of all the synthetic specimens, being a post-processing, takes virtually no time. As explained previously (section \ref{sec:ParamsIdentification_HeterogenousDeterministic}), the inverse function $g^{-1}\left(\frac{\Delta \varepsilon^*}{2}\right)$ needs to be computed for all elements of $n_\textrm{k}$ synthetic specimens, at every iteration of the optimisation process, for computation of the approximated expectation of the fatigue lifetime.
    
    The algorithm explained in Algorithm  \ref{alg:log_likelihood} explains how the log-likelihood is computed for this case, including taking run-outs into account.

\begin{algorithm}[h!tbp]
\caption{Log-Likelihood (LL) Computation}\label{alg:log_likelihood}
\begin{algorithmic}[1]
\State \textbf{Ensure:} \Comment{$N$, $\lambda$ are separated into $N_{\textrm{finite}}$, $\lambda_{\textrm{finite}}$ (where $N<N_{\textrm{max}}$) and $N_{\textrm{inf}}$, $\lambda_{\textrm{inf}}$ (where $N\geq  N_{\textrm{max}}$)}
\State $\textrm{LL} \gets 0$
\State $i \gets 0$
\While{$i < |N_{\textrm{finite}}|$} \label{line:while_not_inf}
    \State $s \gets 0$
    \State $k \gets 0$
    \While{$k < n_\text{k}$}
        \State $s \gets s + f_{N_{\textrm{R}}^{\textrm{s}}}(N_{\textrm{finite}}[i], \lambda_{\textrm{finite}}[k][i], m)$
        \State $k \gets k + 1$
    \EndWhile
    \State $s^{\textrm{avg}} \gets \frac{s}{n_\text{k}}$
    \State $\textrm{LL} \gets \textrm{LL} + \log(s^{\textrm{avg}} + 10^{-10})$
    \State $i \gets i + 1$
\EndWhile
\State
\State $i \gets 0$
\While{$i < |N_{\textrm{inf}}|$}
    \State $m_{\textrm{inf}} \gets 0$
    \State $k \gets 0$
    \While{$k < n_\text{k}$}
        \State $m_{\textrm{inf}} \gets m_{\textrm{inf}} + (1 - F_{N_{\textrm{R}}^{\textrm{s}}}(N_{\textrm{max}}, \lambda_{\textrm{inf}}[k][i], m))$
        \State $k \gets k + 1$
    \EndWhile
    \State $m_{\textrm{inf}}^{\textrm{avg}} \gets \frac{m_{\textrm{inf}}}{n_\text{k}}$
    \State $\textrm{LL} \gets \textrm{LL} + \log(m_{\textrm{inf}}^{\textrm{avg}} +  10^{-10})$
    \State $i \gets i + 1$
\EndWhile
\end{algorithmic}
\end{algorithm}


\section{Calibration and validation on experimental data\label{sec:Results}}
In this section, the behavior of the optimisation process presented in section \ref{sec:Identification} will be studied. The complexity of the fatigue model, i.e. number of required model parameters to fit lifetime data, will also be investigated. The data used is the previously presented experimental high-cycle fatigue data on non-porous and porous specimens (Figure \ref{Fig:FatigueExperiments}). Finally, we will try to identify a fatigue model that is transferable between these two types of pore populations.


The set $\tilde{K}$ is created with a total of 45 synthetically generated porous specimens \footnote{The size of the set $\tilde{K}$ is limited to 45 as meshing and elasto-static finite element computations are required to be performed once for $n_\text{k}$ synthetically generated porous specimens. This step can become computationally expensive as the meshing and elastic computation time grows: each synthetic porous specimen contains between 1.5 to 3 million quadrature points} using the method detailed in section \ref{sec:MesoScale}.

\begin{figure}[h!tbp]
    \centering
    \begin{tikzpicture}
        \node[anchor=south west,inner sep=0] (image) at (0,0) {\includegraphics[width=0.9\textwidth]{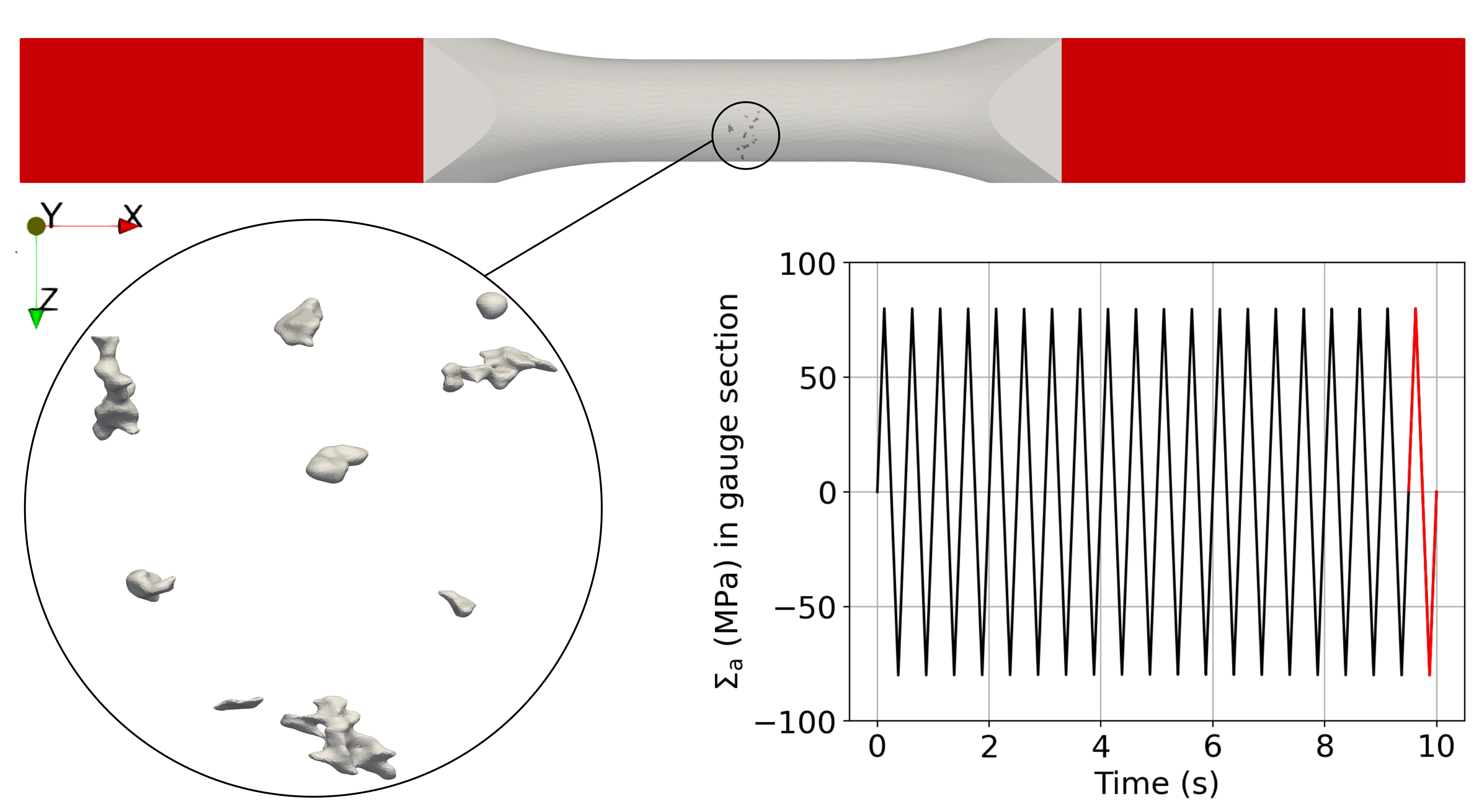}};
        \begin{scope}[x={(image.south east)},y={(image.north west)}]
            \draw[black, ultra thick, <-] (-0.026, 0.86) -- (0.014, 0.86); 
            \node at (-0.004, 0.91) {${\underbar{u}}_a$}; 
            \draw[black, ultra thick, ->] (0.99, 0.86) -- (1.03, 0.86); 
            \node at (1.01, 0.91) {${\underbar{u}}_a$}; 
        \end{scope}
    \end{tikzpicture}
    \caption{\label{fig:BCs_SpecimenRealPores} Boundary conditions (shown in red) for a specimen containing a sub-volume of tomography-informed pores, showing where cyclic displacement ${\underbar{u}}_a$ is applied, to obtain the different levels of loading desired. In the gauge section of the specimen (away from pores), the von Mises stress reaches the desired stress amplitude levels (here, as an example, $\Sigma_a = 80$ MPa). The 20\textsuperscript{th} cycle (shown in red) is chosen for the computation of $\frac{\Delta \varepsilon^*}{2}$.}
\end{figure}
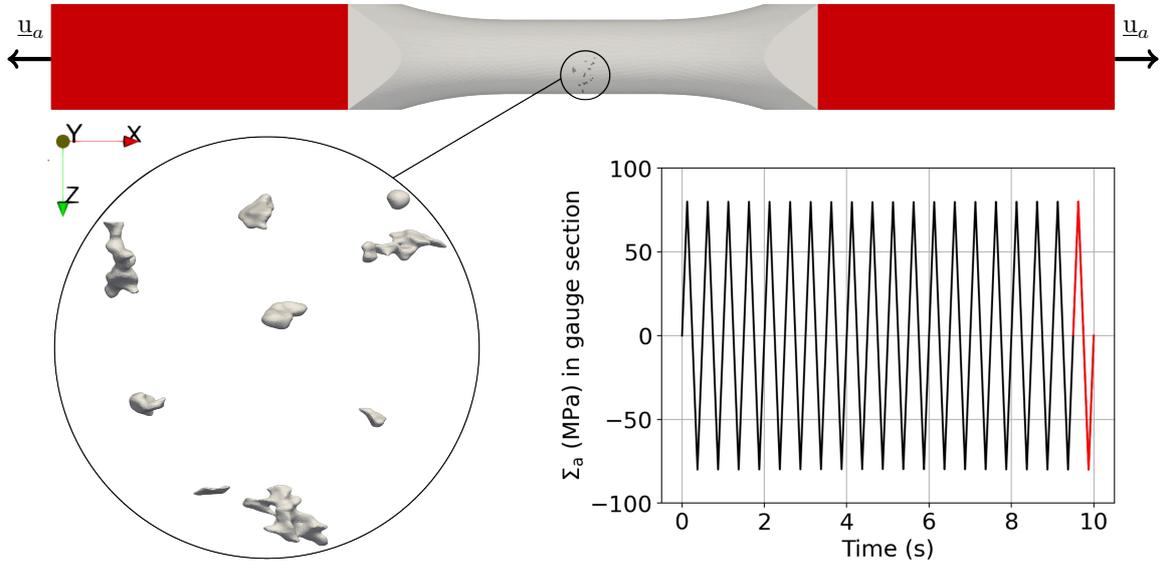

\begin{figure}[h!tbp]
    \centering
    \begin{subfigure}[b]{0.8\textwidth}
        \centering
        \includegraphics[width=\textwidth]{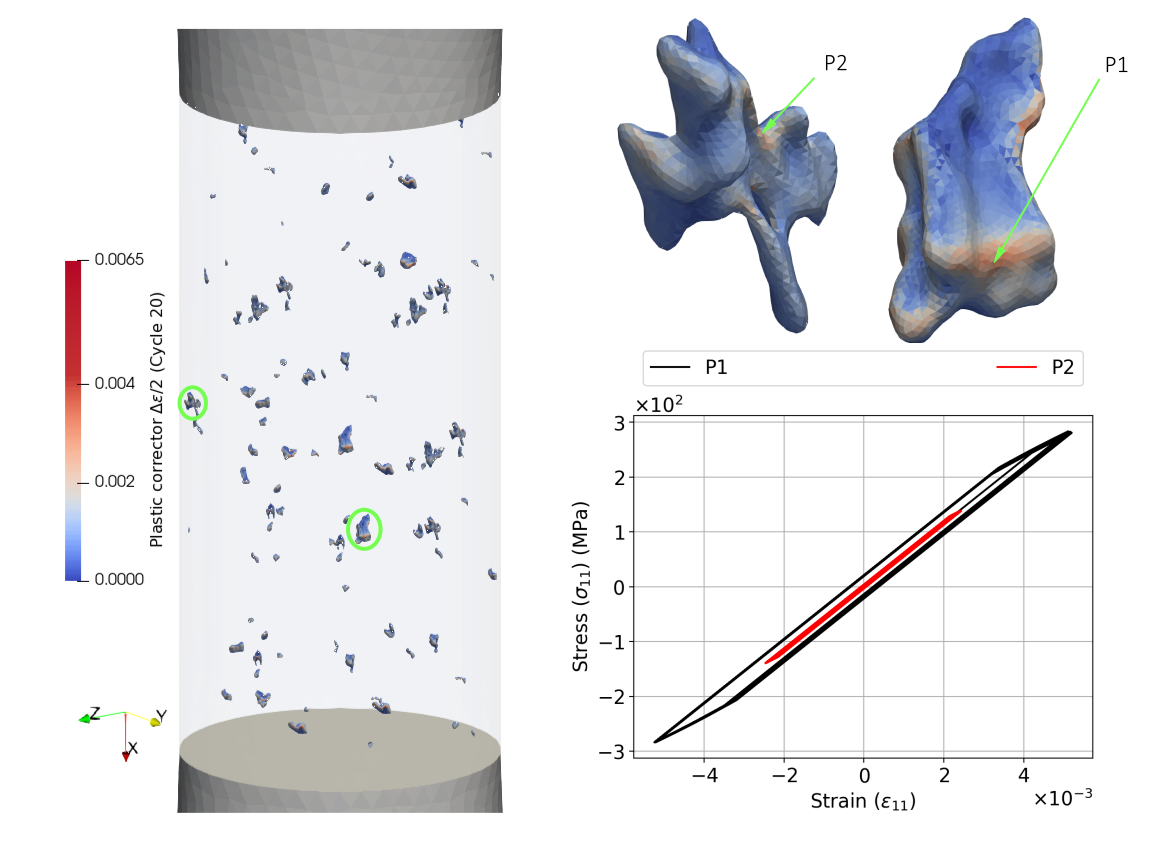}
        \caption{Test specimen 1. $\Sigma_a =$ 80 MPa}
        \label{fig:sp1_criticalpores}
    \end{subfigure}
    \vspace{1em} 
    \begin{subfigure}[b]{0.8\textwidth}
        \centering
        \includegraphics[width=\textwidth]{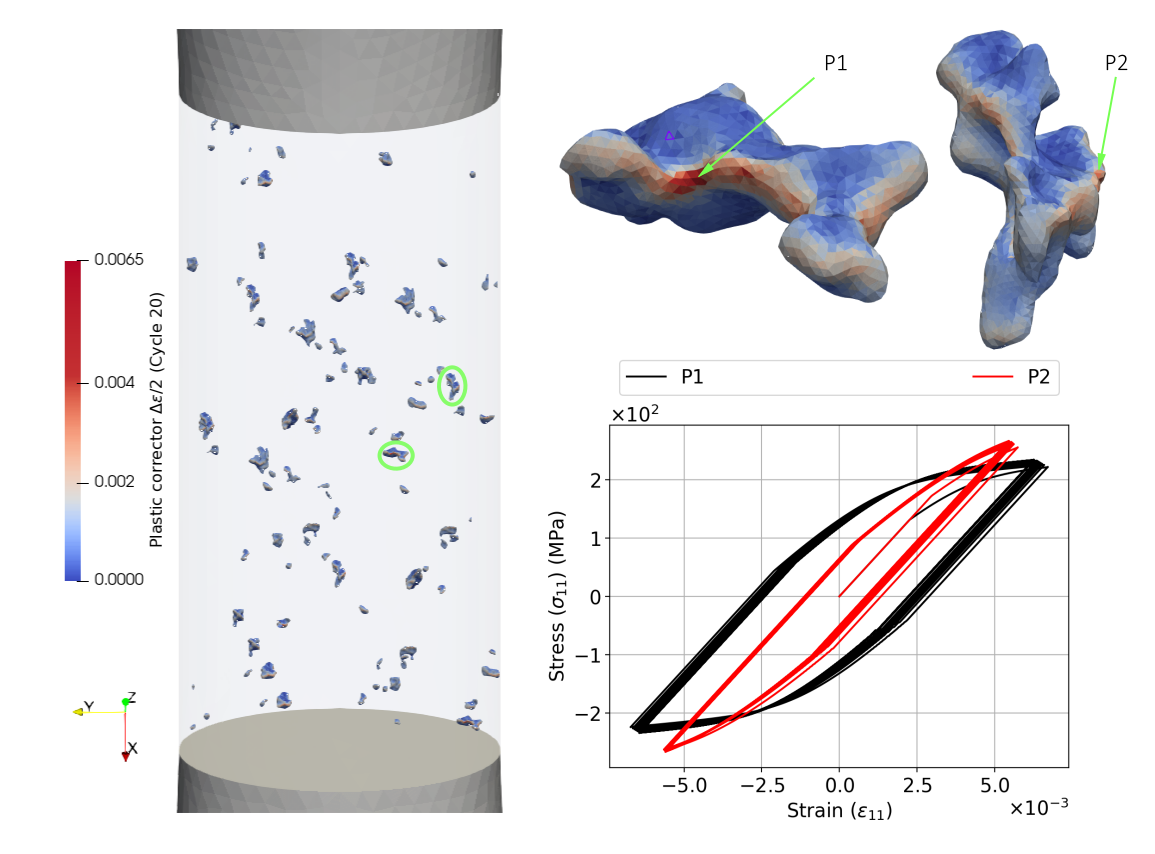}
        \caption{Test specimen 10. $\Sigma_a =$ 80 MPa}
        \label{fig:sp10_criticalpores}
    \end{subfigure}
    
    \caption{ Local criterion ($\frac{\Delta \varepsilon}{2}$) field, in two different synthetically generated porous specimens, named test specimens 1 \& 10. The criterion around the most critical defects and the axial stress-strain components at the highest points are shown. Refer to \hyperref[appendixB]{Appendix B} for an analysis of the accuracy of the fields computed using the plastic correction algorithm compared to that obtained with a full elasto-plastic finite element analysis.}
    \label{Fig:VaryingcriterionDistributionFullField}
\end{figure}

Elasto-plastic computations (elasto-static FEA followed by Neuber-type elasto-plastic correction, as explained in section \ref{sec:MesoScale}) were carried out on each of the synthetically generated porous specimens, for 9 nominal loading levels ($\Sigma_a$ ranging from 20 to 100 MPa). We use the material parameters identified by \cite{VietDucLe2016} for the considered alloy. The parameters of the elasto-plastic model are recalled in table \ref{tab:PlasticityModelParametersAlu}. The boundary conditions (shown in Figure \ref{fig:BCs_SpecimenRealPores}) replicates the experimental conditions under which the high-cycle fatigue data were obtained. Each loading sequence consisted of 20 symmetric tension-compression cycles ($R=-1$), to allow for stabilization of the elasto-plastic response. The fatigue criterion $\Delta \varepsilon^*$ in the stabilized (20\textsuperscript{th}) cycle is extracted.
An example of the criterion computation in two synthetically generated specimens for one of the nominal loading levels ($\Sigma_a = 80$ MPa), is shown in Fig. \ref{Fig:VaryingcriterionDistributionFullField}. In this figure, the axial stress-strain histories for some integration points having high values of $\Delta \varepsilon^*$ are shown, to highlight the stabilization of the stress-strain cycles. 

\begin{table}[h!tbp]
\caption{Parameters of the elasto-plastic model \cite{VietDucLe2016}}
\label{tab:PlasticityModelParametersAlu}
\centering
\begin{tabular}[t]{llllllll}
\hline\noalign{\smallskip}
Parameter & E & $\sigma_y$ & b & Q & C & D  \\
          & (MPa) & (MPa) &   & (MPa) & (MPa) &  \\
\noalign{\smallskip}\hline\noalign{\smallskip}
Value & 75500 & 170 & 19 & 20 & 127499 & 1334\\
\noalign{\smallskip}\hline
\end{tabular}
\end{table}

\clearpage
\subsection{Homogeneous stress distributions: non-porous specimens}\label{sec:results_homogeneous}
The multi-scale methodology is not required for identifying the parameters of the fatigue model on non-porous specimens, as the stress field is homogeneous in the gauge section. The maximum likelihood approach developed in section \ref{sec:ParamsIdentification_Homogenous} is used to identify the parameters $\mu$ of the fatigue model. Optimisation is carried out using the experimental fatigue lifetime data of non-porous specimens, as detailed in section \ref{sec:expdata}.

During the identification of the parameters of the fatigue model, we evaluate the impact of using fatigue models of varying complexity. Three cases are presented:

\begin{equation}\label{eq:casesparams_homo}
    \mu =
\begin{cases}
    [m,A,B,\alpha,\beta,C] & \text{(i)}\\
    [m,A,\alpha,C,\quad B=0,\beta=0] & \text{(ii)}\\
    [m,A,\alpha,\quad B=0,\beta=0, C=0] & \text{(iii)}
\end{cases}
\end{equation}

These three cases presented here correspond to (i) the fatigue model with parameters for both the LCF and HCF lines ($A$, $\alpha$, $B$, $\beta$), and the fatigue limit ($C$) which we term as the 'two-line fatigue model' (ii) the fatigue model with one line and the fatigue limit which we term as the 'one-line fatigue model' (iii) the fatigue model with one line and no fatigue limit.

\paragraph{Optimisation cost and convergence of solution}\mbox{}\\

\begin{figure}[htbp]
    \centering
        \begin{subfigure}[b]{0.49\textwidth}
            \includegraphics[width=1\textwidth]{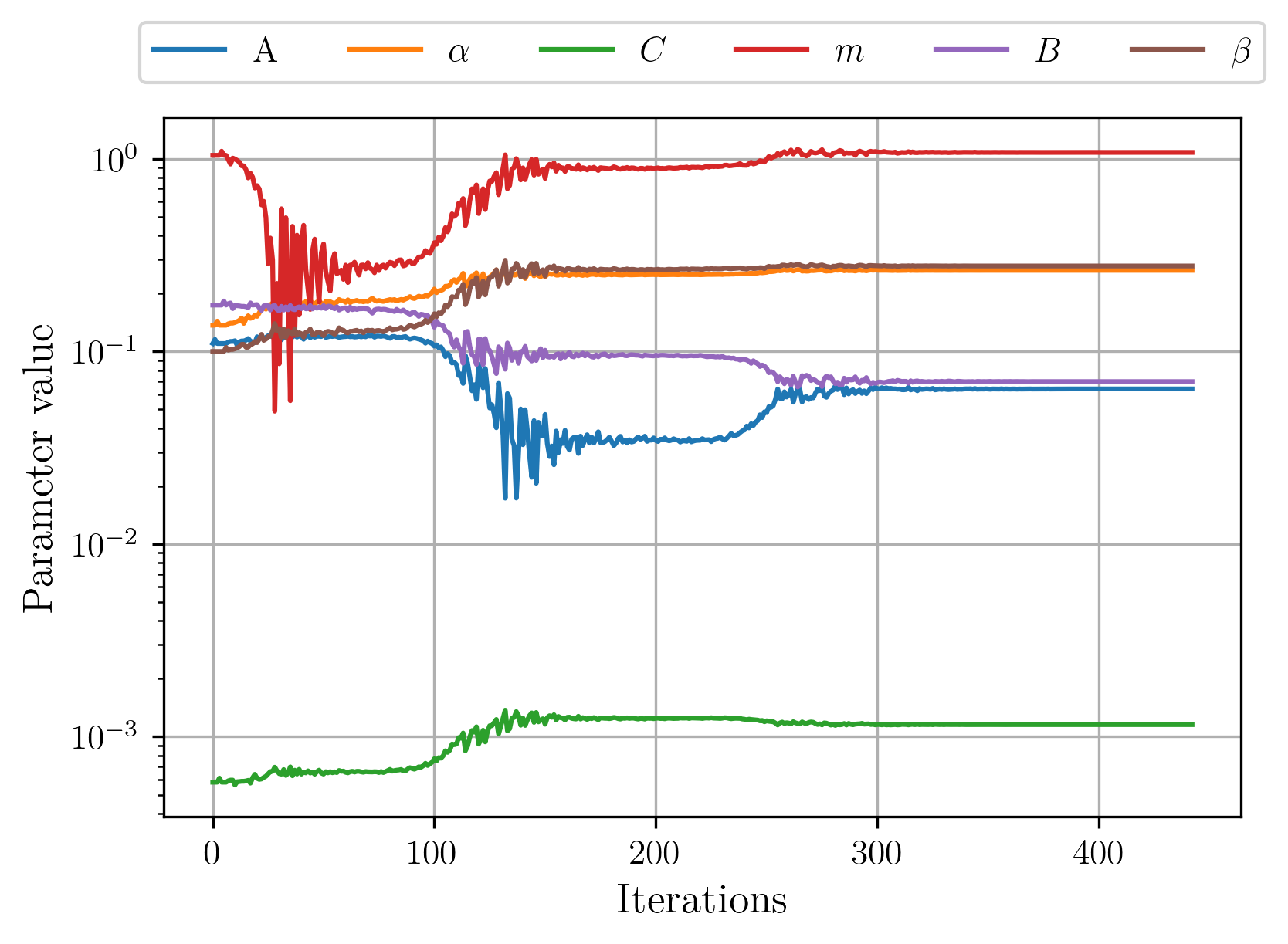}
            \caption{Evolution of parameters}
        \end{subfigure}
        \begin{subfigure}[b]{0.49\textwidth}
            \includegraphics[width=1\textwidth]{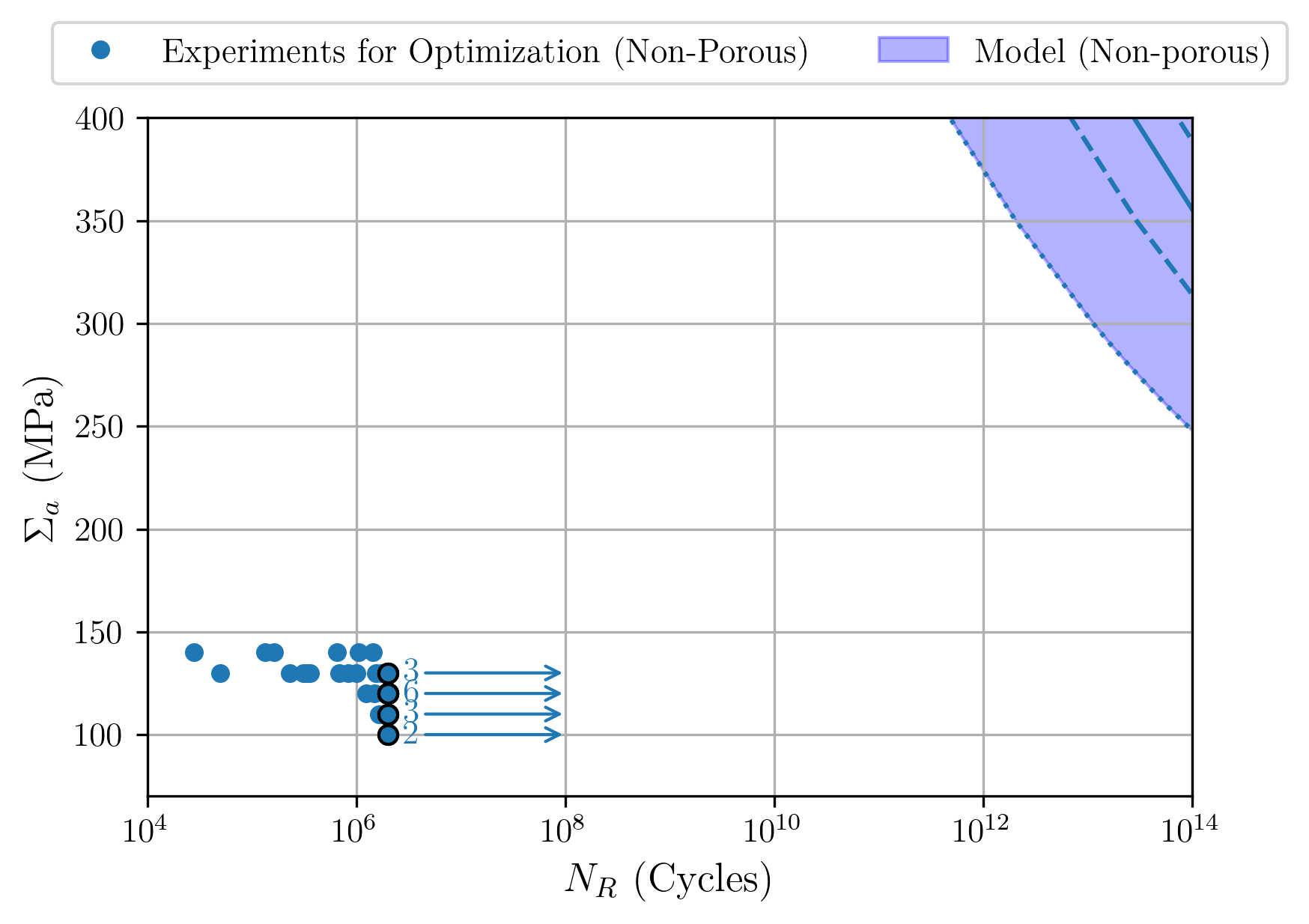}
            \caption{Iteration number 0 (initial guess)}
        \end{subfigure}
        \begin{subfigure}[b]{0.49\textwidth}
            \includegraphics[width=1\textwidth]{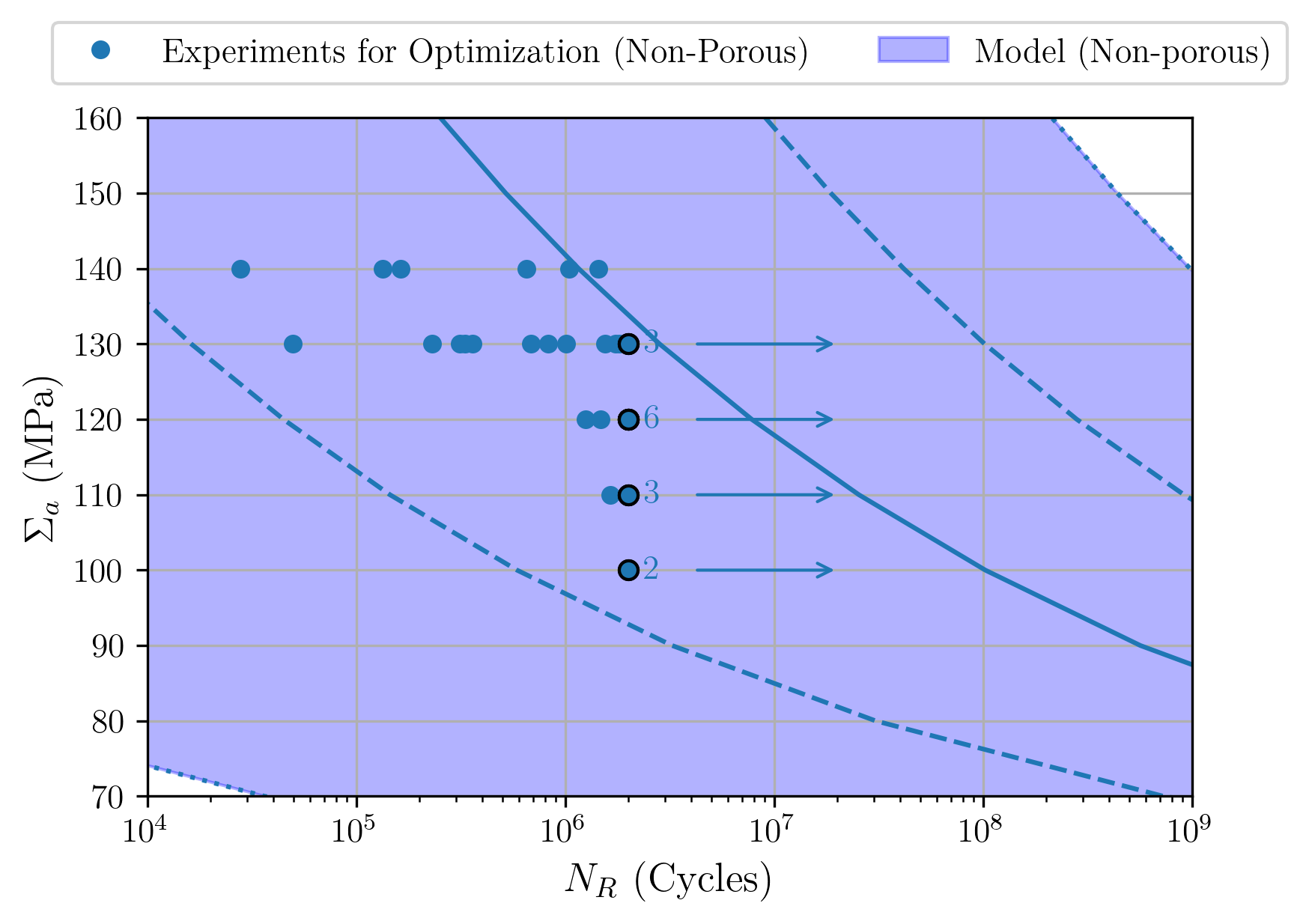}
            \caption{Iteration number 75}
        \end{subfigure}
        \begin{subfigure}[b]{0.49\textwidth}
            \includegraphics[width=1\textwidth]{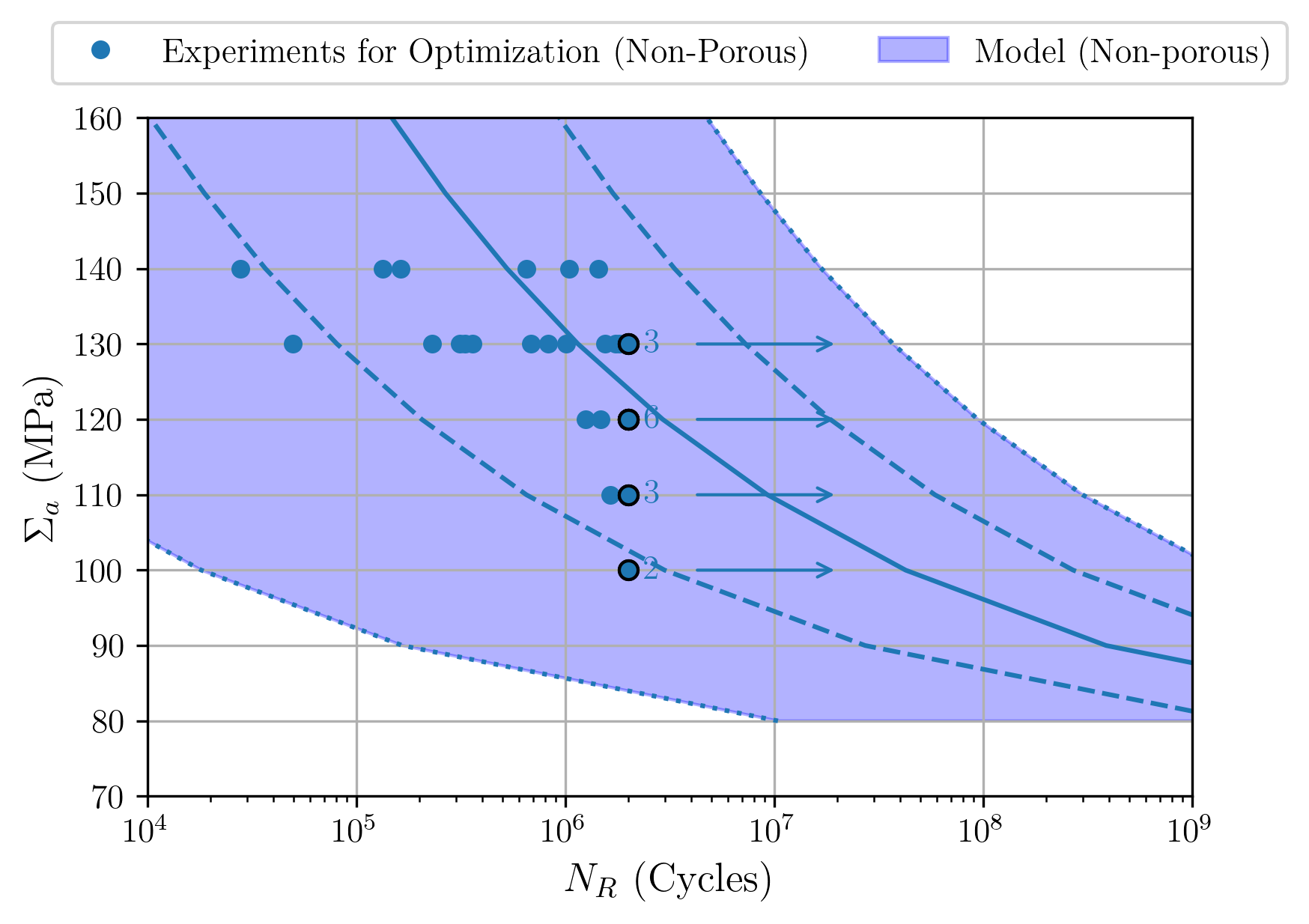}
            \caption{Iteration number 125}
        \end{subfigure}
        \begin{subfigure}[b]{0.49\textwidth}
            \includegraphics[width=1\textwidth]{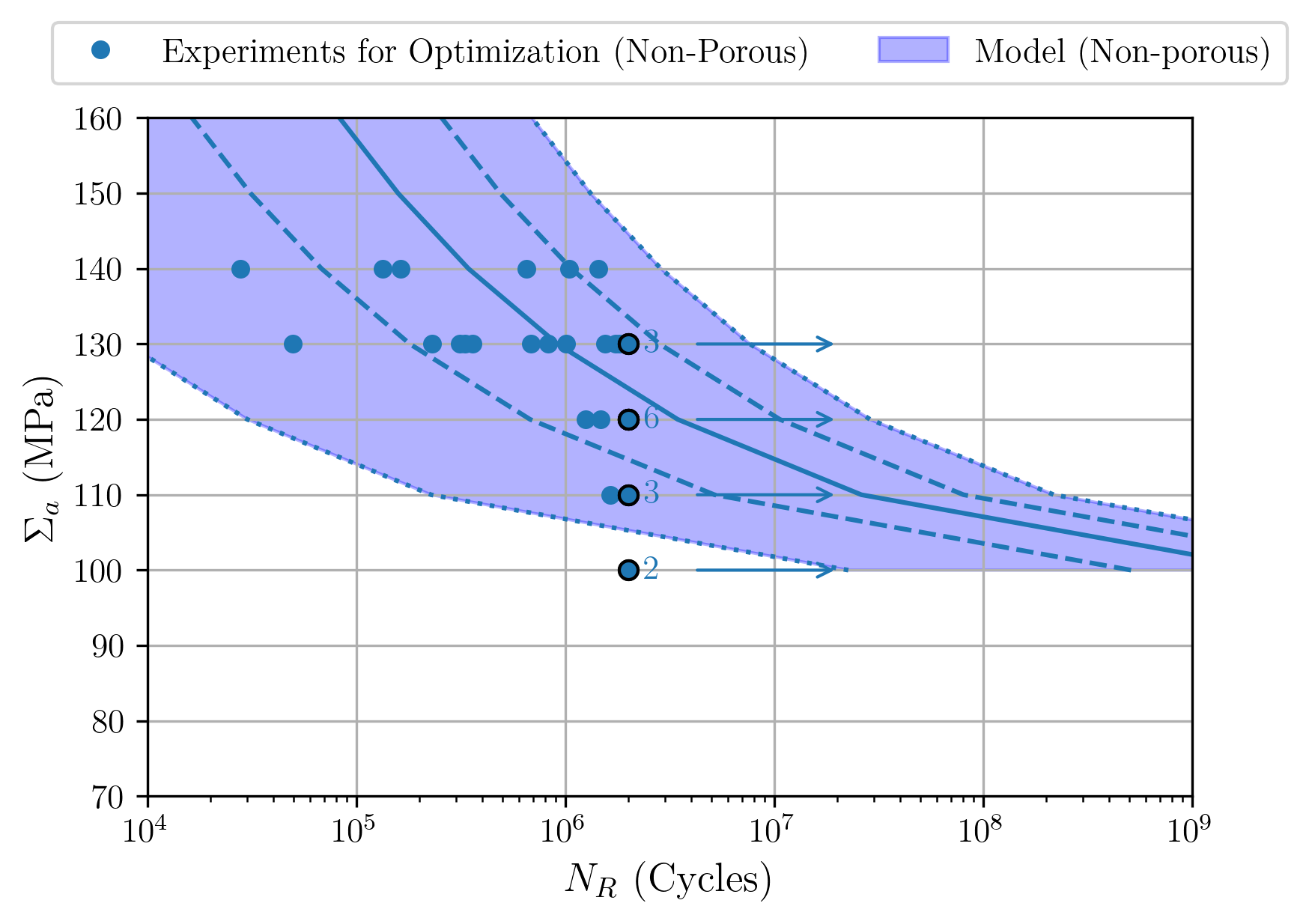}
            \caption{Iteration number 200}
        \end{subfigure}
        \begin{subfigure}[b]{0.49\textwidth}
            \includegraphics[width=1\textwidth]{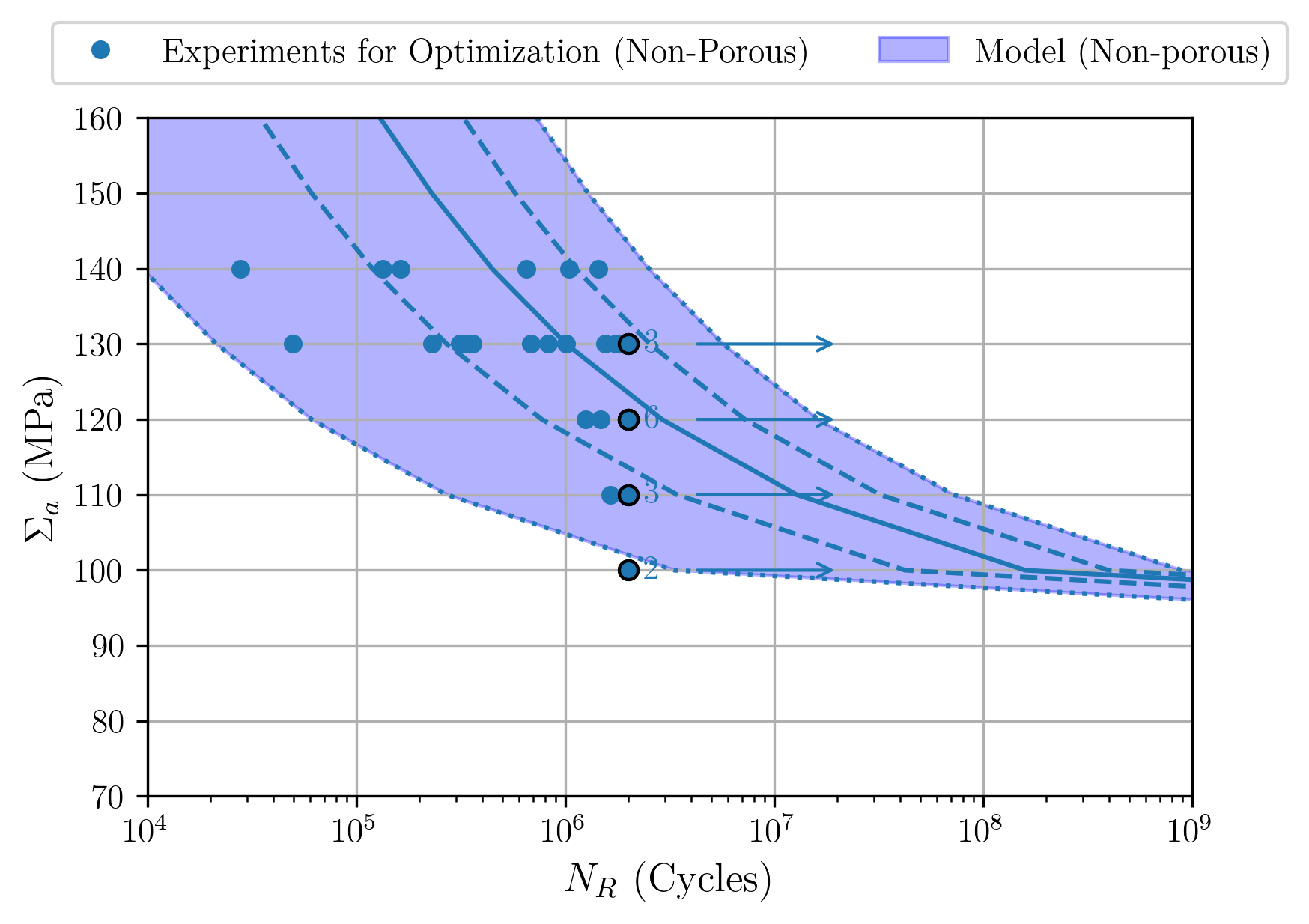}
            \caption{Iteration number 300}
        \end{subfigure}
        
        \caption{ Evolution and stabilization of the model parameters during optimisation for the two-line fatigue model, $\mu = [m,A,B,\alpha,\beta,C]$. The solid line represents the median, the dashed lines represent the 15\% and 85\% quantiles, and the dotted lines represent 1\% and 99\% quantiles.}
        \label{fig:iterations_nonporousonly6params}
\end{figure}

\noindent For all three models, the optimizer (Nelder-Mead) converges towards solutions with very similar likelihood values. The evolution of the parameters and the solution for the two-line fatigue model (equation \eqref{eq:casesparams_homo}(i)) is shown in Fig. \ref{fig:iterations_nonporousonly6params}. Even for an initialisation that is far away from the solution at convergence, the optimisation process is well behaved, as seen by the intermediate and final solutions in Fig. \ref{fig:iterations_nonporousonly6params}(b-f). 

The time required to evaluate all the functions during one iteration is dependent on the number of lifetime data points used for identification. For the considered data-set of 34 fatigue lifetime points, the cost of an iteration is around 0.15 seconds, therefore, the full optimisation procedure over 400 iterations takes around 60 seconds.

\paragraph{Comparison between lifetime models of varying complexity}\mbox{}\\

\begin{figure}[htbp]
    \centering
        \begin{subfigure}[b]{0.49\textwidth}
            \includegraphics[width=1\textwidth]{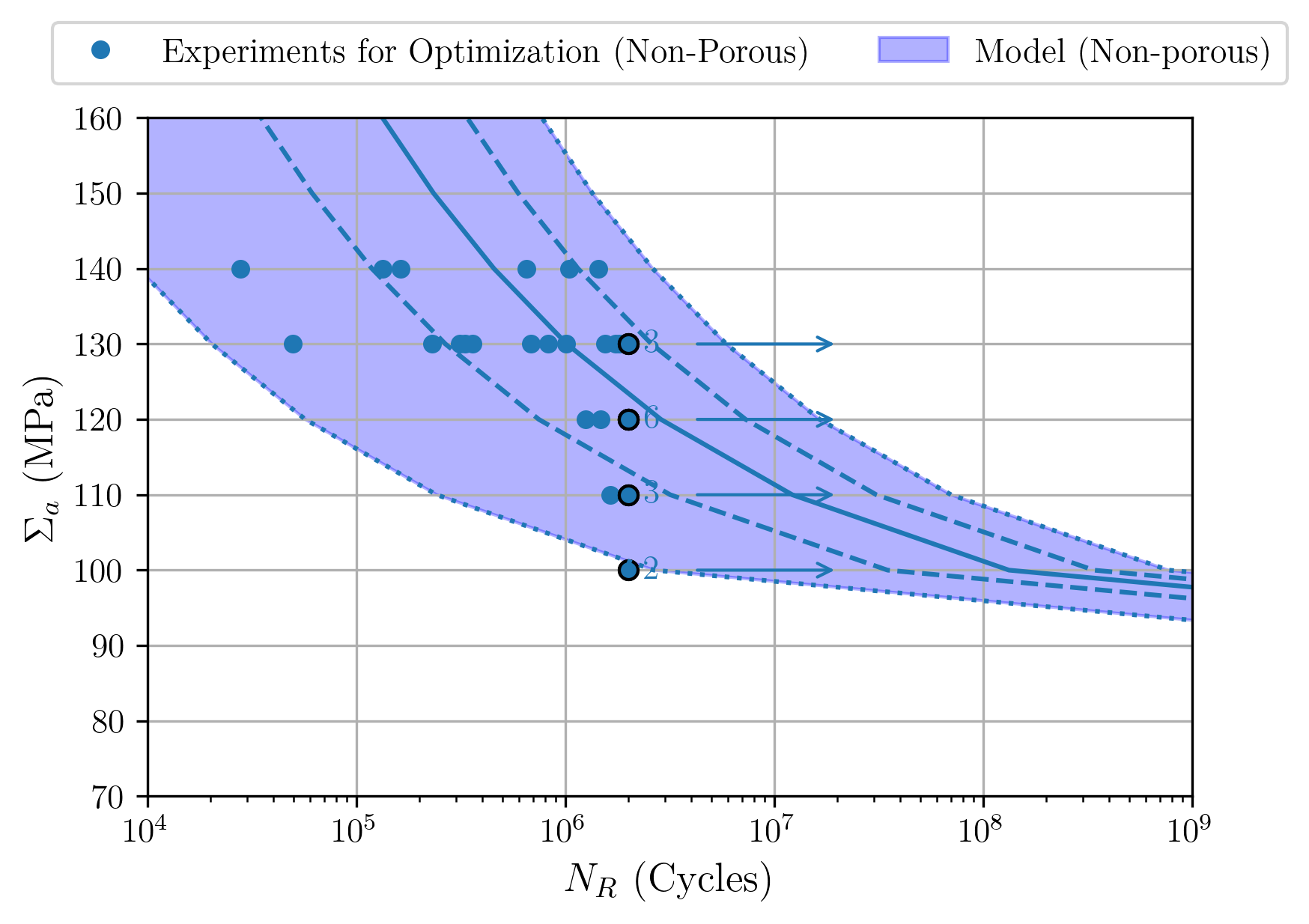}
            \caption{Two-line model, $\mu = [m,A,B,\alpha,\beta,C]$}
        \end{subfigure}           
        \begin{subfigure}[b]{0.49\textwidth}
            \includegraphics[width=1\textwidth]{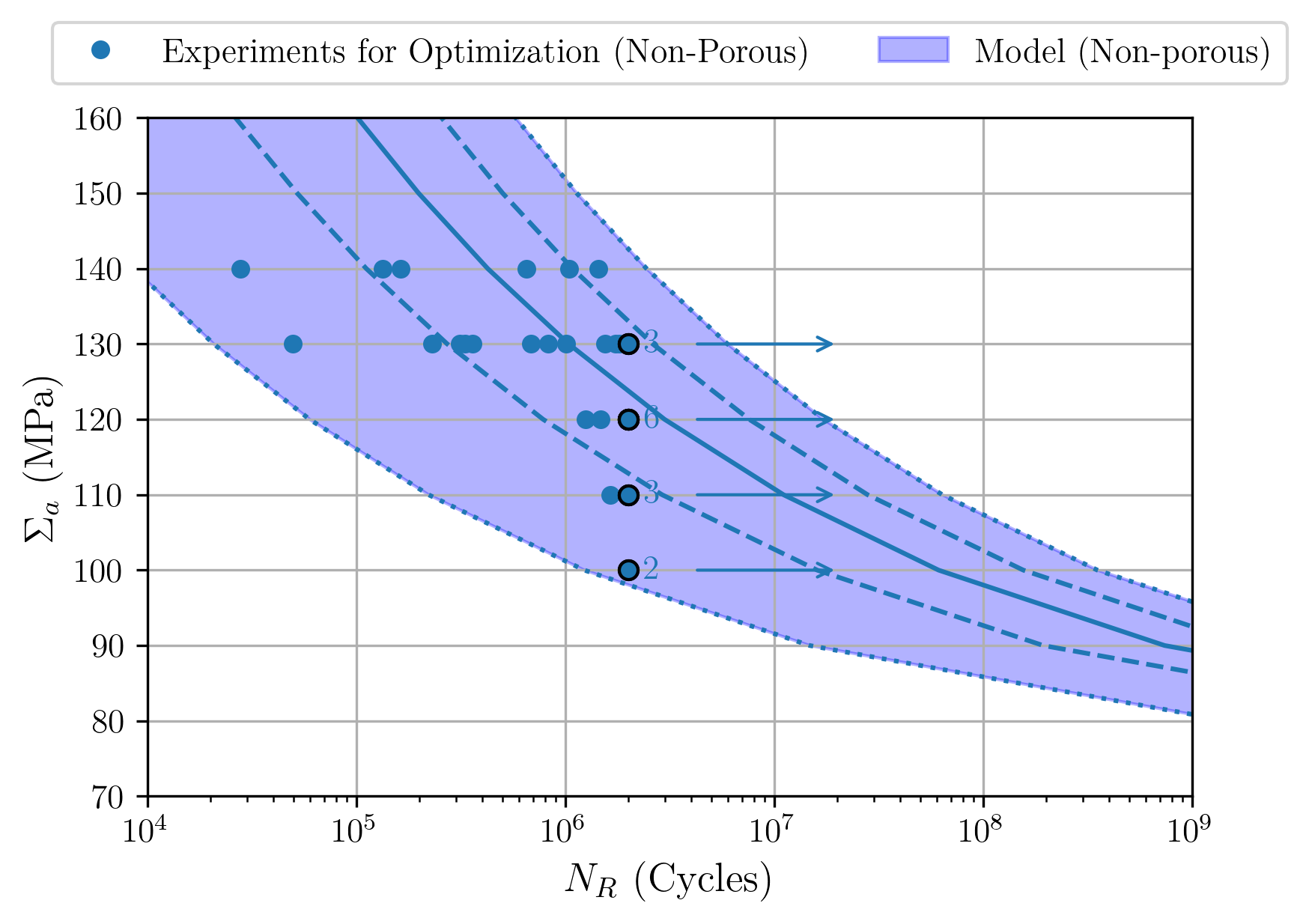}
            \caption{One-line model, $\mu = [m,A,\alpha,C]$}
        \end{subfigure}
        \begin{subfigure}[b]{0.49\textwidth}
            \includegraphics[width=1\textwidth]{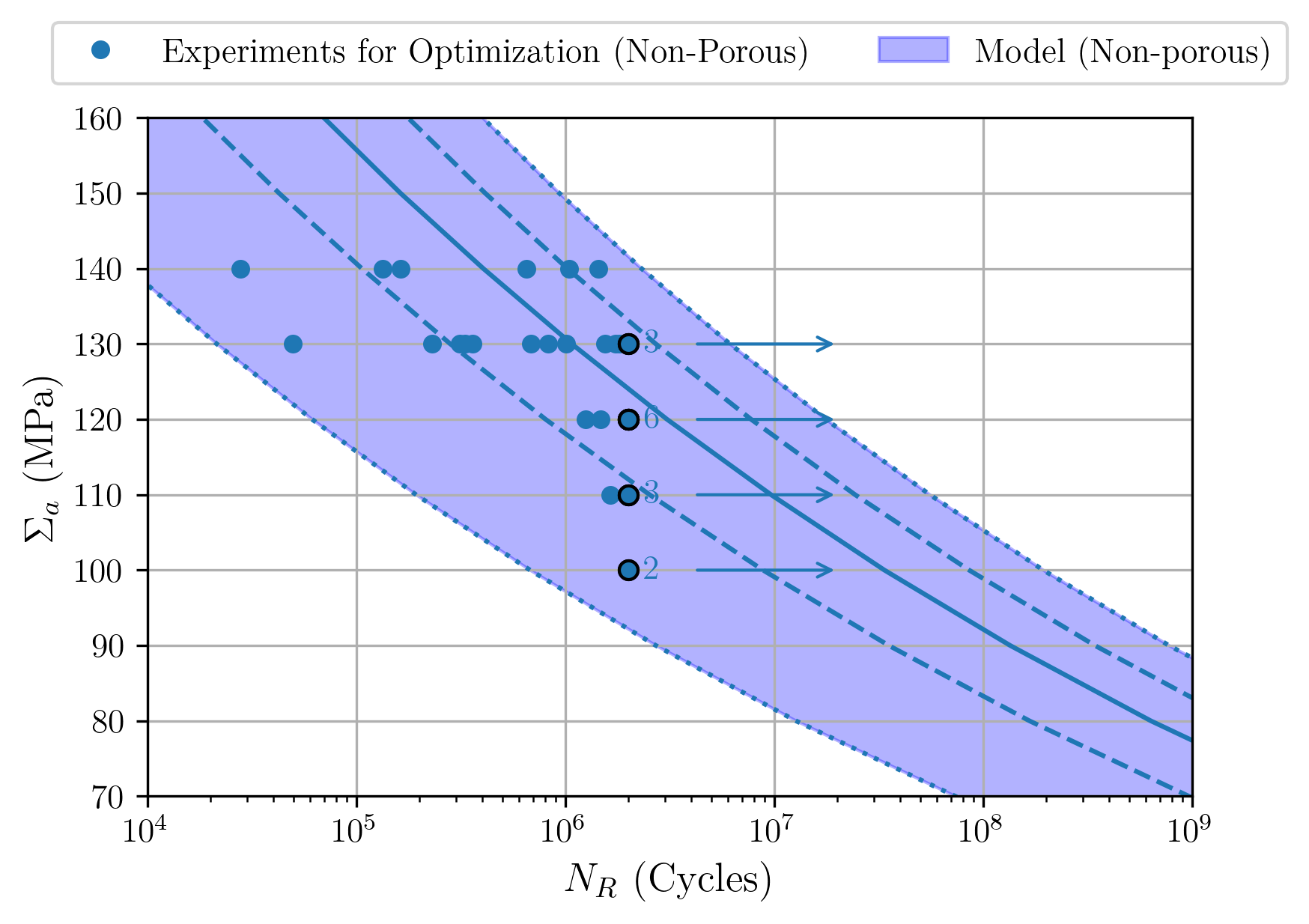}
            \caption{One-line model without the fatigue limit, $\mu = [m,A,\alpha]$}
        \end{subfigure}
        \begin{subfigure}[b]{0.40\textwidth}
            
        \end{subfigure}
        \caption{ Results of the probabilistic strain-life model, with three different fatigue models. The solid line represents the median, the dashed lines represent the 15\% and 85\% quantiles, and the dotted lines represent 1\% and 99\% quantiles.}
        \label{Fig:Homogenous_cases}
\end{figure}

\noindent The results of optimisation are presented in Fig. \ref{Fig:Homogenous_cases}. For all the models,  it is observed that the solutions for the first and second fatigue models (equation \eqref{eq:casesparams_homo}(i) and (ii)) are similar, which suggests that the parameters $B$ and $\beta$, i.e. the second line in the fatigue model, are not required. In our view, this conclusion would have been different had we had access to a richer data-set with more experiments across more loading levels. The third fatigue model (equation \eqref{eq:casesparams_homo}(iii)) shows a significant decrease in the quality of the fit, indicating the requirement for the parameter $C$ for getting a meaningful fatigue limit beyond which the lifetime is infinite.

Furthermore, the results show that the model is capable of taking into account run-outs. Indeed, all the quantiles of the model continue past the point $N_R = 2 \times 10^6$ cycles, which indicates the capability of the model to assign a finite probability of failure for some number of cycles greater than $N_R = 2 \times 10^6$ cycles.

\clearpage
\subsection{Structures with pores whose exact distribution is unknown: porous specimens}\label{sec:results_twoscale_porousonly}
The fatigue model is now identified on the experimental fatigue lifetime data of porous material presented in Section \ref{sec:expdata}. The multi-scale methodology developed in section \ref{sec:TwoScaleModel} for structures with heterogeneous stress distributions due to pores whose exact
geometries and locations are unknown is used. The maximum likelihood approach developed in section \ref{sec:ParamsIdentification_Heterogenous} was used as the identification procedure. From the previously defined set $\tilde{K}$, a total of $n_{\text{k}}=10$ synthetic specimens were randomly selected for each ($N^i , \Sigma_a^i$) pair in order to approximate the expectation operator in the expression \eqref{eq:approxexpectation} of the likelihood function.\footnote{The value of $n_{\text{k}}=10$ is chosen to balance robustness and computational cost in the model identification process. The robustness increases with  $n_{\text{k}}$: each experimental point has a likelihood of failing from a larger variety of pore configurations, which is ideal as the exact pore distribution associated to the failure point is unknown. However, as will be discussed later, the computational cost increases linearly with $n_{\text{k}}$.}

During the identification of the parameters of the strain-life model, we evaluate the impact of using fatigue models of varying complexity. Two cases are presented (as $C$ has previously been proven necessary for obtaining a fatigue limit):

\begin{equation}
    \mu =
\begin{cases}
    [m,A,B,\alpha,\beta,C] & \text{(i)}\\
    [m,A,\alpha,C,\quad B=0,\beta=0] & \text{(ii)}\\
\end{cases}
\end{equation}

As defined previously (section \ref{sec:results_homogeneous}), the cases correspond to (i) the 'two-line fatigue model' and (ii) the 'one-line fatigue model'.

\paragraph{Optimisation cost and convergence of solution}\mbox{}\\

\begin{figure}[htbp]
    \centering
        \begin{subfigure}[b]{0.49\textwidth}
            \includegraphics[width=1\textwidth]{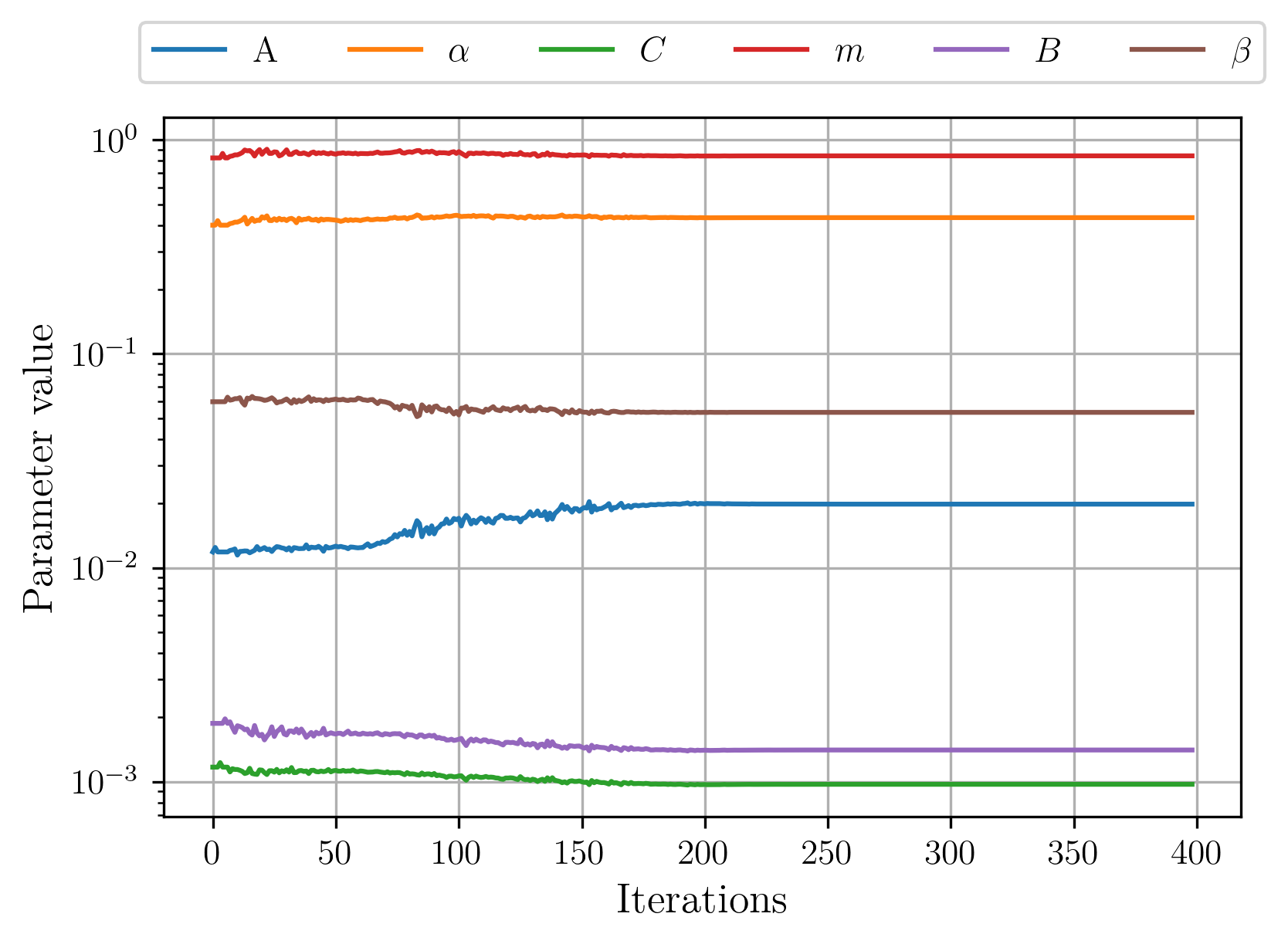}
            \caption{Evolution of parameters}
        \end{subfigure}
        \begin{subfigure}[b]{0.49\textwidth}
            \includegraphics[width=1\textwidth]{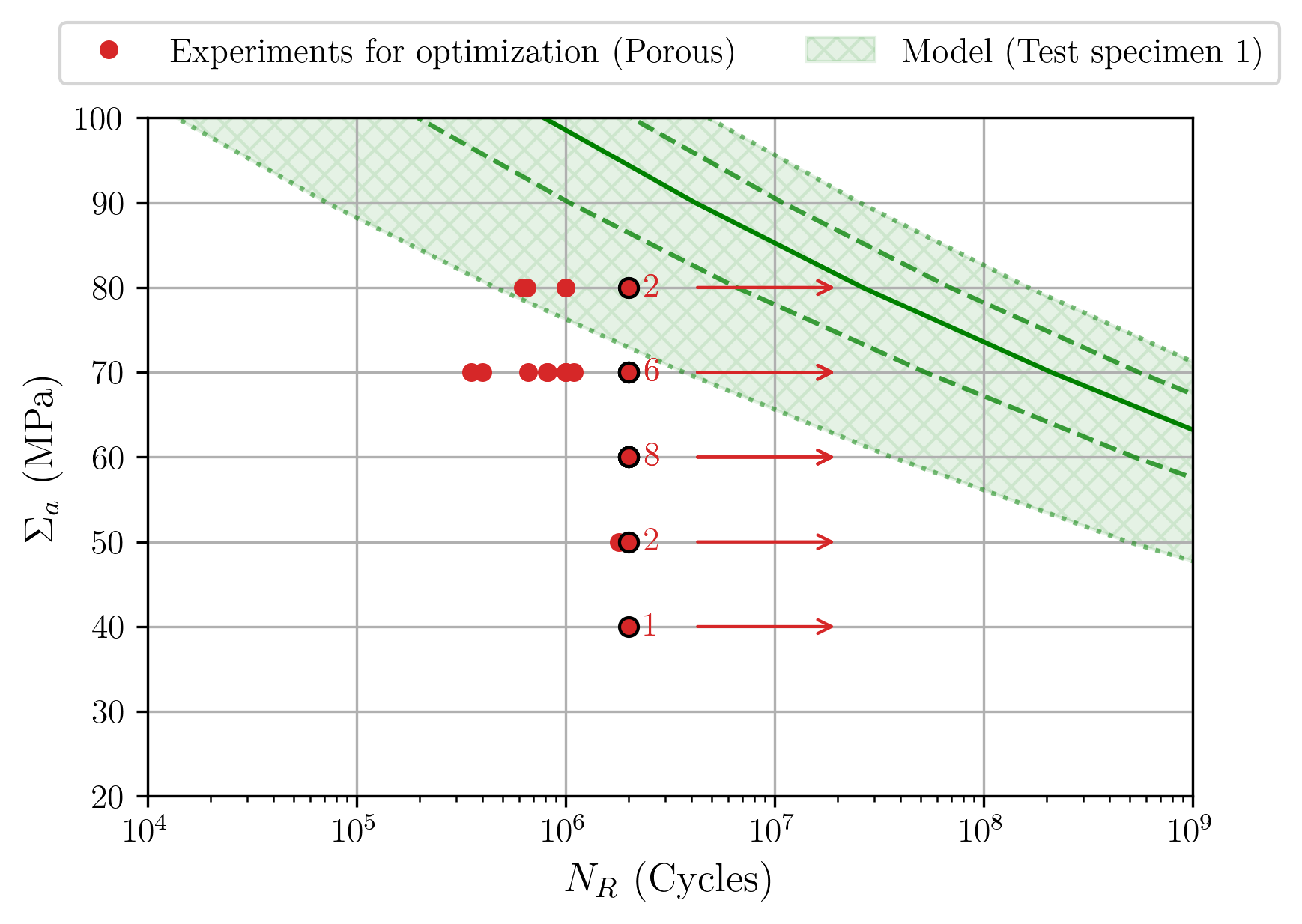}
            \caption{Iteration number 0 (initial guess)}
        \end{subfigure}
        \begin{subfigure}[b]{0.49\textwidth}
            \includegraphics[width=1\textwidth]{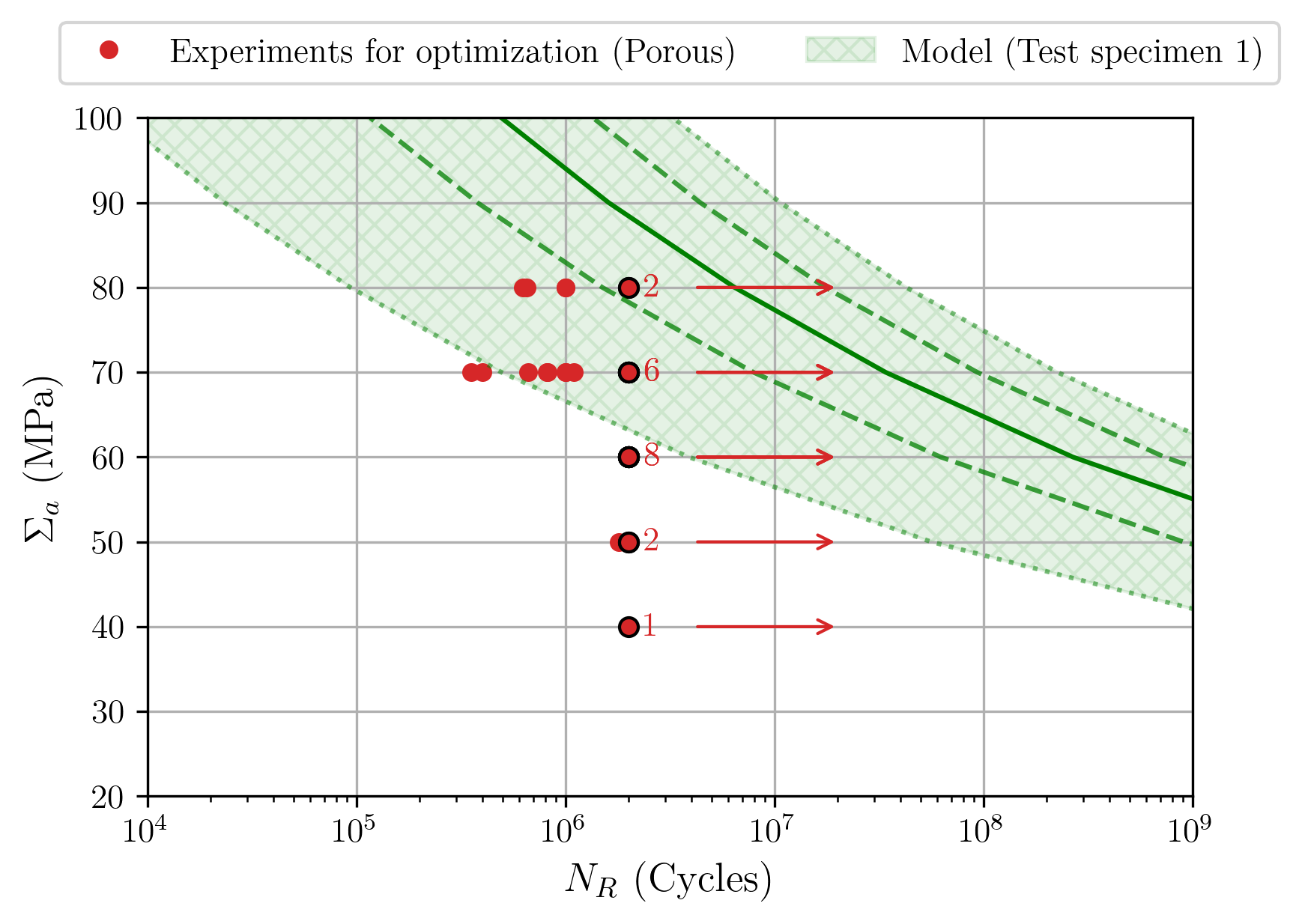}
            \caption{Iteration number 10}
        \end{subfigure}
        \begin{subfigure}[b]{0.49\textwidth}
            \includegraphics[width=1\textwidth]{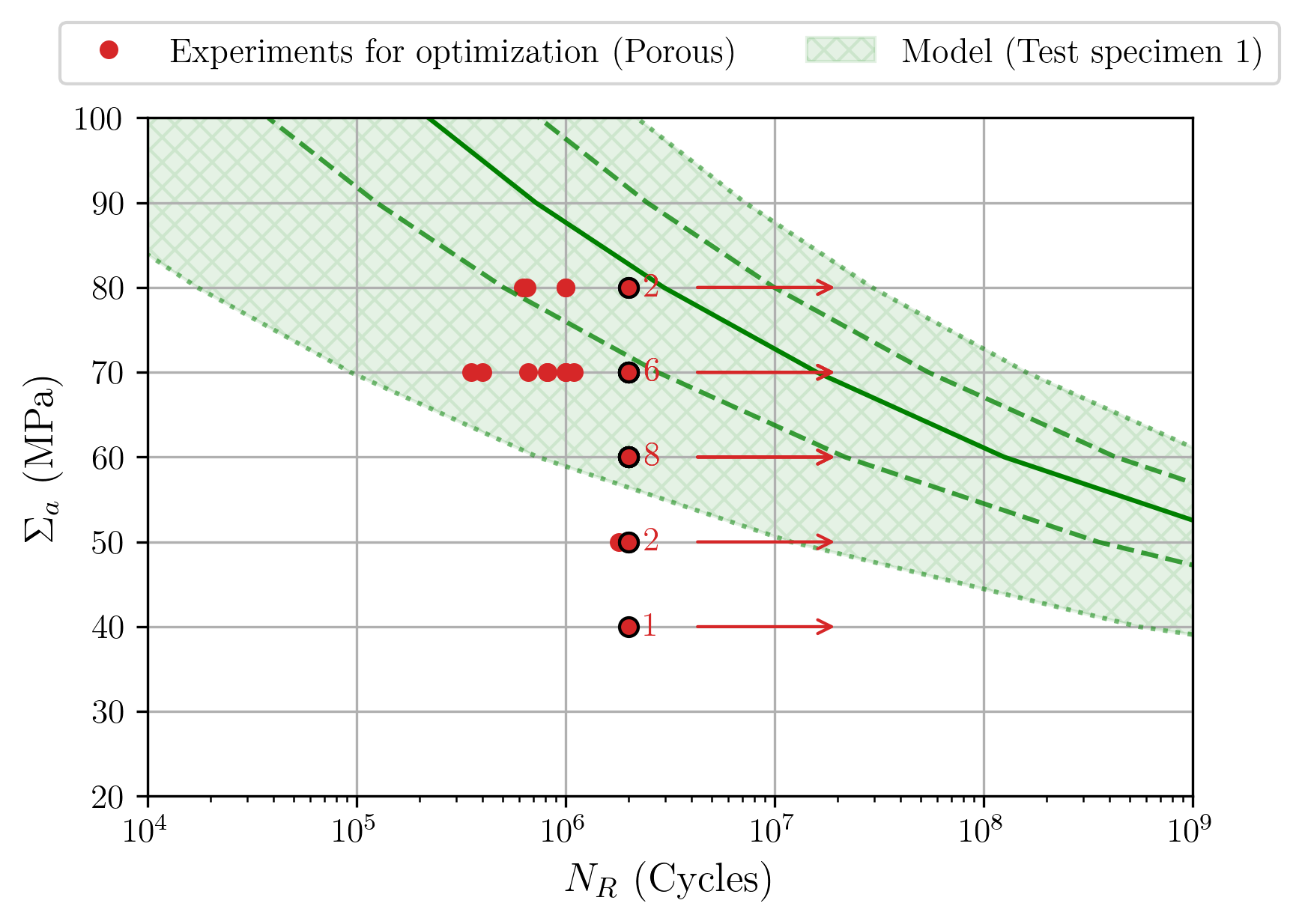}
            \caption{Iteration number 50}
        \end{subfigure}
        \begin{subfigure}[b]{0.49\textwidth}
            \includegraphics[width=1\textwidth]{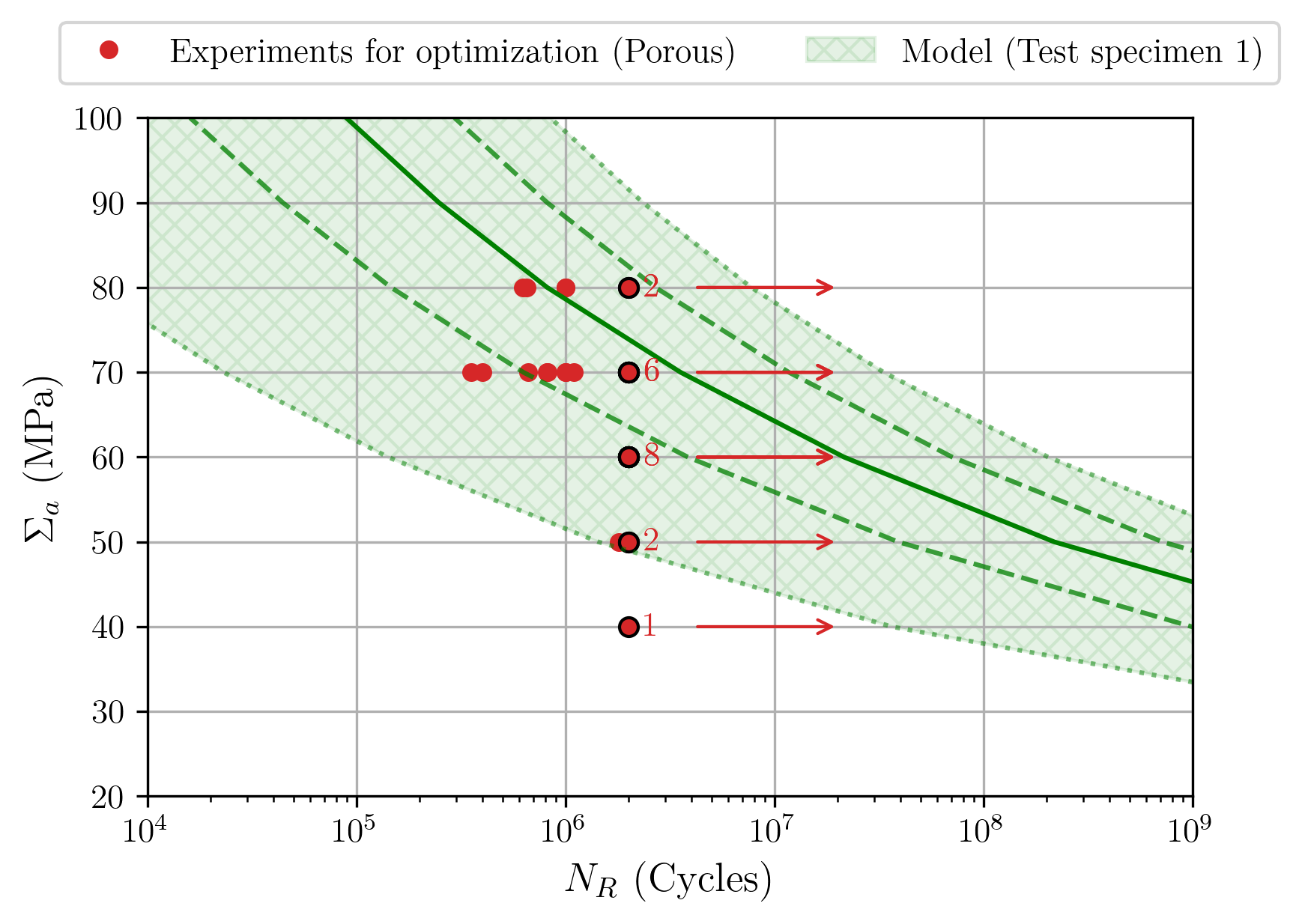}
            \caption{Iteration number 100}
        \end{subfigure}
        \begin{subfigure}[b]{0.49\textwidth}
            \includegraphics[width=1\textwidth]{images/iter200_Iterations_porousonly6params.png}
            \caption{Iteration number 200}
        \end{subfigure}
        
        \caption{ Evolution and stabilization of the model parameters during optimisation for the two-line fatigue model, $\mu = [m,A,B,\alpha,\beta,C]$ during optimisation on the porous experimental data using the multi-scale method}
        \label{fig:iterations_porousonly6params}
\end{figure}

\noindent The parameters stabilise after a certain number of iterations of optimisation, as shown in the Fig. \ref{fig:iterations_porousonly6params}(a) for the two-line fatigue model. In Fig. \ref{fig:iterations_porousonly6params} we show an optimisation process that converges well due to good initialisation. Indeed, if the initialisation is very far away from the expected solution, the optimisation process does not yield satisfactory results.

The cost of evaluating all the functions during one iteration is dependent on the number of synthetically generated specimens $n_k$ taken to compute the expectation of the fatigue lifetime distribution and the number of lifetime data points. It also depends on the number of elements in each synthetically generated specimen, which is around 1.5-2.5 million elements each. For the considered data-set of 32 fatigue lifetime points, the cost of one iteration is around 42 seconds if $n_k = 1$ and 425 seconds for $n_k = 10$. For the presented case ($n_k = 10$), the cost of optimisation over 400 iterations is thus 47 hours.

\paragraph{Wöhler curve representing uncertainty due to pore distributions}\mbox{}\\

\begin{figure}[h!tbp]
  \centering
    \begin{subfigure}[b]{0.49\textwidth}
        \includegraphics[width=\textwidth]{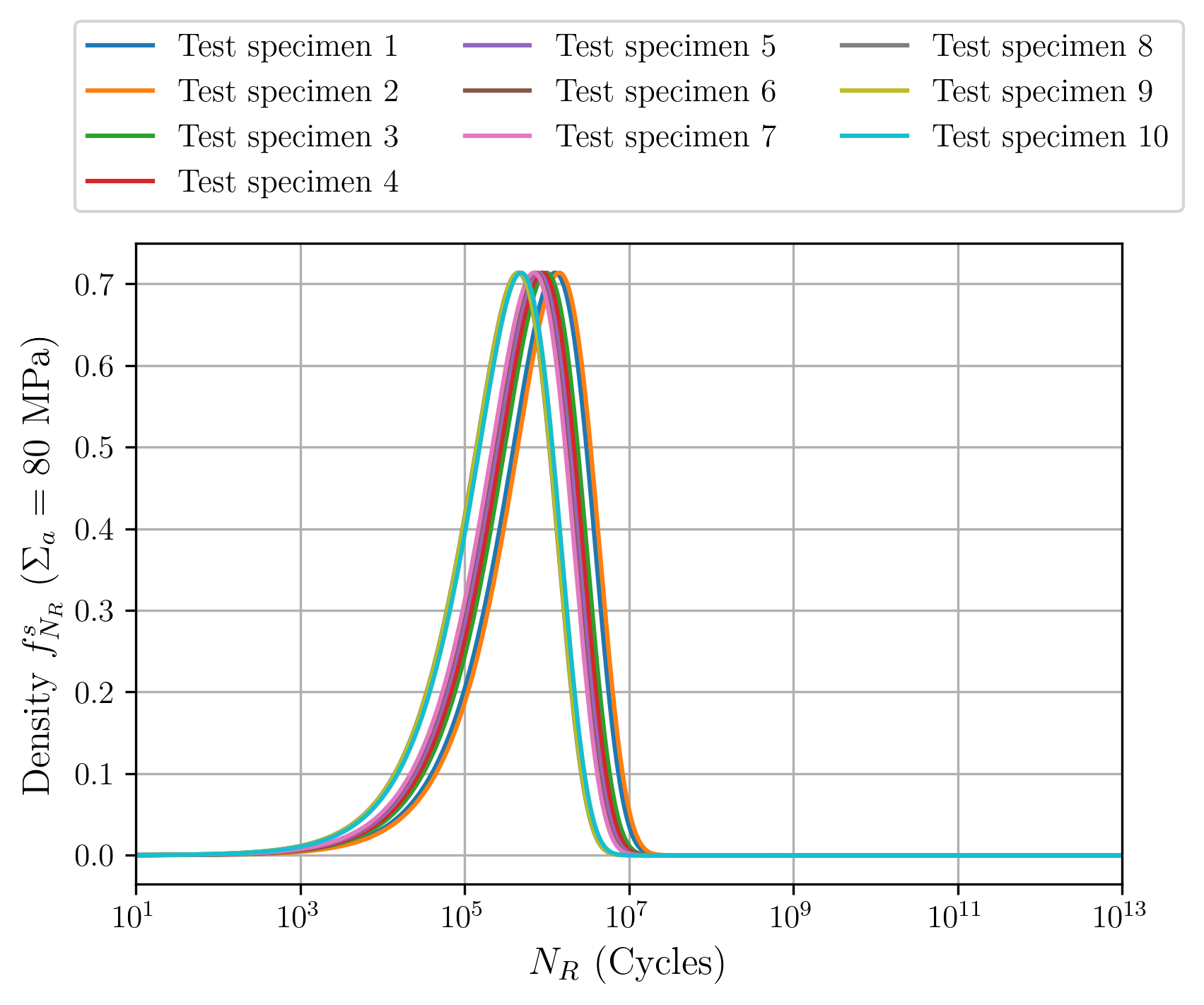}
        \caption{}
    \end{subfigure}%
    \hfill
    \begin{subfigure}[b]{0.49\textwidth}
        \includegraphics[width=\textwidth]{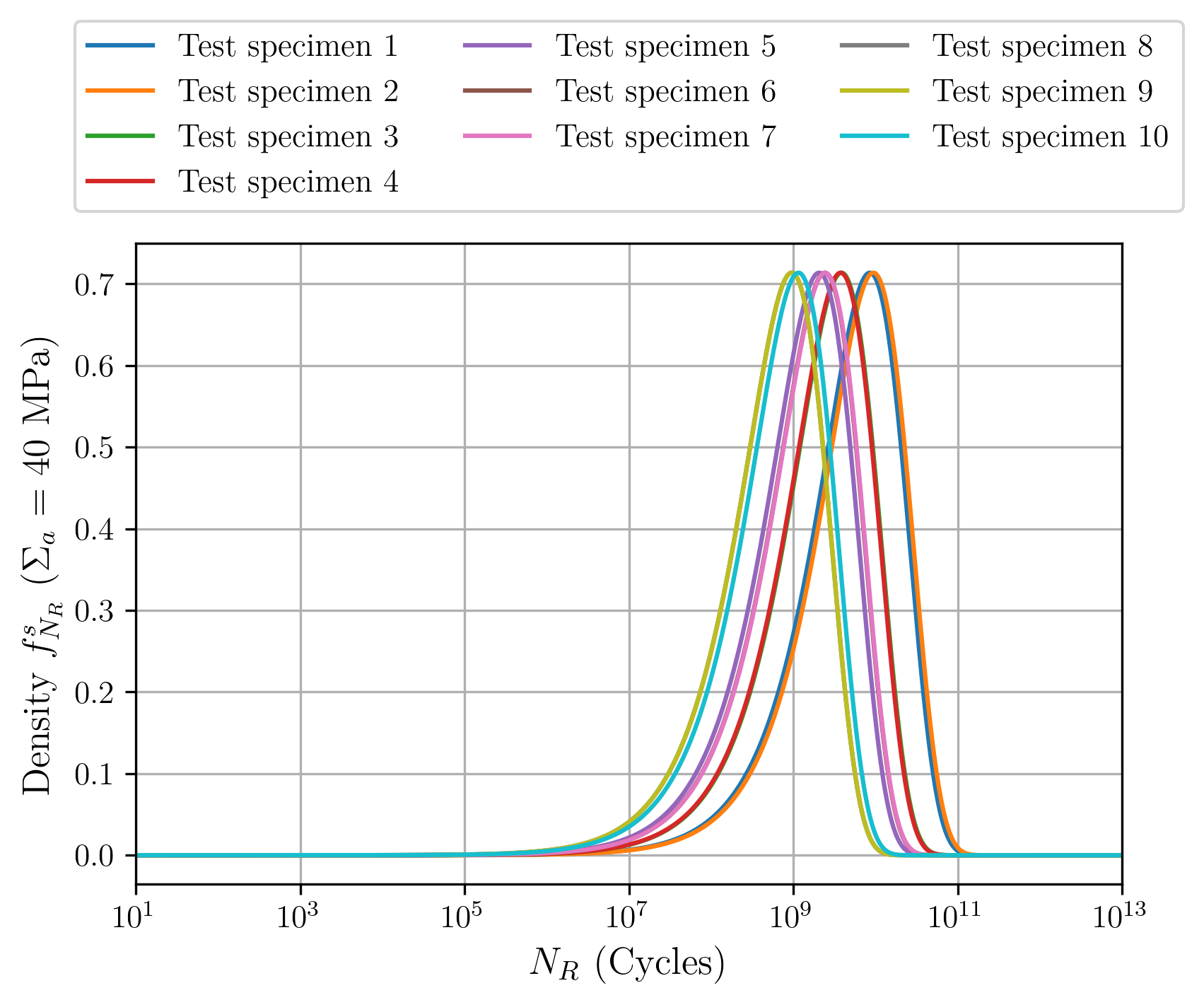}
        \caption{}
    \end{subfigure}
    
    \begin{subfigure}[b]{0.49\textwidth}
        \includegraphics[width=\textwidth]{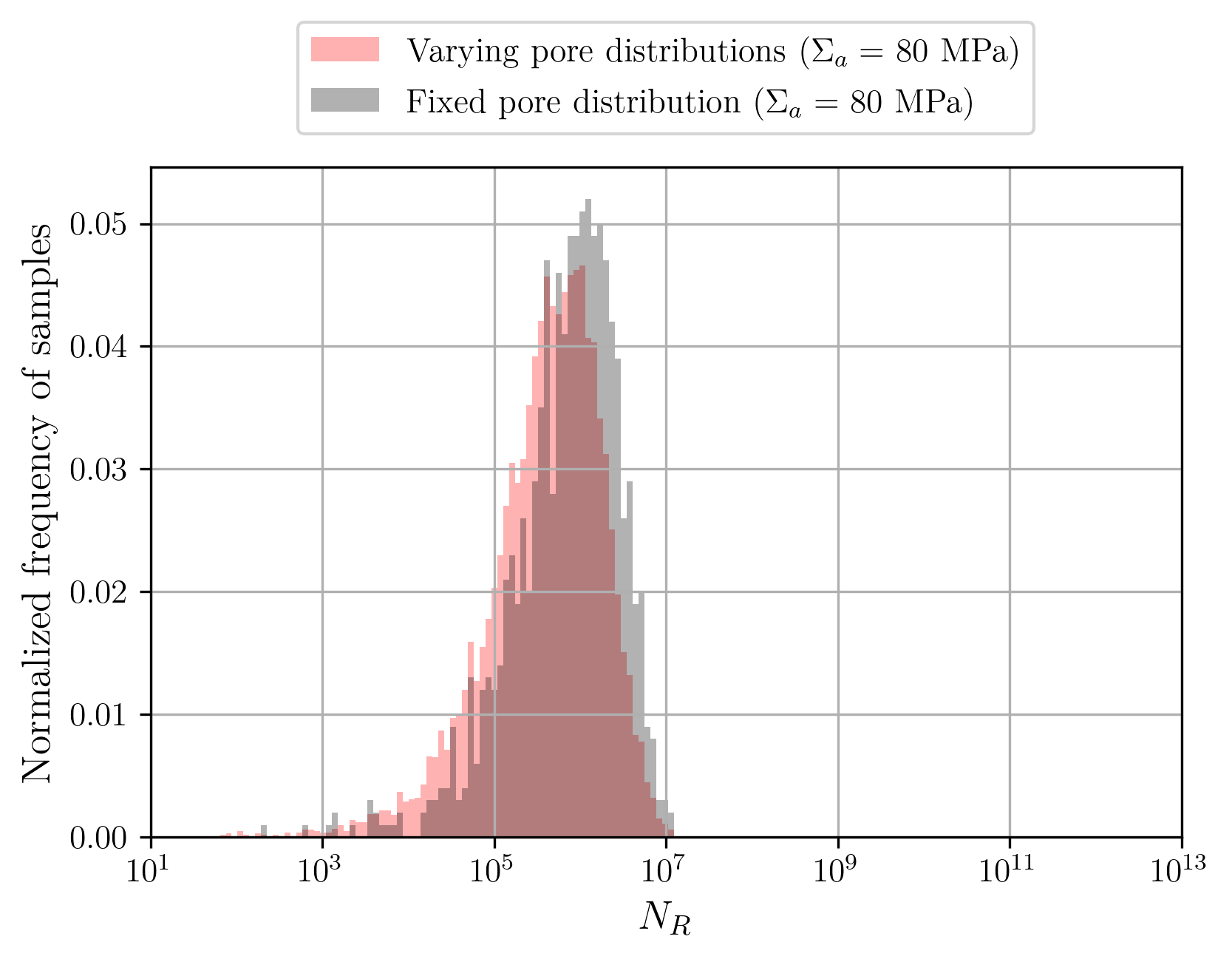}
        \caption{}
    \end{subfigure}%
    \hfill
    \begin{subfigure}[b]{0.49\textwidth}
        \includegraphics[width=\textwidth]{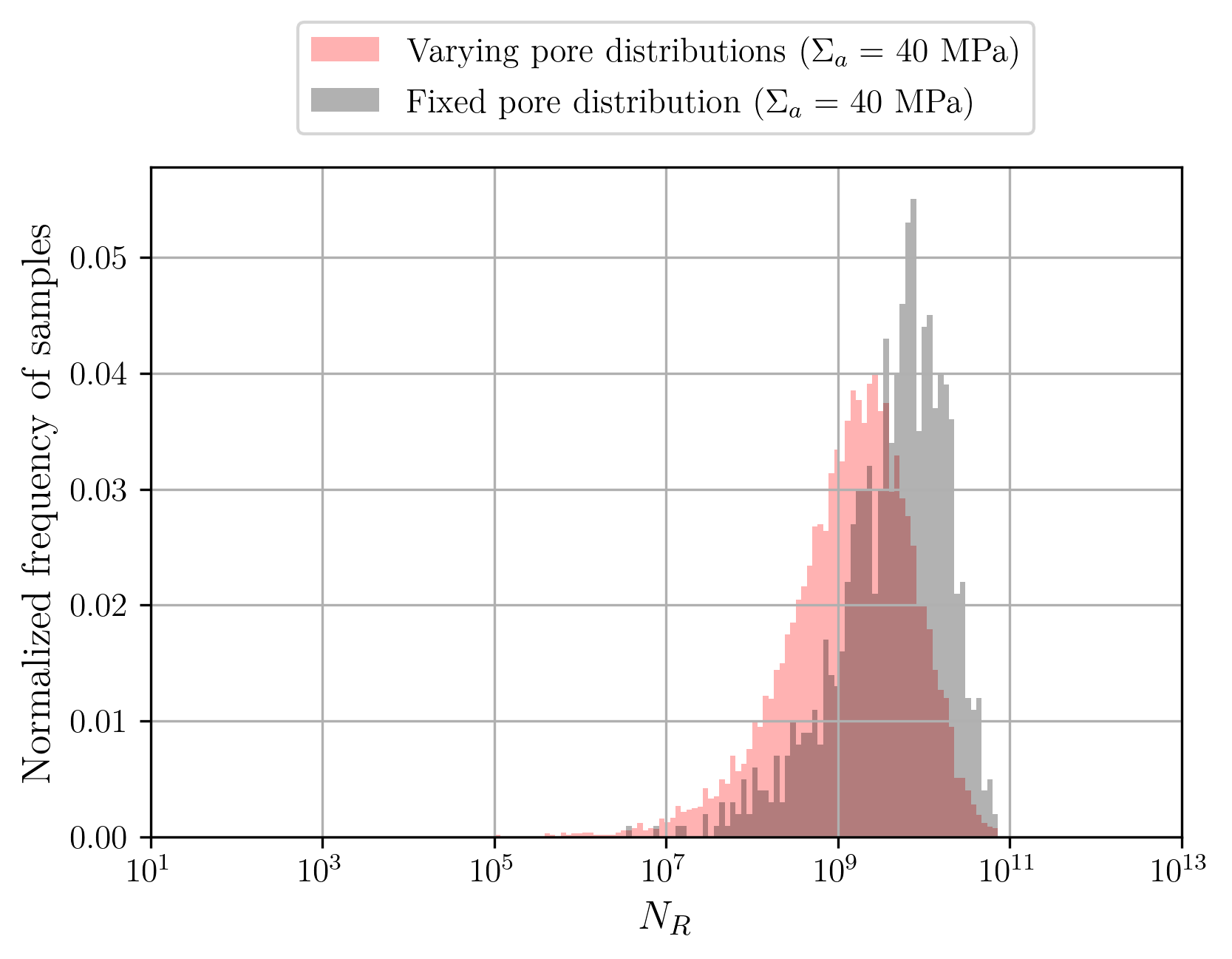}
        \caption{}
    \end{subfigure}
        
    \caption{\label{Fig:ConstructionMultiplePorous_OptimPorous_pdfs} For the two-line strain-life model (parameters $\mu = [m,A,B,\alpha,\beta,C]$)  identified on porous data: Failure densities for 10 different synthetically generated porous test specimens at (a) $\Sigma_a = $ 80 MPa (b) $\Sigma_a = $ 40 MPa.  Proportion of uncertainty coming from a specimen with a fixed pore distribution (black) compared to specimens with varying pore distributions (red histogram constructed by combining 1000 samples each from 10 different porous test specimens) at (c) $\Sigma_a = $  80 MPa (d) $\Sigma_a = $ 40 MPa}
\end{figure}

\begin{figure}[!htbp]
    \centering
        \begin{subfigure}[b]{0.8\textwidth}
            \includegraphics[width=1\textwidth]{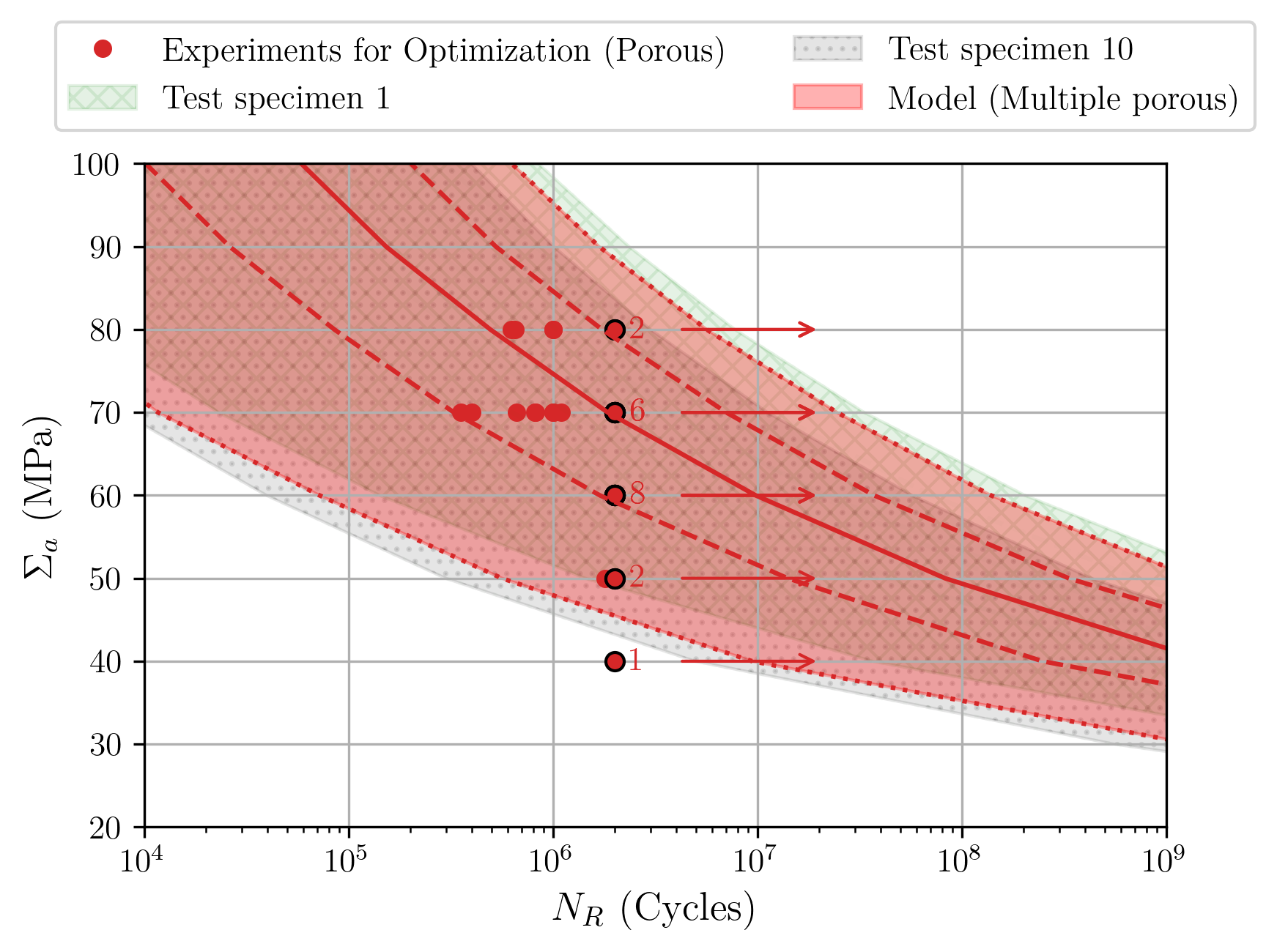}
            \caption{}
        \end{subfigure}
        
        \caption{ Results of the multi-scale probabilistic strain-life model identified on all available porous data, with all parameters active ($\mu = [m,A,B,\alpha,\beta,C]$). Proportion of uncertainty coming from fixed pore distributions (two test specimens shown in green and grey hatch) to varying pore distributions (red master curve constructed by combining 1000 samples each from 10 different porous test specimens at every applied stress amplitude level). The solid line represents the median, the dashed lines represent the 15\% and 85\% quantiles, and the dotted lines represent 1\% and 99\% quantiles.
 }
        \label{Fig:multipleporouswohler_optimporous}
\end{figure}

\noindent Fatigue lifetime distributions for multiple realizations of synthetically generated porous specimens of the same volume but varying pore distributions (examples previously shown in \ref{Fig:VaryingcriterionDistributionFullField}) were obtained, at various nominal loading levels. The synthetically generated porous specimens show variations in their fatigue lifetime distributions, as shown in Fig. \ref{Fig:ConstructionMultiplePorous_OptimPorous_pdfs} for two different nominal applied stress amplitude levels, or in Fig. \ref{Fig:multipleporouswohler_optimporous} for the entire Wöhler curve for a couple of specimens.

A fatigue lifetime distribution or Wöhler curve representing uncertainty due to pore distributions can be constructed by combining samples from a certain number of porous specimens. The failure distribution representing this uncertainty in pore distributions is constructed by combining 1000 samples each from 10 different randomly generated porous test specimens (shown by the red histogram in Fig.  \ref{Fig:ConstructionMultiplePorous_OptimPorous_pdfs}(c,d)). The quantiles are then computed on 10000 samples for each nominal stress level, which are joined to construct the quantiles of the porous Wöhler curve (shown by the red area in Fig.  \ref{Fig:multipleporouswohler_optimporous}). This failure distribution or Wöhler curve represents uncertainty due to defects at two scales - micro-heterogeneity and varying pore distributions. We note that the uncertainty in fatigue lifetime due to varying pore distributions in the specimen is relatively low as compared to the uncertainty due to the micro-heterogeneity. In other words, at the scale of this specimen, the impact of varying pore distributions on the lifetime is not high. This is because of the volume of the specimen, which is large as compared to the pore size, which increases the likelihood of finding critical pore configurations in all the specimens, and makes all the synthetic specimens have similar stress distributions.

%

\paragraph{Comparison between lifetime models of varying complexity}\mbox{}\\

\begin{figure}[htbp]
    \centering
        \begin{subfigure}[b]{0.49\textwidth}
            \includegraphics[width=1\textwidth]{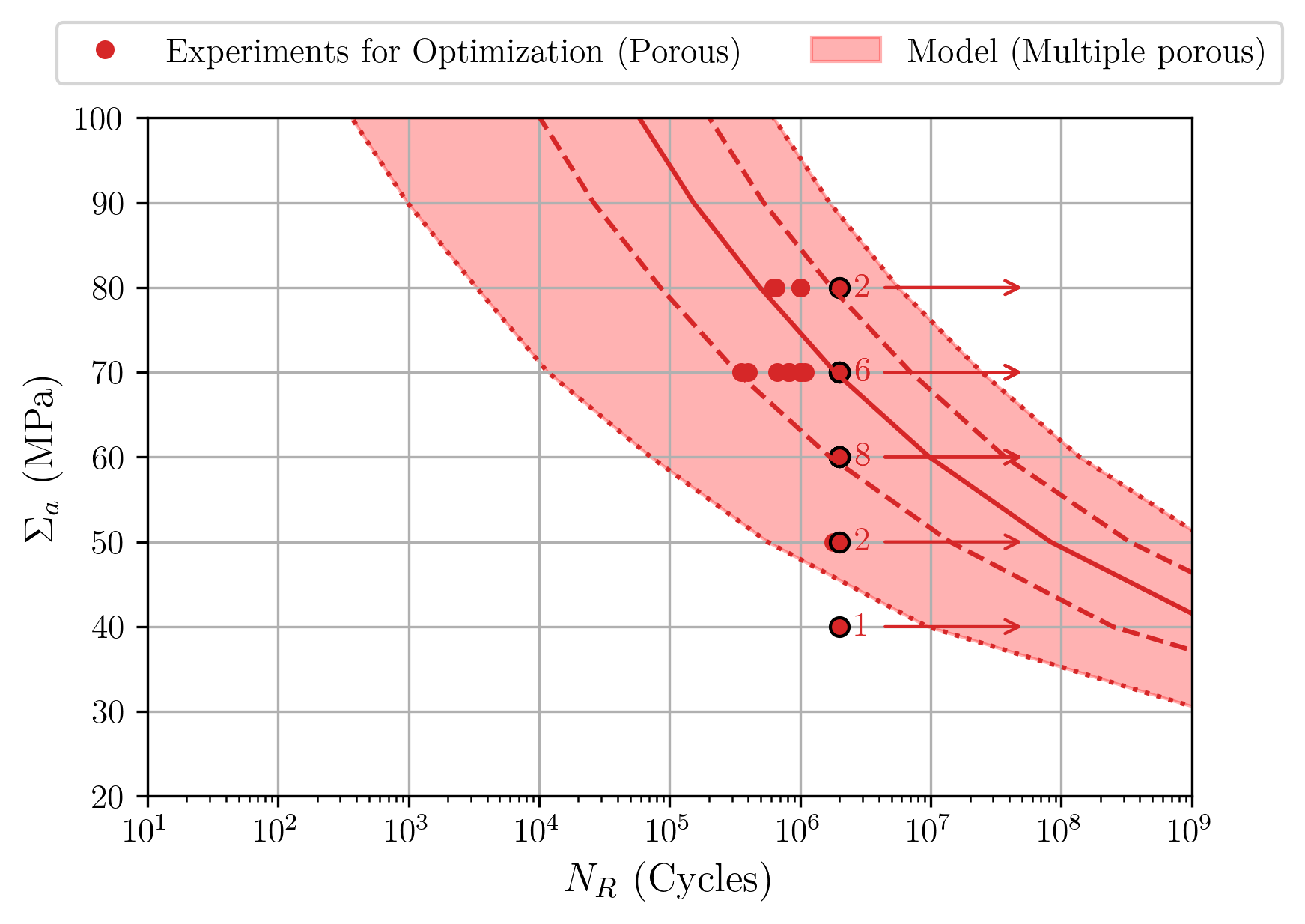}
            \caption{Two-line model, $\mu = [m,A,B,\alpha,\beta,C]$}
        \end{subfigure}
        \begin{subfigure}[b]{0.49\textwidth}
            \includegraphics[width=1\textwidth]{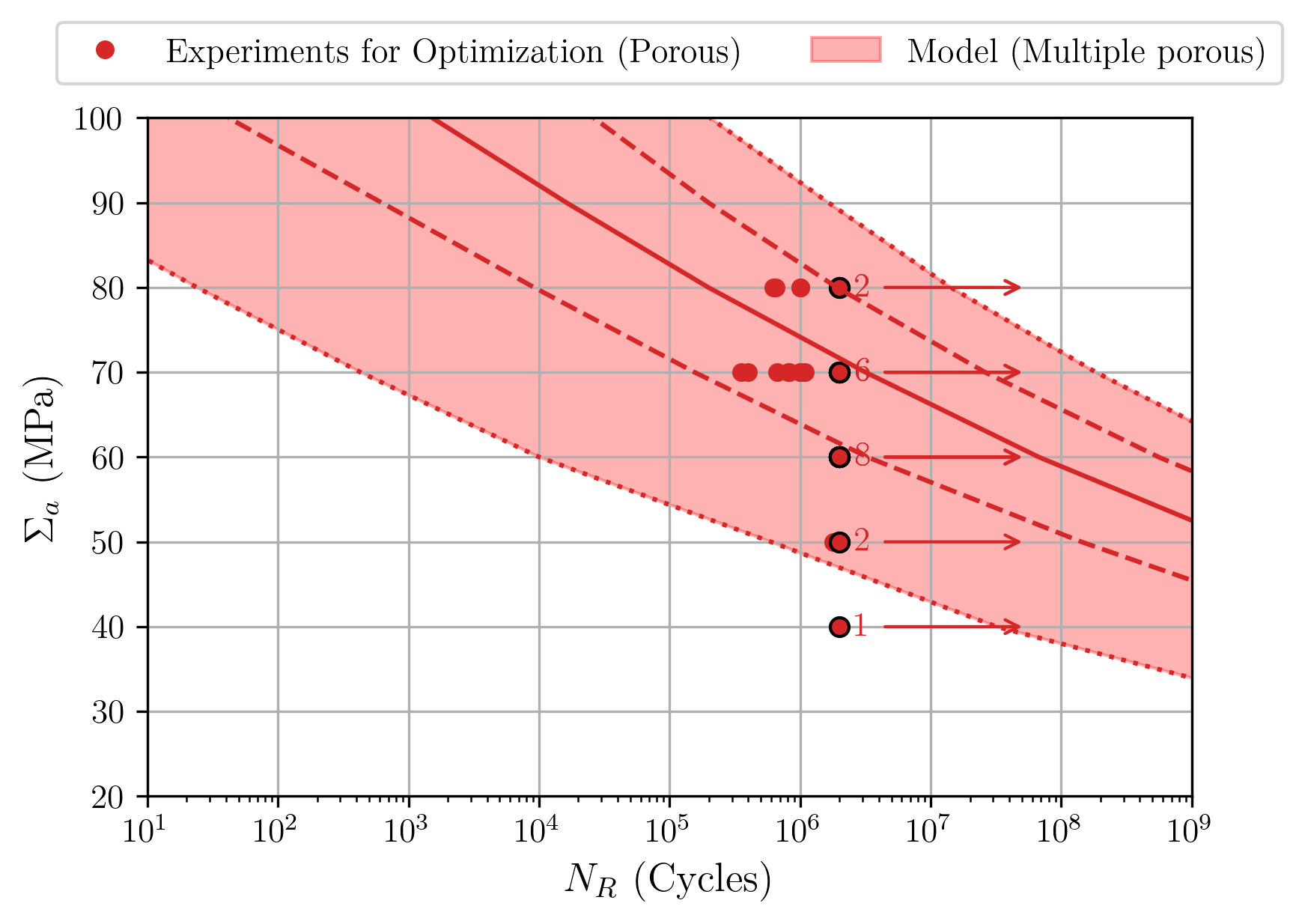}
            \caption{One-line model, $\mu = [m,A,\alpha,C]$}
        \end{subfigure}

        \caption{ Results of the probabilistic strain-life model using the multi-scale approach, identified on all available porous data, with two fatigue lifetime models. The solid line represents the median, the dashed lines represent the 15\% and 85\% quantiles, and the dotted lines represent 1\% and 99\% quantiles. }
        \label{Fig:Heterogenous_cases}
\end{figure}

\noindent The comparison between a two-line strain life model and one line strain life model is shown in Fig.  \ref{Fig:Heterogenous_cases}. The two models (with four and six parameters) yield qualitatively similar probability distributions for the specimen fatigue lifetime (with the 6 parameter model having only a very slightly lower likelihood value than the one obtained using the 4 parameter model). We will further investigate the more general two-line strain-life model with 6 parameters, to allow for the possibility of two distinct regimes.

\paragraph{Lifetime prediction on a non-porous test specimen}\mbox{}\\

\begin{figure}[!htbp]
    \centering
        \begin{subfigure}[b]{0.8\textwidth}
            \includegraphics[width=1\textwidth]{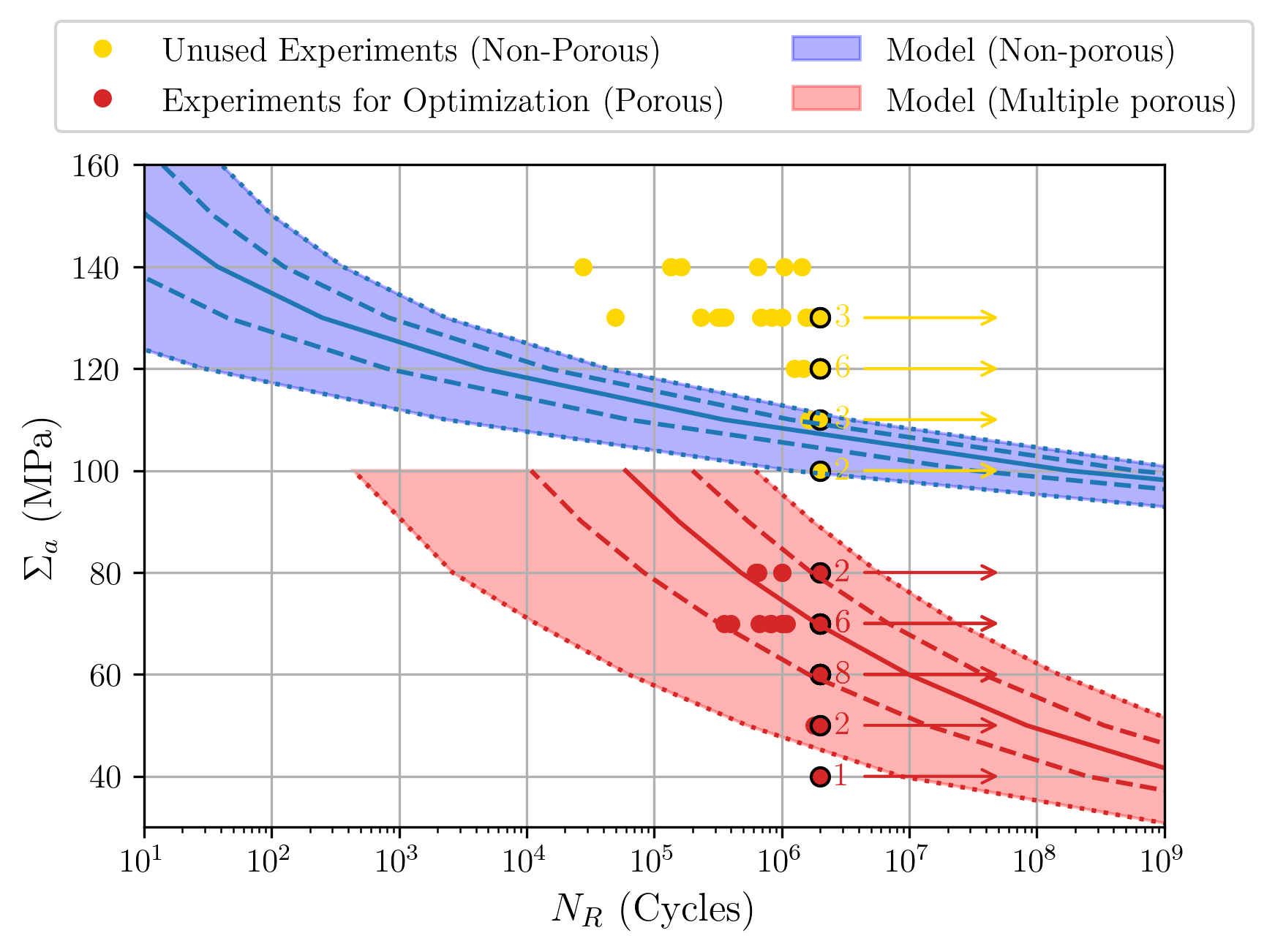}
            \caption{}
        \end{subfigure}
        
        \caption{ Prediction of the multi-scale probabilistic strain-life model identified on all available porous data, with all parameters active ($\mu = [m,A,B,\alpha,\beta,C]$), on a non-porous specimen. The solid line represents the median, the dashed lines represent the 15\% and 85\% quantiles, and the dotted lines represent 1\% and 99\% quantiles.
 }
        \label{Fig:onefitwohler_6params_predictnonporous_optimporous}
\end{figure}

\noindent Ideally we should also be able to predict the fatigue lifetime of porous materials with different pore distributions than the ones that were experimentally tested. Using the two-line fatigue model identified on the porous data, we try to predict the fatigue behavior of a non-porous specimen. The results are shown in Fig.  \ref{Fig:onefitwohler_6params_predictnonporous_optimporous}. We see that the prediction on the non-porous specimens is incorrect. The predicted lifetime is lower than what experiments suggest. This could be due to three reasons:
\begin{description}
    \item[1.] Misrepresentation of pores: Several uncertainties and assumptions have been propagated while meshing the pores (smoothing the surfaces, removing smaller pores, etc.). This could explain the discrepancy; the real state of stress due to the pores may not be correctly modelled. It is also possible that the surface roughness plays a role, at a scale that is smaller than that of the tomographic scan of the pores, and is not taken into account in the mesh.
    
    \item[2.] Lack of data: The current data-set for porous materials lacks fatigue lifetime data for when the material around the pores is loaded at the same stress levels (100-140 MPa) as in the data-set for non-porous specimens. To address this, we require experiments in one of two ways. One possible solution is obtaining further experimental data on porous specimens at very low stress levels, such that the material around the pores reaches the required stress levels (100-140 MPa) that the non-porous specimens were subject to. Another solution is to obtain experimental data on non-porous specimens at the required (100-140 MPa) levels. Using this additional data during the optimisation process, we can identify a model that is transferable between material with these two levels of porosity.
    
    \item[3.] It could also be possible that the two base materials are different, or that the material in the vicinity of pores behaves differently under fatigue loading. Also, the hydro-static pressure may have an effect on fatigue behavior, but is not taken into account in the fatigue model criterion.
\end{description}

\clearpage
\subsection{Towards a model transferable between different pore populations using lesser experimental data}\label{sec:jointresults_reduceddata}

In this section, we demonstrate that identifying a model that is transferable between different types of pore populations is facilitated by the multi-scale methodology proposed in this paper. 

A simplified method to model the fatigue behavior of materials with different pore populations is to treat them as separate materials with different fatigue properties, and each material is thus fit separately with separate fatigue models. To do this, a large number of data-points would be required for identifying the parameters of the fatigue models for each of these materials.

The multi-scale approach, however, considers different pore populations to have the same base material. Fatigue lifetime data associated with different data-sets pertaining to different pore populations therefore share the same underlying statistics. Thus, the multi-scale approach is expected to require less data for identifying a model that is transferable between different types of pore populations.

Only the two-line fatigue model will be considered for the following numerical examples, in order to allow for the presence of two distinct regimes:

\begin{equation}
    \mu = [m,A,B,\alpha,\beta,C]
\end{equation}

\begin{figure}[h!tbp]
  \centering
    \begin{subfigure}[b]{0.705\textwidth}
        \includegraphics[width=\textwidth]{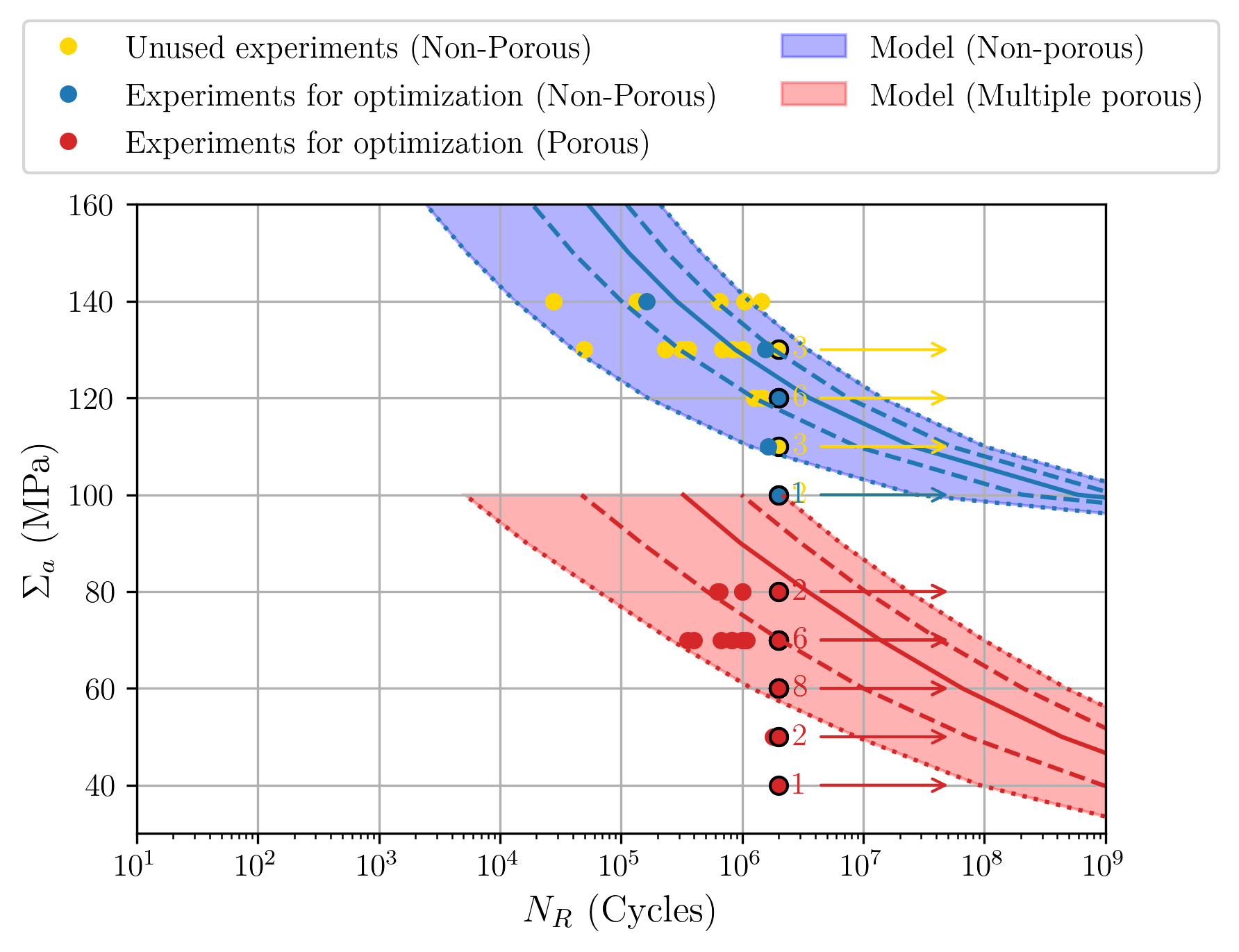}
    \end{subfigure}
    \caption{\label{fig:StrainLife_LesserExps_Case1} Identified fatigue model on the reduced data-set $\mathcal{I^{\text{red}}}$ of 5 random non-porous data points and the data-set $\mathcal{J}$ of all the available 32 porous data points, using the multi-scale method}
\end{figure}

We now consider the fatigue lifetime data-set associated with non-porous and porous material presented in section \ref{sec:expdata}. A reduction of the non-porous fatigue lifetime data-set $\mathcal{I}$ is done by randomly picking one experimental point for each applied stress amplitude level. The resulting reduced non-porous dataset $\mathcal{I^\text{red}}$ is shown in Fig. \ref{fig:StrainLife_LesserExps_Case1} (a).

\paragraph{Identification of lifetime model on reduced data-set: multi-scale method}\mbox{}\\

\noindent Here, we are identifying the fatigue lifetime model on both structures with homogeneous stress distributions (non-porous specimens) and structures with pores whose exact distribution is unknown (porous specimens) simultaneously. The function to minimize is taken as a sum of the functions defined in section \ref{sec:Identification}:

\begin{equation}
        \mu =
        \argmax_{\tilde{\mu}} {\left( \sum_{i\in \mathcal{I^\text{red}}} \ln f_{N_{\textrm{R}}^{\textrm{s}}}(N^{i}; \tilde{\mu}, \Sigma_a^{i})  + \sum_{i\in \mathcal{J}} \ln \left(  \frac{1}{n_{\textrm{k}}}\sum_{k\in \tilde{K}} f^k_{N_{\textrm{R}}^{\textrm{s}}}(N^{i}; \tilde{\mu}, \Sigma_a^{i})  \right)\right)}
    \end{equation}

The reduced set of non-porous fatigue experiments $\mathcal{I^\text{red}}$ is used in the optimisation process. The same set $\tilde{K}$ used in the previous section is taken for the synthetically generated porous specimens, and $n_\text{k}=10$.

The model parameters are jointly identified on the reduced non-porous data-set $\mathcal{I^\text{red}}$ and the porous data-set $\mathcal{J}$. The result of the optimisation process is shown in Fig.  \ref{fig:StrainLife_LesserExps_Case1}. The fit is consistent on both the non-porous as well as porous cases, despite the reduction of non-porous data. This indicates that fewer fatigue experiments are required to identify the parameters when using the multi-scale approach. This is because the multi-scale approach considers porous and non-porous specimens as the same material, and information from the two sets of data are  statistically correlated.

\paragraph{Identification of lifetime model on reduced data-set: homogenised method}\mbox{}\\

\begin{figure}[h!tbp]
  \centering
    \begin{subfigure}[b]{0.765\textwidth}
        \includegraphics[width=\textwidth]{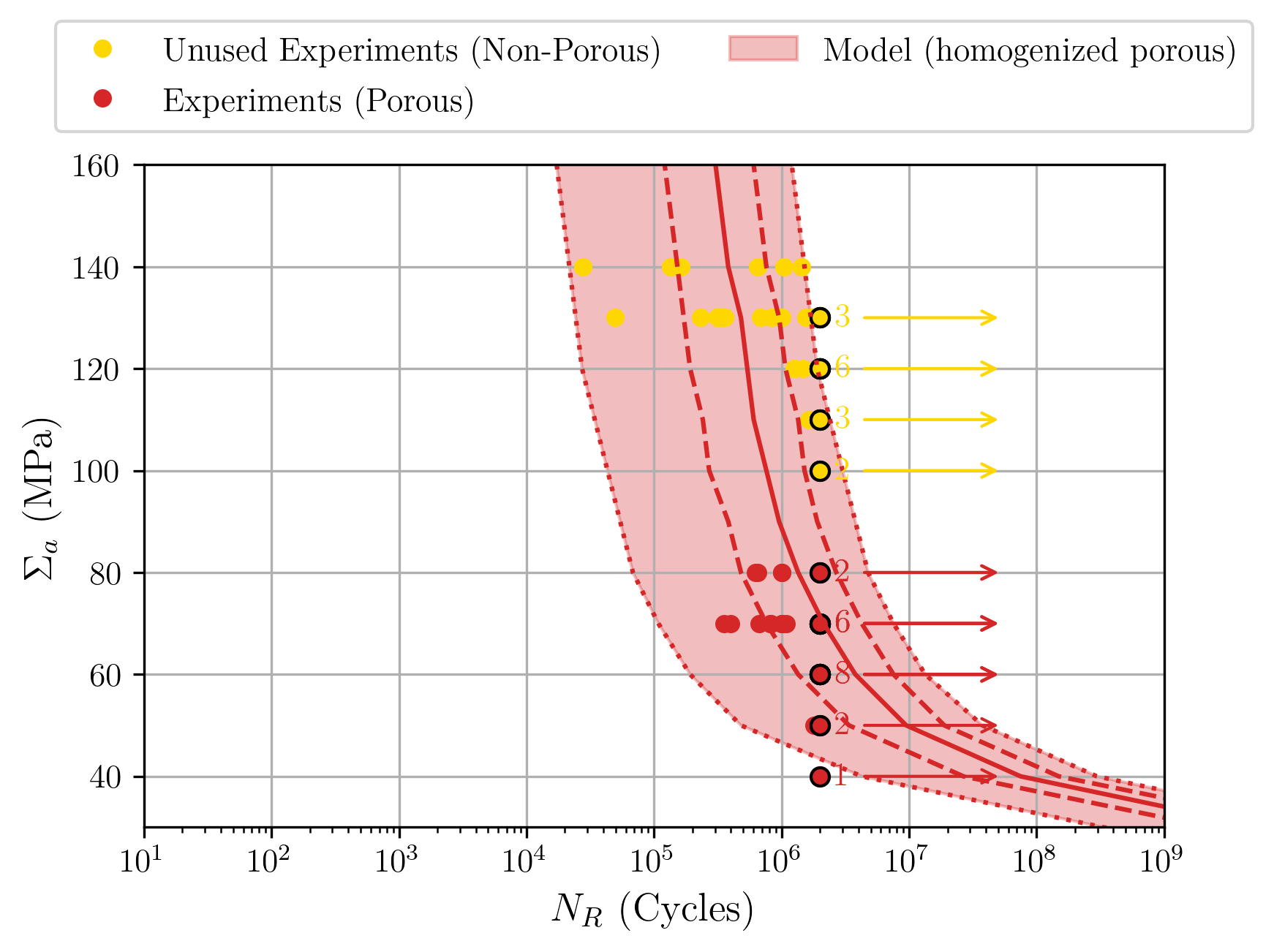}
    \end{subfigure}

    \caption{\label{fig:StrainLife_LesserExps_Case2} Identified fatigue model on the data-set $\mathcal{J}$ of all the available 32 porous data points, considering the porous material as a homogenised material, using a 0D identification method}
\end{figure}

\noindent Identifying a probabilistic fatigue model on the reduced data-set $\mathcal{I^\text{red}}$ is difficult due to an insufficient data-set, which makes several solutions possible.
When considering the porous material as a new 'homogenised' material, porous specimens may be considered to contain homogenised material with equivalent fatigue properties. A homogenised model is identified without using the multi-scale method, using the 0D identification procedure detailed in section \ref{sec:ParamsIdentification_Homogenous} on the set $\mathcal{J}$ of porous experimental data. This gives the fit shown by the red curve in Fig. \ref{fig:StrainLife_LesserExps_Case2}. While the fit is good on the porous specimens given the data, the model remains non-transferable to other types of pore populations. This is seen when extrapolating the homogenised porous model to the stress levels at which non-porous specimens were tested. The homogenised porous model overlaps with the non-porous experimental data, which is against the generally observed trend of porous material failing sooner than non-porous material.

\clearpage
\section{Numerical investigations of multi-scale method properties}

In this section, certain properties of the multi-scale method will be investigated. The dependence of the mesh size on the fatigue lifetime curve will be shown. The pore-surface interaction effect under iso-volume conditions will be investigated, as well as the volumetric size effect. Finally, an attempt to replace the multi-scale method by using a naive homogenization technique will be presented.

\subsection{Mesh convergence}

\begin{figure}[htbp]
    \centering
    \begin{subfigure}[b]{0.35\textwidth}
        \includegraphics[width=\textwidth]{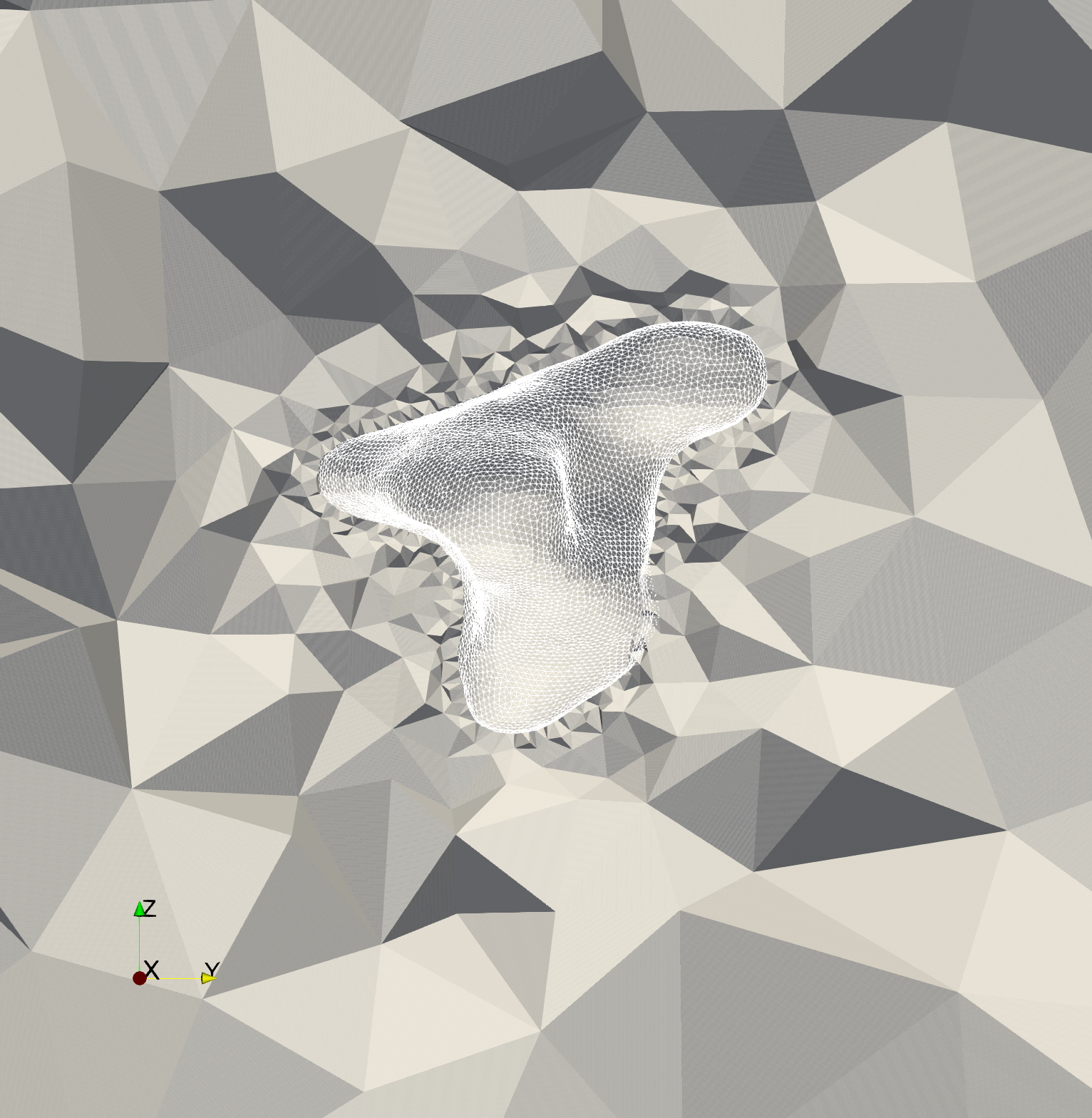}
        \caption{}
    \end{subfigure}
    \begin{subfigure}[b]{0.35\textwidth}
        \includegraphics[width=\textwidth]{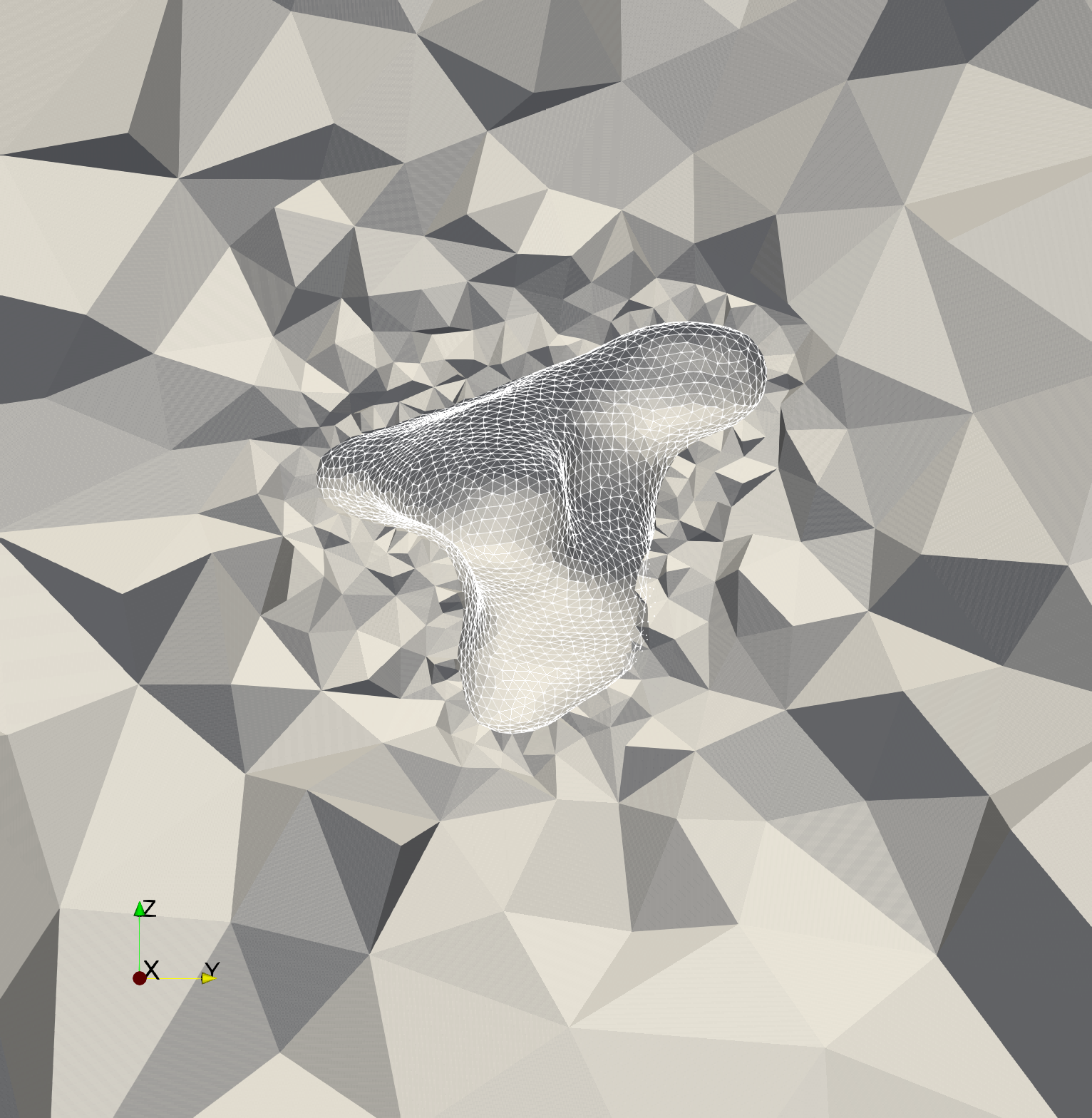}
        \caption{}
    \end{subfigure}
    \begin{subfigure}[b]{0.35\textwidth}
        \includegraphics[width=\textwidth]{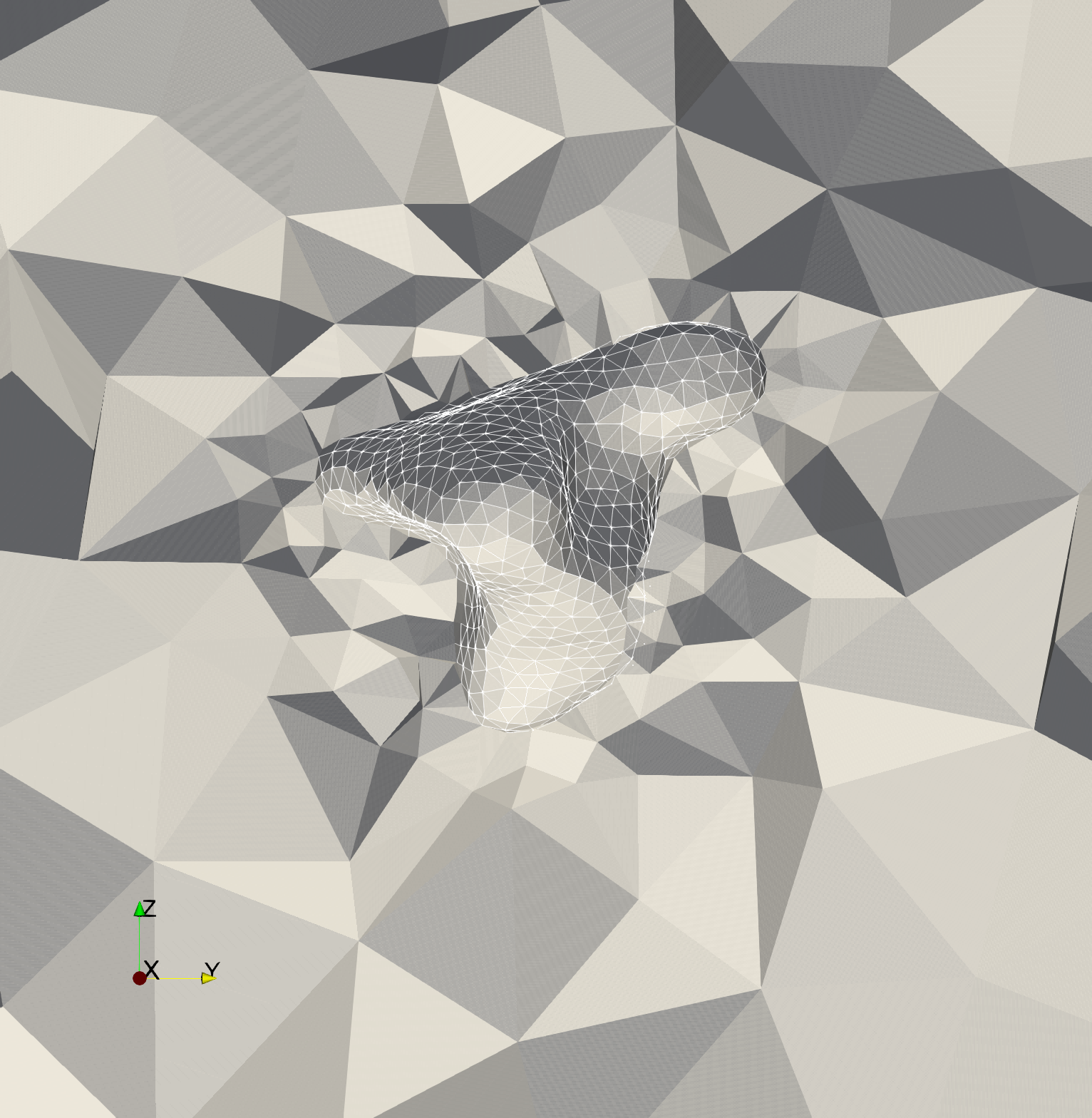}
        \caption{}
    \end{subfigure}
    \begin{subfigure}[b]{0.35\textwidth}
        \includegraphics[width=\textwidth]{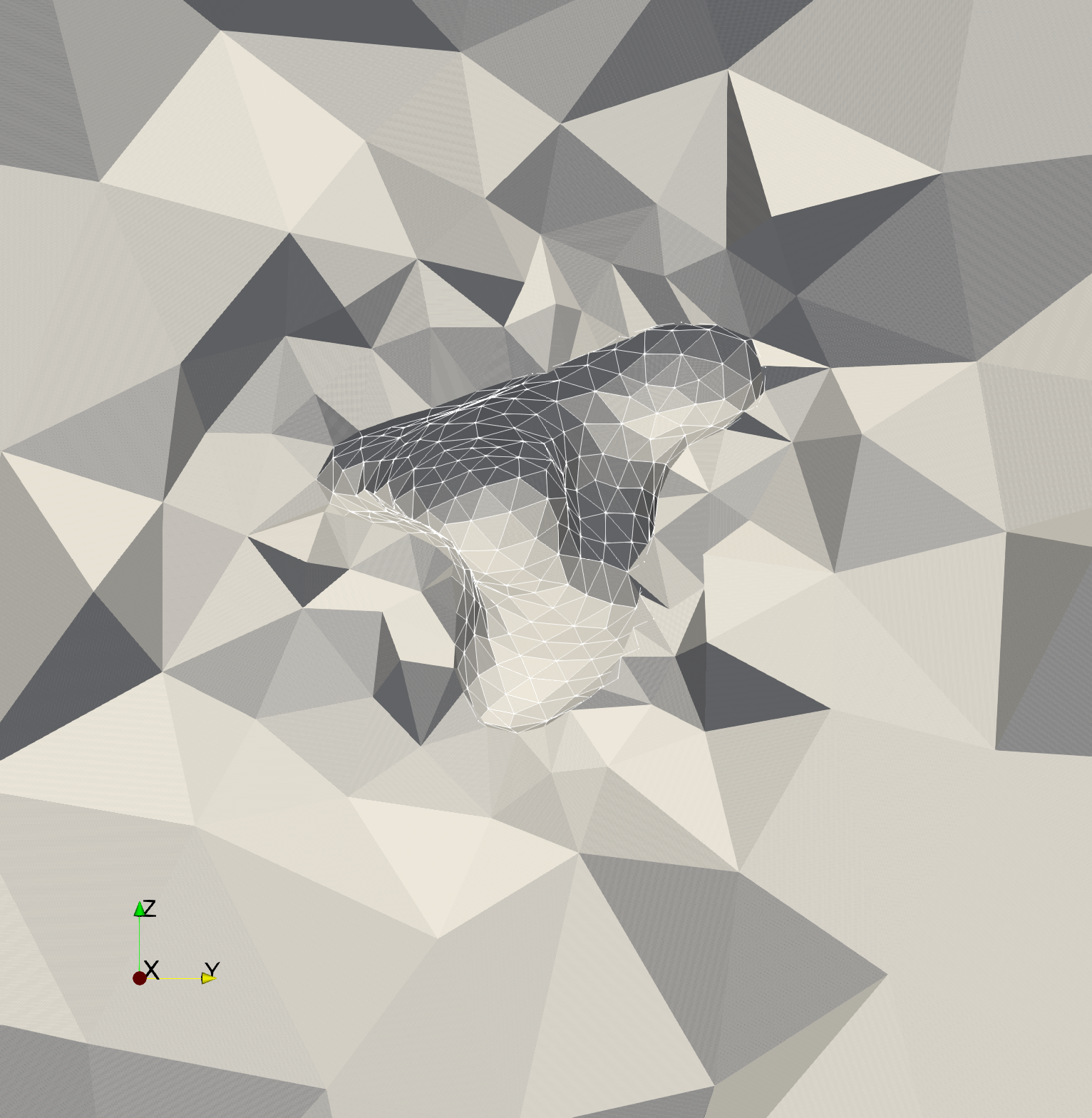}
        \caption{}
    \end{subfigure}
    
    \caption{\label{fig:meshrefinements} Mesh refinement around a pore, with the smallest element having size (a) 5 $\mu m$ (b) 10 $\mu m$ (c) 19 $\mu m$ (d) 29 $\mu m$}
\end{figure}

\begin{figure}[htbp]
        \centering
        \includegraphics[width=0.8\textwidth]{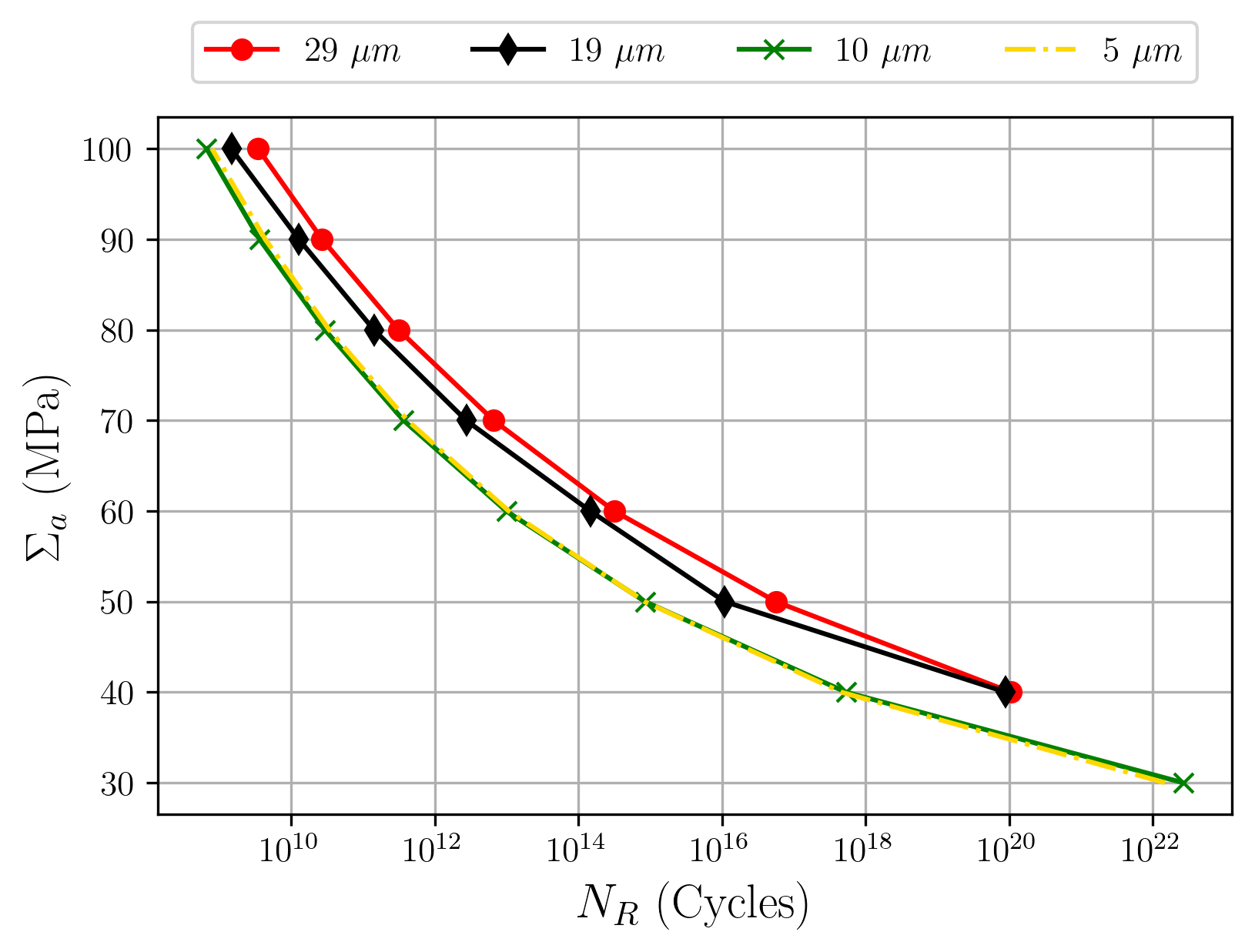}
        \caption{\label{fig:meshrefinements_solution} Effect of mesh size on the median lifetime for a given set of two-line strain life model parameters}
\end{figure}

The influence of the mesh size on the median of the lifetime predicted by the model is studied in this subsection. A synthetic specimen with one pore was meshed at different levels of refinement. The two-line fatigue model with parameters taken from section \ref{sec:jointresults_reduceddata} is taken. Figure \ref{fig:meshrefinements} shows four different refinement levels, with the size of the smallest element going from 5 microns to 29 microns. Figure \ref{fig:meshrefinements_solution} shows the effect of these mesh sizes on the median lifetime, across several load levels. As seen in this figure, the median fatigue lifetime curve exhibits a dependence on mesh size. The lifetime curve for the mesh with the smallest element size of 5 $\mu$m is identical to that for the 10 $\mu$m mesh. The optimal mesh size, beyond which further refinement does not change the solution, therefore lies between 10 and 19 $\mu$m.

\subsection{Pore-surface interaction effect (iso-volume condition)}\label{sec:isovolumeeffect}

\begin{figure}[h!tbp]
\centering
\begin{subfigure}[b]{1\textwidth}
    \centering
    \centering
    \includegraphics[width=0.8\textwidth]{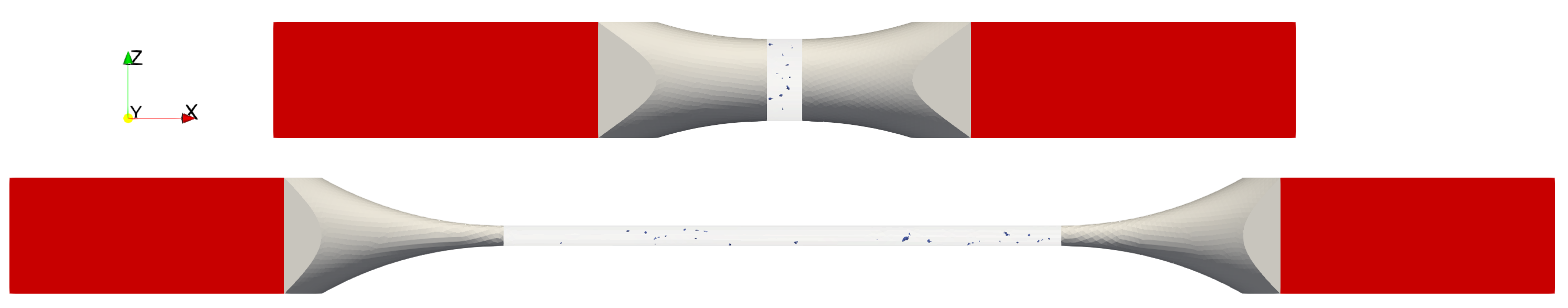}
    \caption{Examples of one realization of synthetically generated specimens: one with radius $r$ and length $L$, the other with radius $r/4$ and length $4^2 L$, thus having the same volume but different levels of pore-surface interaction.}
\end{subfigure}
\begin{subfigure}[b]{1\textwidth}
    \centering
    \centering
    \includegraphics[width=0.8\textwidth]{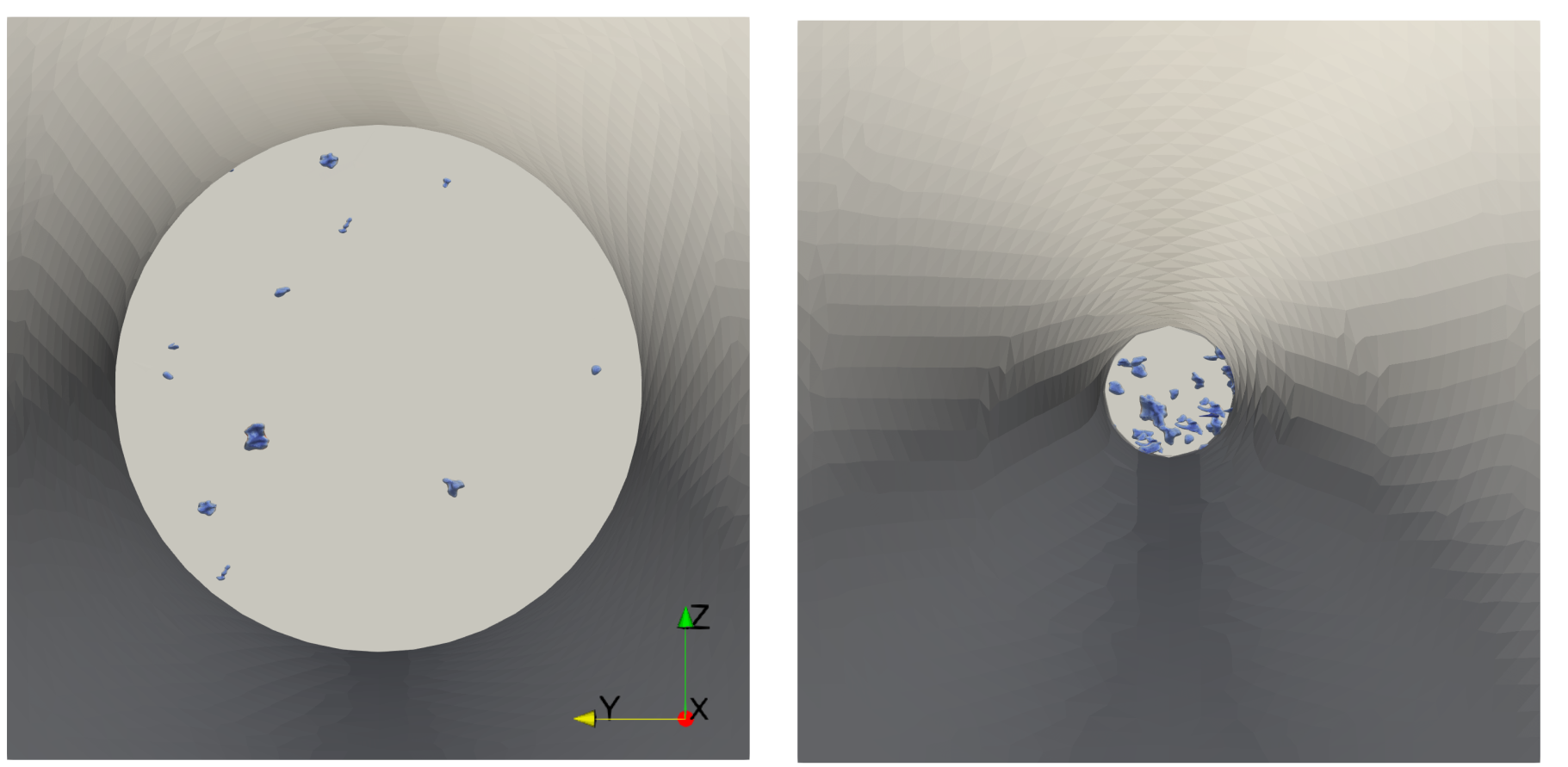}
    \caption{Projection of all the pores in the two specimens into the YZ plane}
\end{subfigure}
\begin{subfigure}[b]{1\textwidth}
    \centering
    \includegraphics[width=0.8\textwidth]{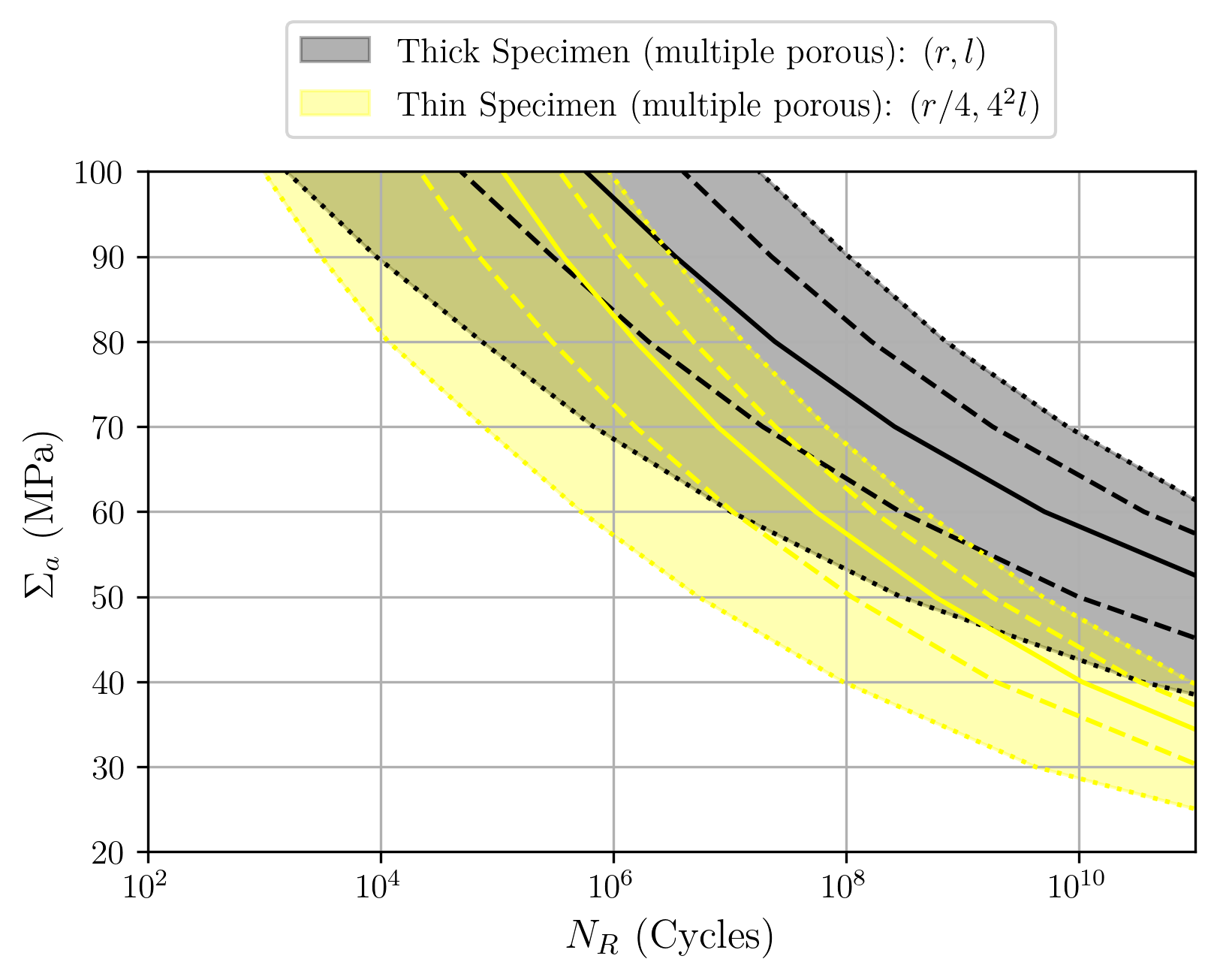}
    \caption{Wöhler curves for the original and thin specimens}
\end{subfigure}
\caption{Effect of varying pore-surface interaction under iso-volume conditions on fatigue lifetime, as demonstrated on two specimens}
\label{Fig:SizeEffectThinspecimen}
\end{figure}

Thin specimens, transformed from a thicker specimen to have more surface area to volume ratio, while keeping the volume of the two specimens equal, have higher pore-surface interaction, and consequently have a lower lifetime - this effect has been reported before by other authors \cite{Pegues2017, romano2019}. This effect is an automatic output of the two scale method introduced in this paper. Figure \ref{Fig:SizeEffectThinspecimen}(a) shows the geometries of two specimens: one specimen has radius $r$ and length $L$, the other specimen has radius $r/4$ and length $4^2 L$ in order to keep the volume constant. Both of them are populated with pores using the same method presented in section \ref{sec:MesoScale}. For the thick specimen, the projected image of the pores (see Figure \ref{Fig:SizeEffectThinspecimen}(b)) shows a few pores interacting with the surface. The thin specimen, owing to lower cross-sectional area, has more pores interacting with the surface.

To avoid the effect of randomness in pore distributions, the expected lifetime needs to be computed for both specimens. Similar to section \ref{sec:results_twoscale_porousonly}, a Wöhler curve representing uncertainty due to varying pore distributions is constructed by combining samples from a certain number of realizations of synthetically generated pore fields in the two specimens. For each nominal stress level, the failure distribution representing this uncertainty in pore distributions is constructed by combining 1000 samples each from 10 different randomly generated realizations. The quantiles are then computed on 10000 samples for each nominal stress level, which are joined to construct the quantiles of the porous Wöhler curve.

The predictions of the fatigue lifetime model, shown in Fig. \ref{Fig:SizeEffectThinspecimen}(c), shows a decrease in the fatigue lifetime of a thinner specimen  over all applied stress amplitude levels.

\clearpage
\subsection{Volumetric size effect (iso-stress condition)}\label{sec:volumeeffect}

\begin{figure}[h!tbp]
  \centering
  \begin{subfigure}[b]{0.8\textwidth}
        \includegraphics[width=\textwidth]{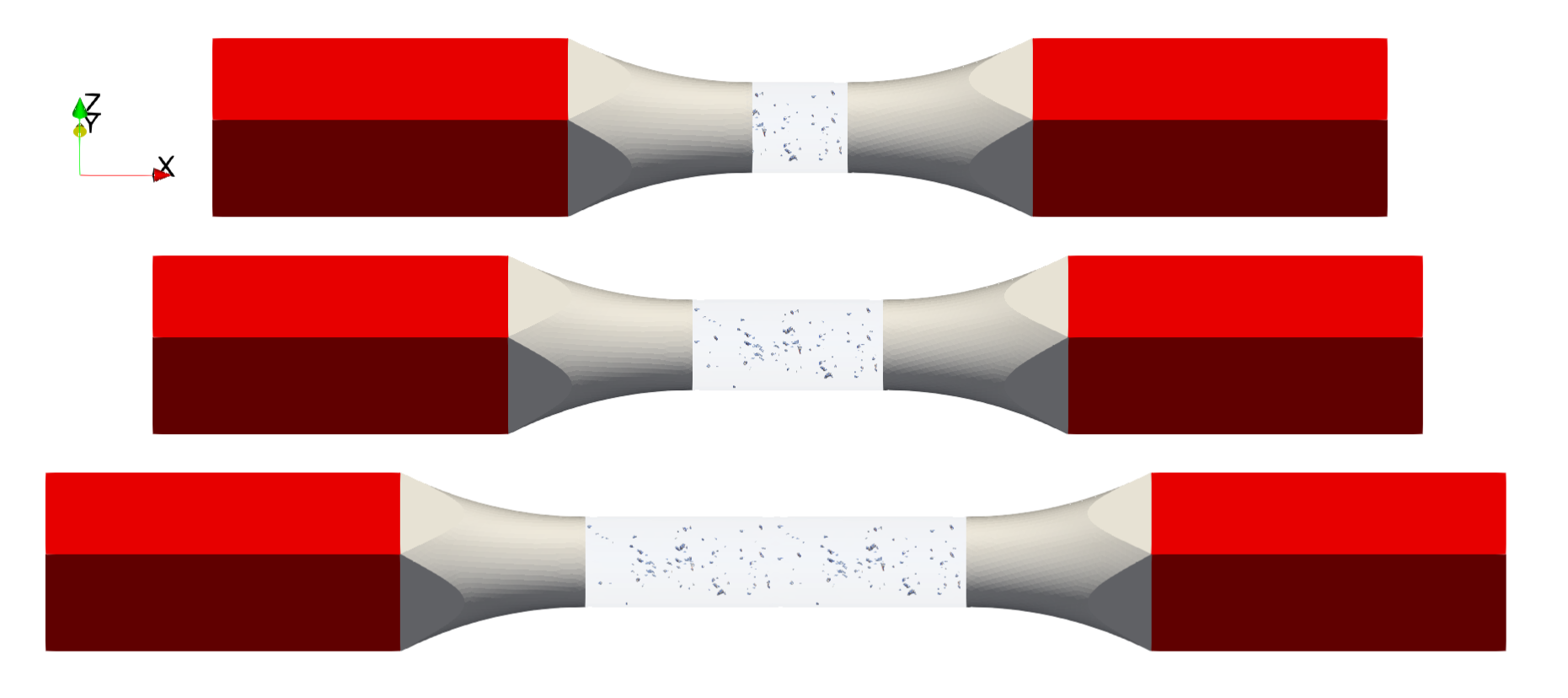}
        \caption{Synthetically generated porous specimens of volume $V/2$, $V$ and $2V$ (owing to different lengths of gauge sections but same cross sectional area), undergoing the same loading (same applied stress amplitudes)}
    \end{subfigure}
    \begin{subfigure}[b]{0.8\textwidth}
        \includegraphics[width=\textwidth]{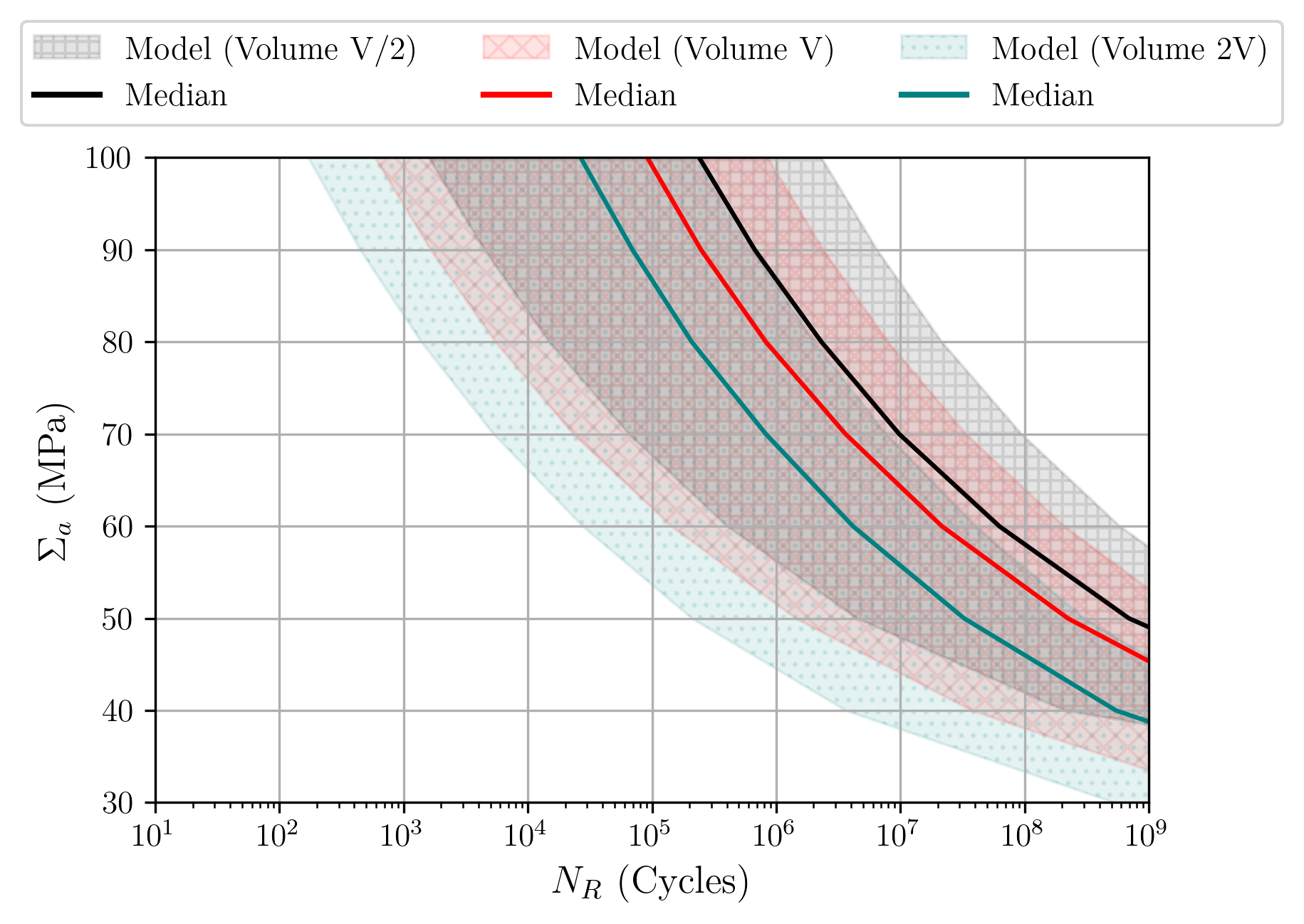}
        \caption{Wöhler curves obtained by the model on the three specimens, with solid lines showing the median fatigue lifetimes and hatch filled areas demarcating the area contained between 1\% and 99\% quantiles of the fatigue lifetime distributions}
    \end{subfigure}
    \caption{\label{fig:volumesizeeffect} Demonstration of the statistical size effect due to varying volumes}
\end{figure}

As an automatic outcome of the model, specimens with different volumes but loaded under the same stress amplitude show a statistical size effect. Fig. \ref{fig:volumesizeeffect} shows the Wöhler curves of three synthetically generated porous specimens with volumes $V/2$, $V$ and $2V$. The same pore population in the smallest specimen was 'tiled' multiple times to get bigger specimens, to isolate only the specimen volume size effect (and not add lifetime variability due to changing pore distributions as demonstrated in section \ref{sec:results_twoscale_porousonly}). Across all load levels, specimens with successively higher volumes show a decreasing fatigue lifetime. This is the well known statistical size effect in fatigue, i.e. bigger specimens have a larger probability of finding a critical flaw \cite{Makkonen2001, ElKhoukhi2021}.

\clearpage
\subsection{Naive homogenization to replace multi-scale method}\label{sec:transferability}

The multi-scale method developed in this paper is expensive to compute for entire structures, as several configurations of pore distributions need to be evaluated. In this section, we will try to replace the multi-scale method, termed $\textbf{A}$, with a simplified, homogenised fatigue model, termed $\textbf{B}$. For the identification of the homogenised model, we will generate synthetic data using the multi-scale method, for several realizations of pore distributions in a given geometry. The question is whether the homogenised model will accurately approximate the fatigue lifetime solution when applied to different geometries. The two-line fatigue model will be used for this study, with a fixed set of parameters (similar to those in section \ref{sec:jointresults_reduceddata}). 



\paragraph{Generation of synthetic fatigue lifetime data using the multi-scale method}\mbox{}\\

\begin{figure}[h!tbp]
  \centering
    \begin{subfigure}[t]{0.27\textwidth}
        \includegraphics[width=\textwidth]{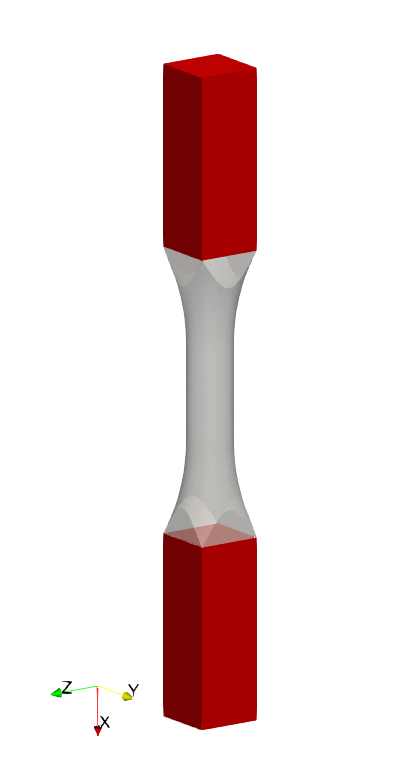}
        \caption{3D mesh of a pore-free cylindrical specimen}
    \end{subfigure}
    \begin{subfigure}[t]{0.6\textwidth}
        \includegraphics[width=\textwidth]{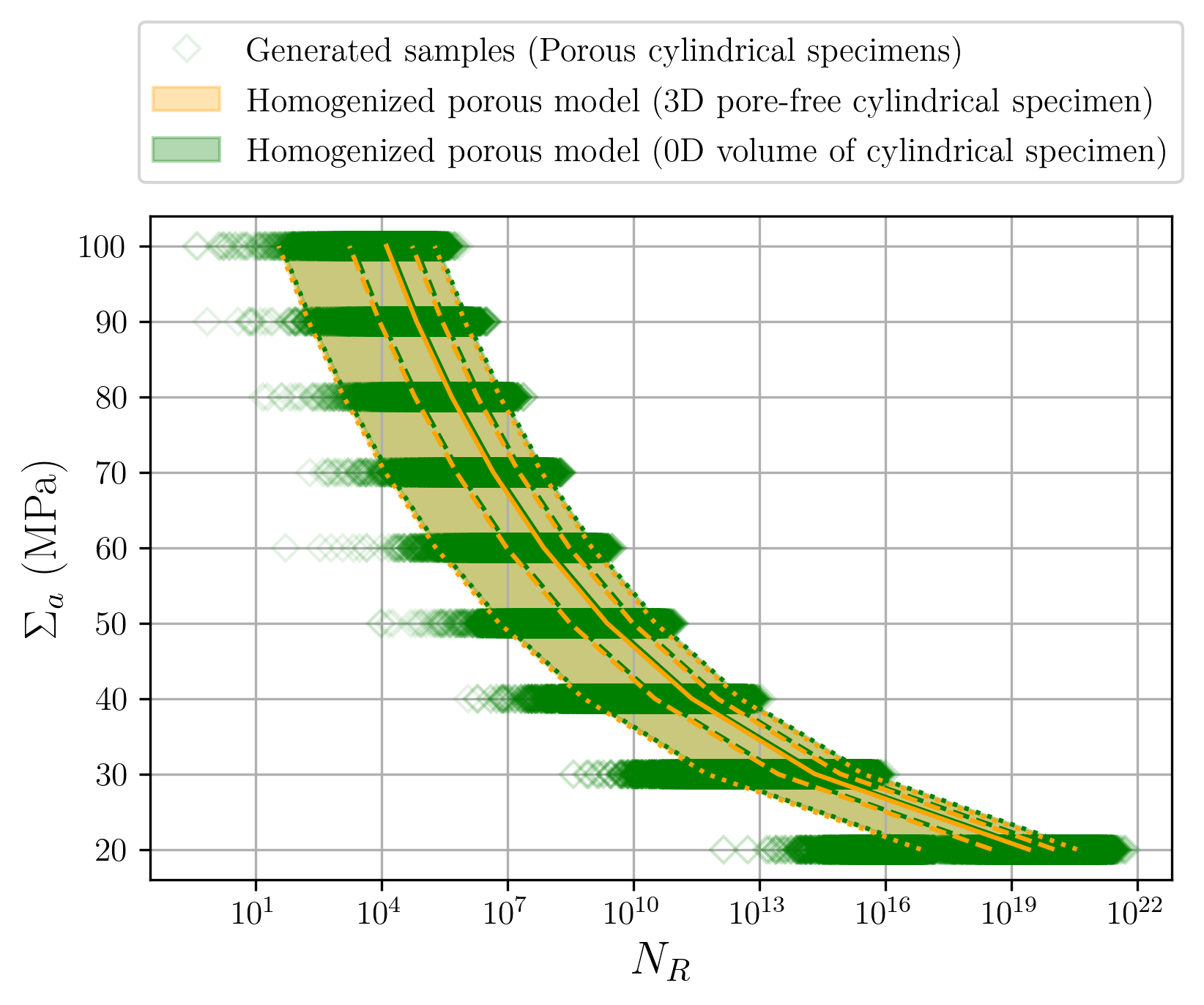}
        \caption{Generated synthetic lifetime data from cylindrical porous specimens, identified homogenised model and predictions of the homogenised model on the 3D mesh of the pore-free cylindrical specimen}
    \end{subfigure}
    \caption{\label{fig:transferability1} Identification of the parameters of a 0D homogenised porous fatigue model $\textbf{B}$ on generated synthetic lifetime data from the model $\textbf{A}$, and validation on a 3D FE pore-free cylindrical specimen of the same shape and volume as the original cylindrical porous specimens}
\end{figure}

\noindent We will now generate synthetic data using the multi-scale method $\textbf{A}$. A total of 10 cylindrical synthetically generated porous specimens were considered for data generation (examples of the geometries have previously been shown in section \ref{Fig:VaryingcriterionDistributionFullField}). For each synthetically generated porous specimen, 1000 samples each across 9 nominal loading levels were generated using this model, as shown by the green markers in Figure \ref{fig:transferability1}.

\paragraph{Identification of a homogenised 0D fatigue model}\mbox{}\\

\noindent Using the 0D identification procedure as described in section \ref{sec:ParamsIdentification_Homogenous}, the parameters of a fatigue lifetime model that maximized the likelihood of the synthetic generated data were obtained. This model $\textbf{B}$ is a homogenised porous fatigue lifetime model. To check the quality of this homogenised model, a 3D mesh of a homogeneous pore-free cylindrical specimen of the same size and geometry as the porous cylindrical specimens was created (shown in Figure \ref{fig:transferability1}(a)), and the model $\textbf{B}$ was used to obtain predictions on it using the weakest link method developed in section \ref{sec:structurelevel}. The results, shown in Figure \ref{fig:transferability1}(b), indicate that the homogenised model $\textbf{B}$ works well to replace the multi-scale method for the given geometry.

\paragraph{Notched geometry}\mbox{}\\

\begin{figure}[h!tbp]
  \centering
    \begin{subfigure}[b]{0.20\textwidth}
        \includegraphics[width=\textwidth]{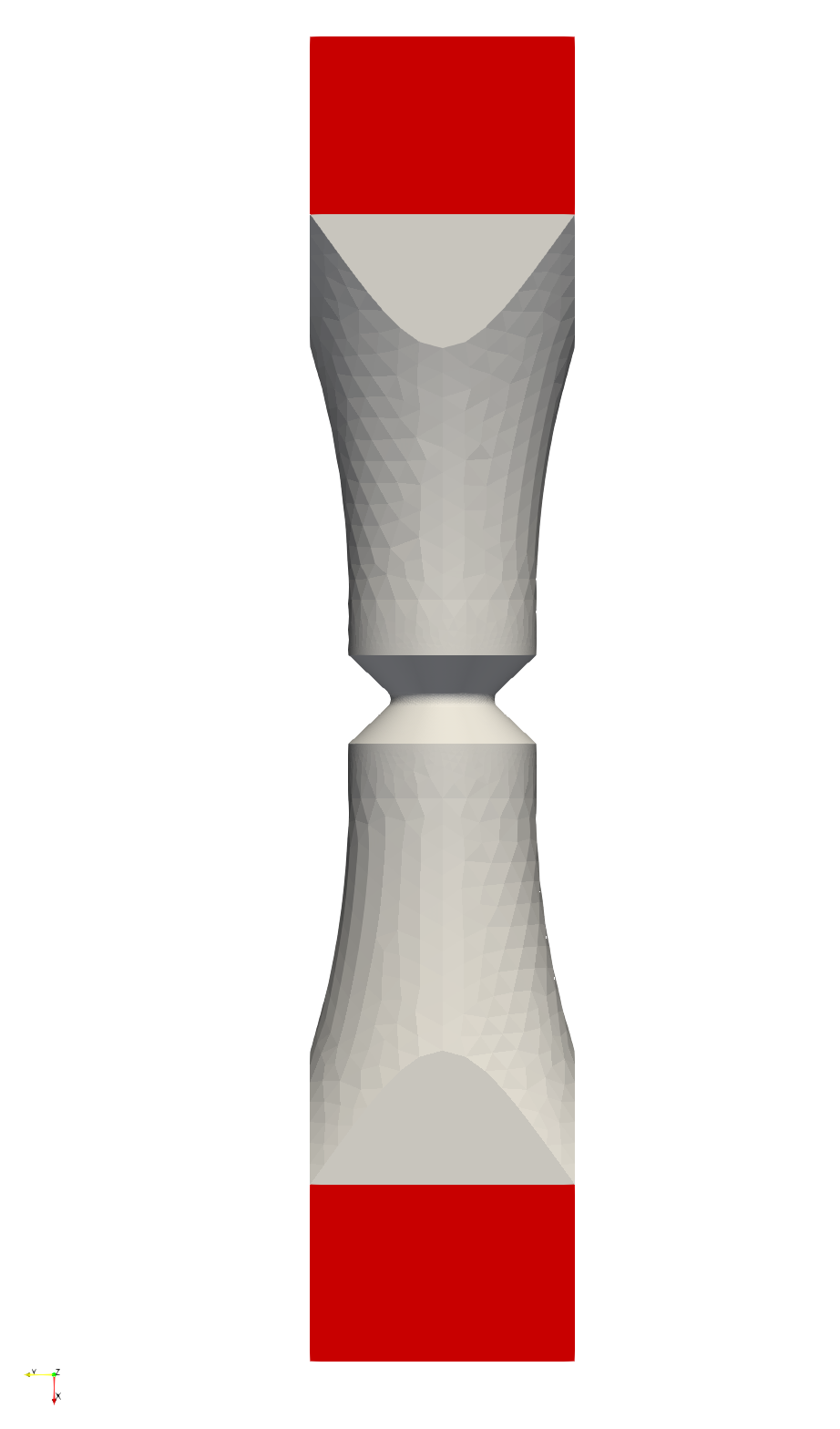}
        \caption{}
    \end{subfigure}
    \begin{subfigure}[b]{0.35\textwidth}
        \includegraphics[width=\textwidth]{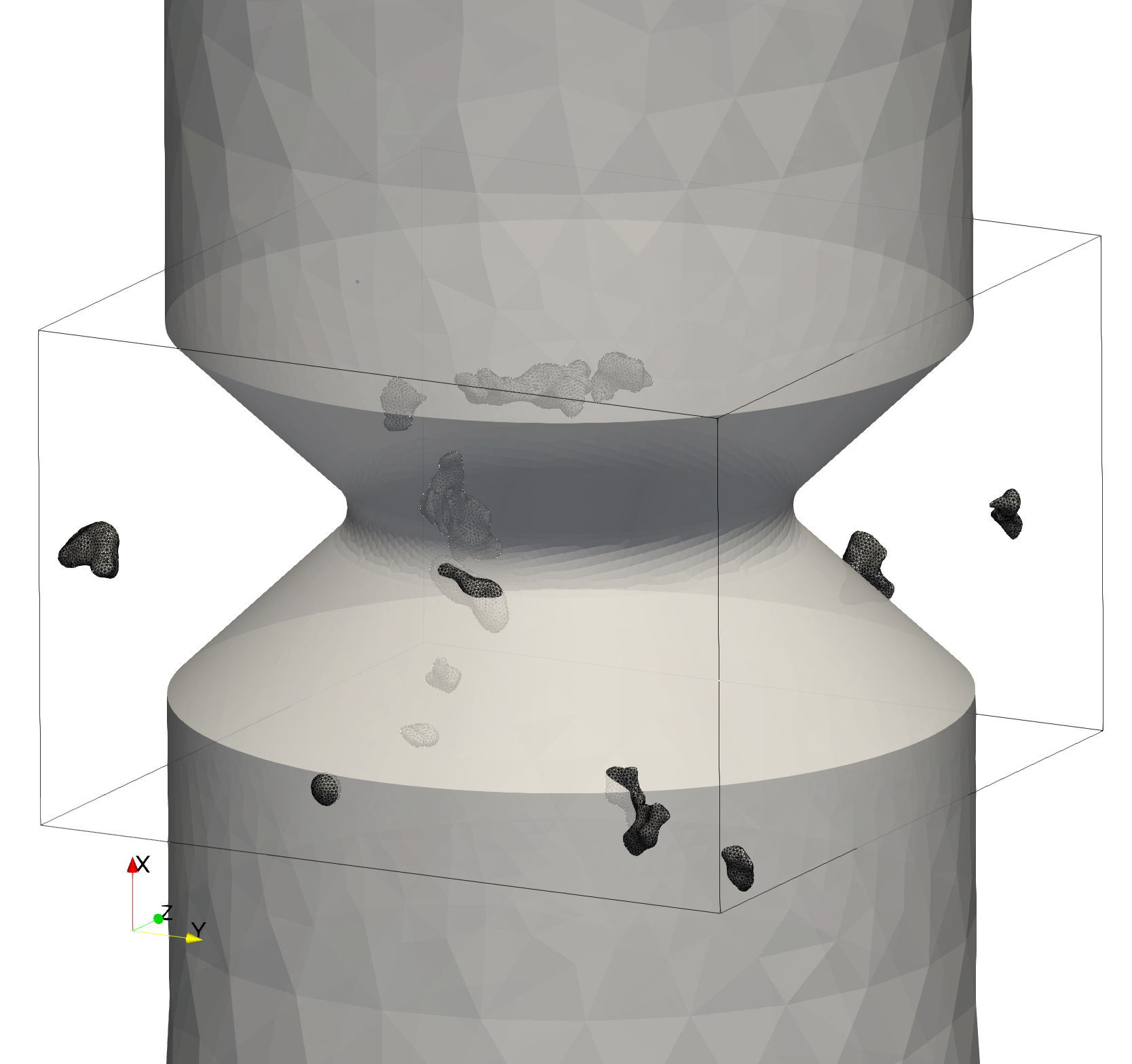}
        \caption{}
    \end{subfigure}
    \begin{subfigure}[b]{0.35\textwidth}
        \includegraphics[width=\textwidth]{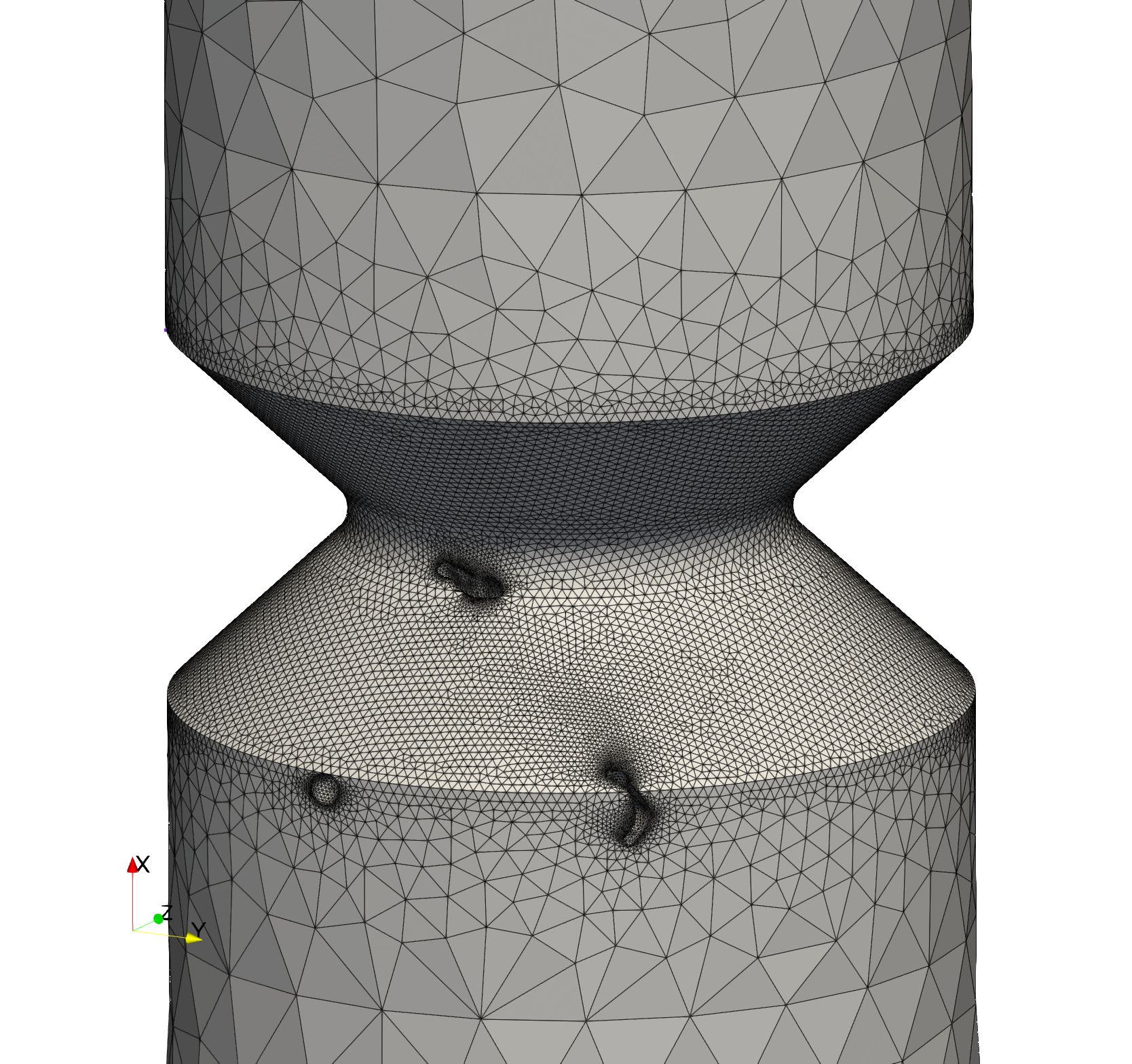}
        \caption{}
    \end{subfigure}
    
    \caption{\label{fig:transferability2} Creation of a notched specimen (a) boundary conditions (b) sub-volume of pores used for creation of a porous notched specimen (c) resulting mesh of the porous notched specimen}
\end{figure}

\noindent A geometry with a macroscopic notch was created, as shown in Fig. \ref{fig:transferability2}. The sub-volume of pores used to create a porous version of this geometry was taken from the same tomography used to create the cylindrical porous specimens. Next, several configurations of the notched specimen were created: 

\begin{description}
    \item[1.] Porous notched specimen with the normal expected pore density, with pores interacting with the notch
    \item[2.] Porous notched specimen with a surface-breaking pore interacting with the notch artificially removed
    \item[3.] Notched specimen without pores
\end{description}

\begin{figure}[h!tbp]
  \centering
    \begin{subfigure}[b]{0.48\textwidth}
        \includegraphics[width=\textwidth]{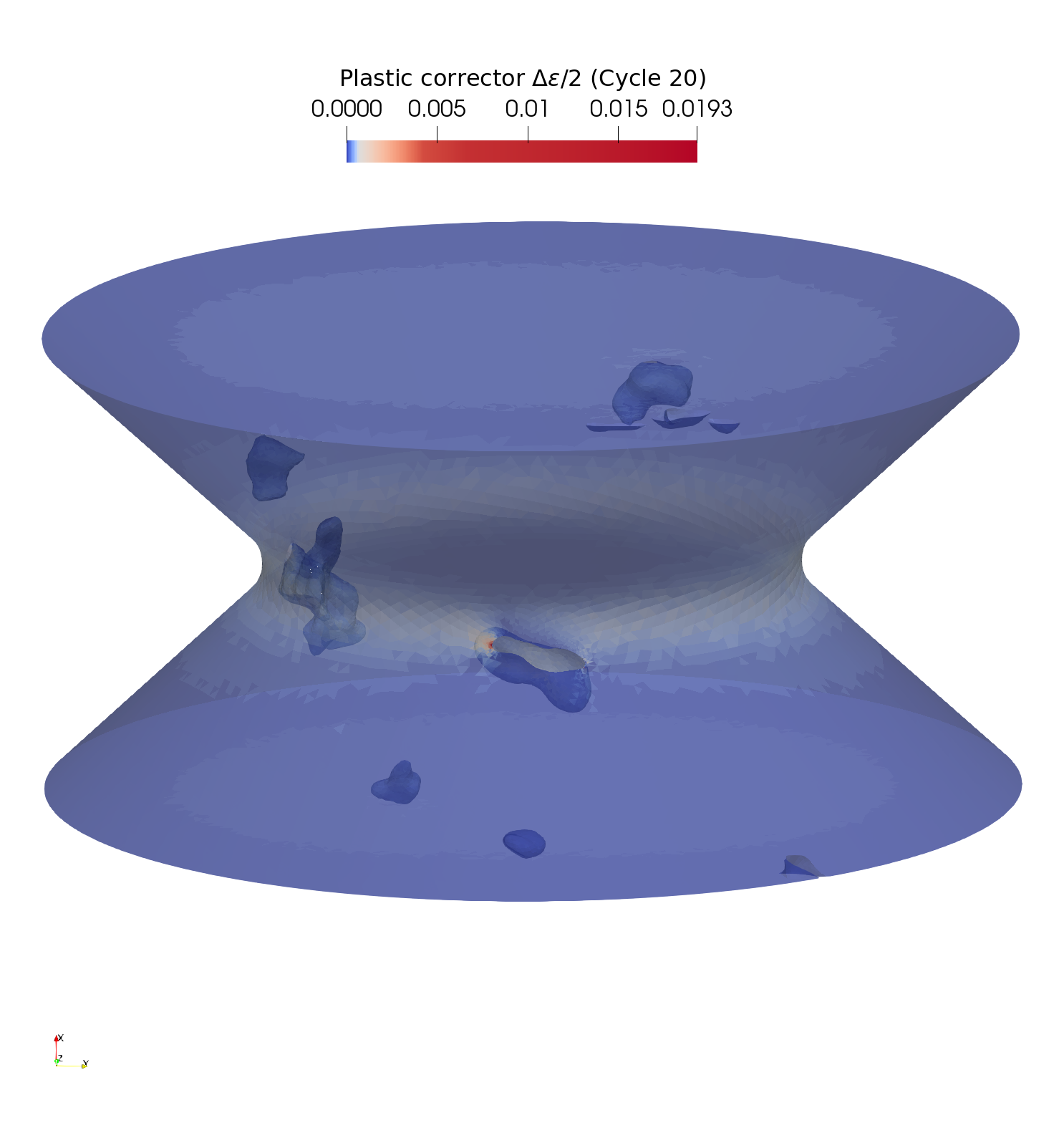}
        \caption{}
    \end{subfigure}
    \begin{subfigure}[b]{0.48\textwidth}
        \includegraphics[width=\textwidth]{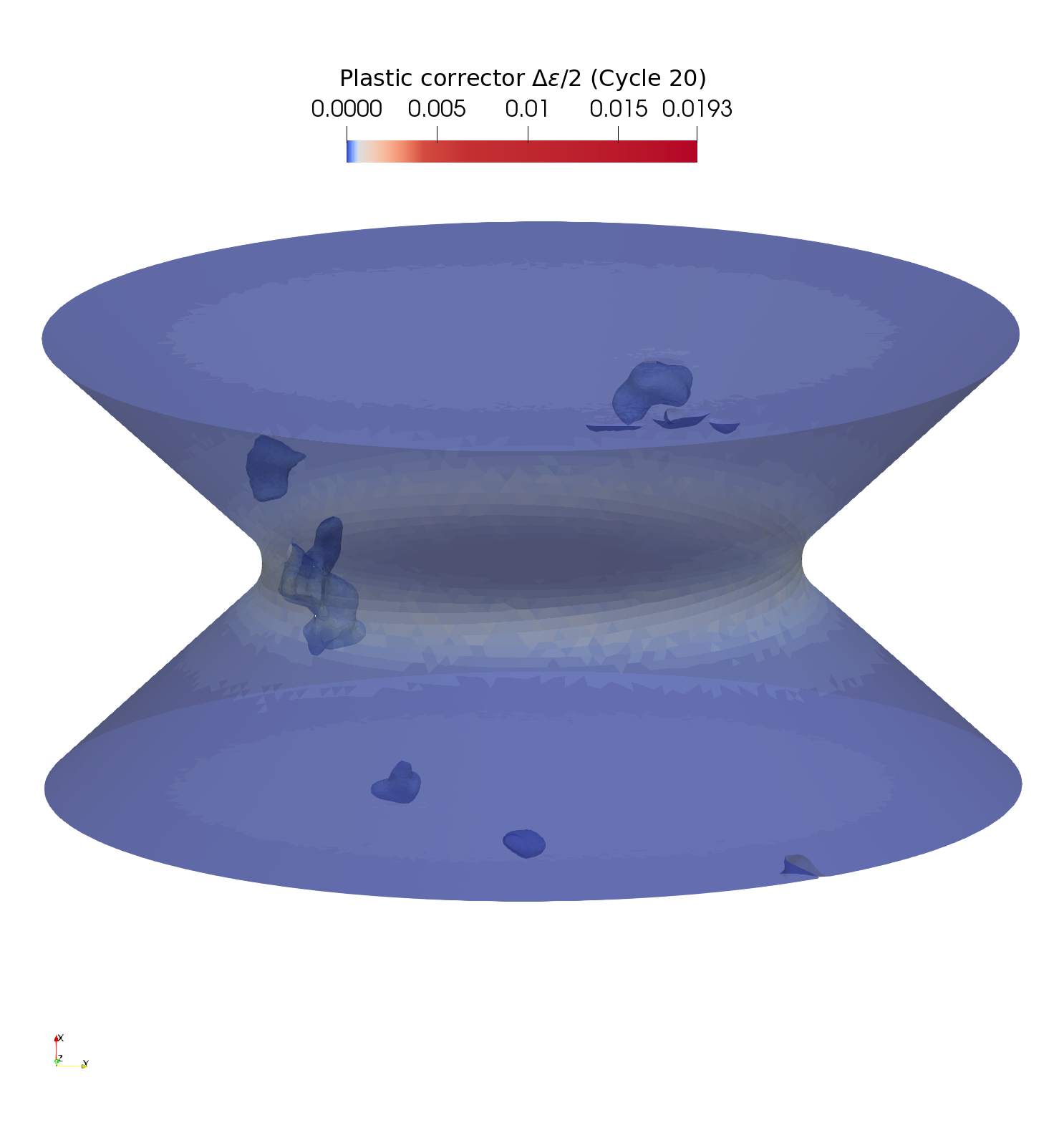}
        \caption{}
    \end{subfigure}
    \begin{subfigure}[b]{0.48\textwidth}
        \includegraphics[width=\textwidth]{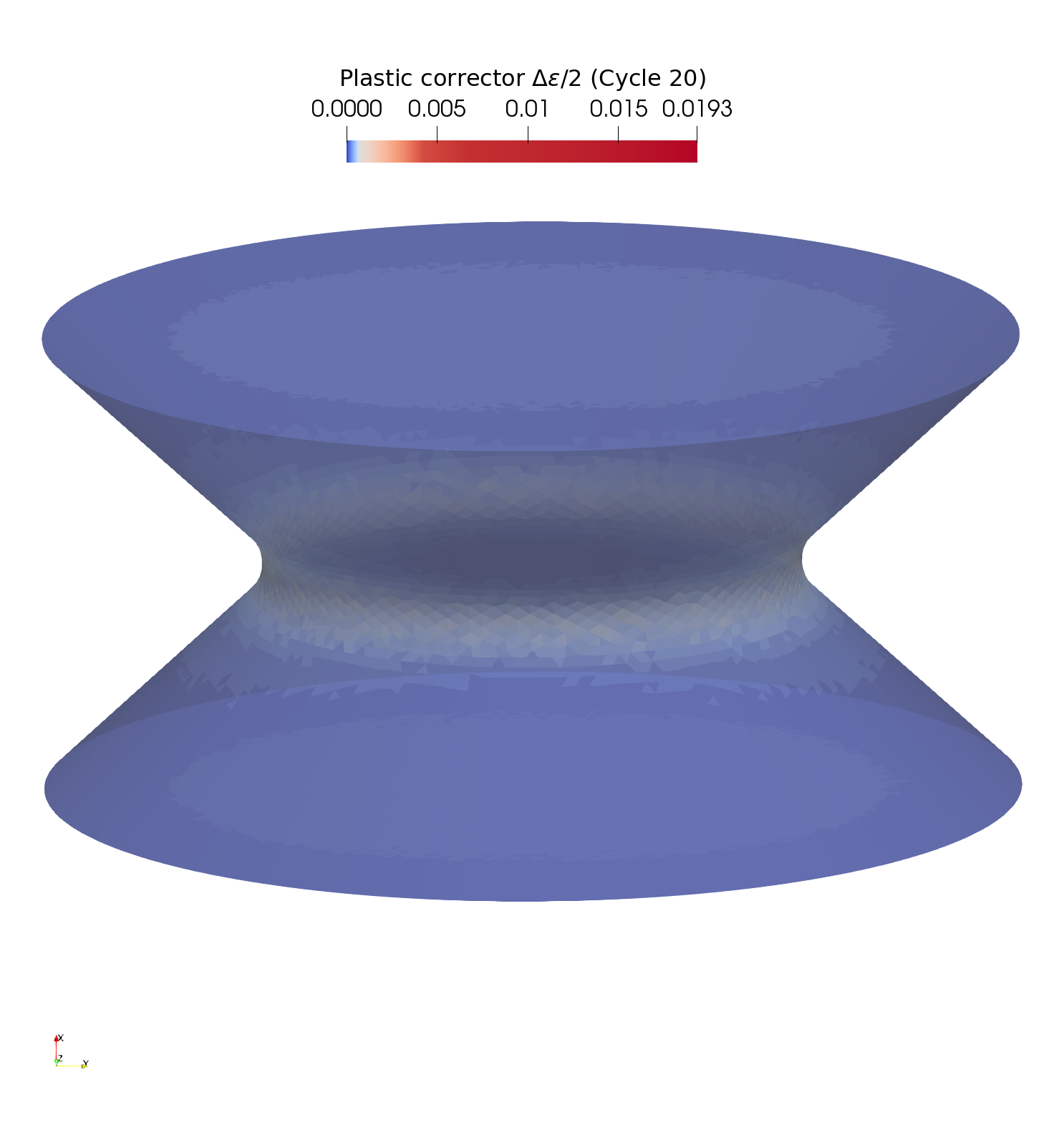}
        \caption{}
    \end{subfigure}
    
    \caption{\label{fig:transferability3} Extracted volume (for lifetime prediction) of different versions of the notched specimen (a) normal density of pores (b) reduced pores interacting with notch (c) no pores}
\end{figure}

The criterion of the lifetime model was obtained on all these geometries via the plastic corrector methodology. The elements in the notched region were extracted for lifetime prediction, with the assumption that fatigue lifetime failure is influenced only by this volume. The extracted volumes of the different versions are shown in Fig. \ref{fig:transferability3}. In the specimen with the normal expected pore density, there are two pores interacting with the notch that give rise to very high levels of plasticity. Concerning the specimen with removed surface-breaking pore interacting with the notch, there is a lesser amount of plasticity due to reduced pore-notch interaction. The specimen without pores has the least amount of plasticity, developing only due to the notch.

\paragraph{Transferability of homogenised model to notched geometry}\mbox{}\\

\begin{figure}[h!tbp]
  \centering
    \begin{subfigure}[b]{0.6\textwidth}
        \includegraphics[width=\textwidth]{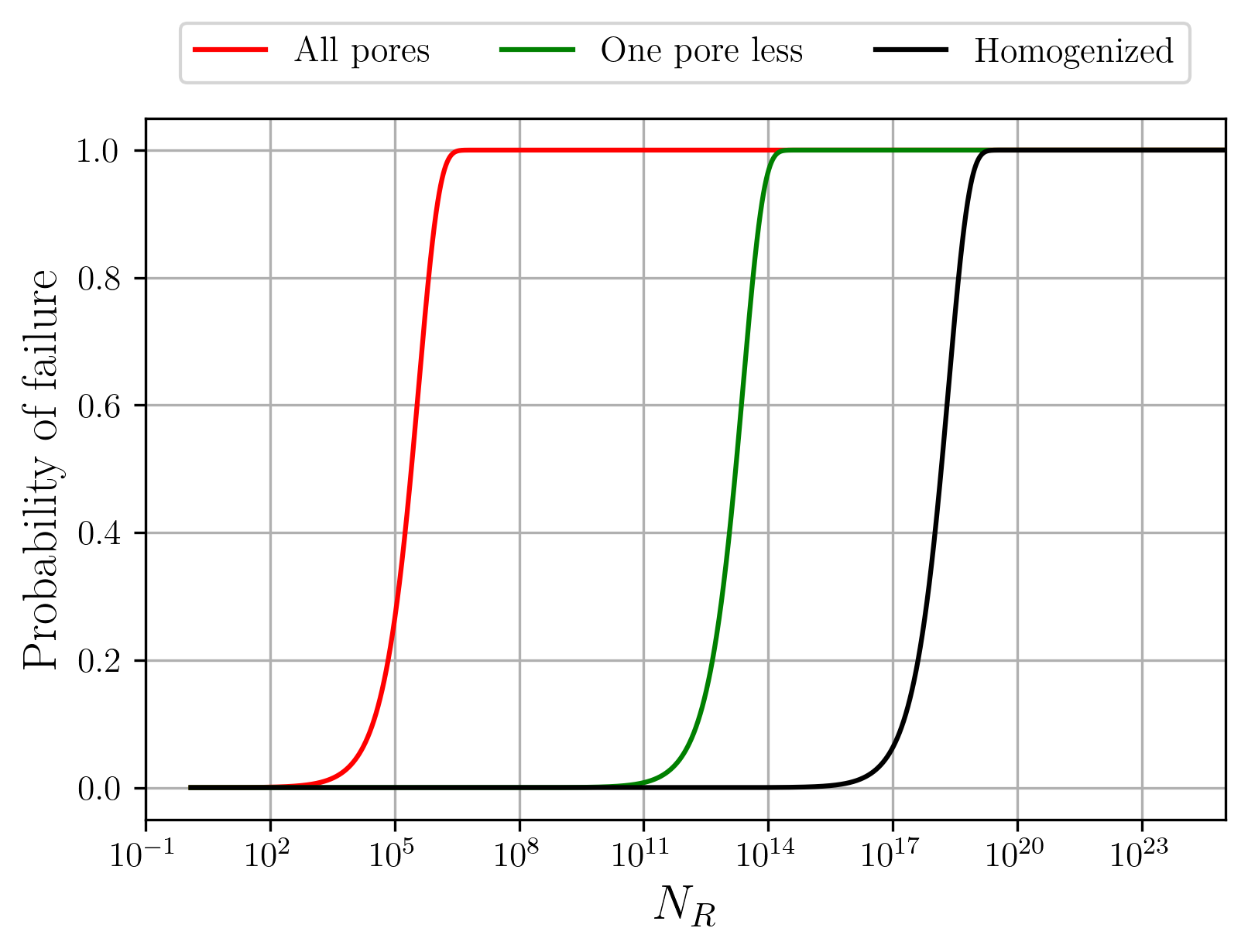}
        \caption{}
    \end{subfigure}
    \caption{\label{fig:transferability4} Lifetime predictions using model $\textbf{A}$ on the notched volume with normal density of pores (red) model $\textbf{A}$ on the notched volume with one less pore interacting with the notch (green) homogenised model $\textbf{B}$ on a notched volume with no pores (black)}
\end{figure}

\noindent We are now in a position to check the transferability of the homogenised model $\textbf{B}$ to the notched geometry. Lifetime predictions (shown in \ref{fig:transferability4}) were obtained using the following configurations:
\begin{description}
    \item[1.] Model $\textbf{A}$ on the notched volume with normal density of pores (red)
    \item[2.] Model $\textbf{A}$ on the notched volume with one less surface-breaking pore in the notch region (green) 
    \item[3.] homogenised model $\textbf{B}$ on the notched volume with no pores (black)
\end{description}

The first case i.e. obtained using the multi-scale method $\textbf{A}$ on the volume with the normal expected pore density, can be considered as the reference solution. The predictions using the same method $\textbf{A}$ on the volume with a surface-breaking pore interacting with the notch artificially removed shows a increase in the fatigue lifetime, due to lesser stresses in the notch region. However, both of these predictions show lesser lifetimes of the porous notch volume as compared to the homogenised porous model $\textbf{B}$. This indicates that the homogenised model $\textbf{B}$ identified on cylindrical specimens is not transferable to other geometries.

\clearpage
\section{Overview of findings and elements of discussion \label{sec:discussion}}

\subsection{Need for meshing pores \label{disc:needfortwoscale}}
The multi-scale method can be applied to arbitrary geometries for fatigue lifetime predictions. However, there is a need to explicitly mesh the pores in these geometries. This is because a homogenised model (simulating porous material behavior on the fatigue lifetime) that remains transferable across varying geometries of structures is difficult to obtain. The reasons for this difficulty are that (a) the core and surface materials exhibit different fatigue behaviour, as shown in section \ref{sec:isovolumeeffect}, and (b) the homogenised model identified for one geometry cannot capture the pore-notch interaction in the notched geometries, as discussed in section \ref{sec:transferability}. Thus, there is a need to mesh the pores when using the multi-scale method for fatigue lifetime predictions on arbitrary geometries containing porous material. 

\subsection{Choice of fatigue model complexity \label{disc:choiceofmodel}}
When optimisation is performed on only the non-porous data points (results shown in Fig. \ref{Fig:Homogenous_cases}), we observe that the strain-life model with two lines (parameters $A$, $\alpha$, $B$, $\beta$ and $C$) and the model with one line (parameters $A$, $\alpha$  and $C$) have the same quality of fit, i.e. there is no need for the second line when fitting data in only one regime (HCF regime). However, if more data is present, especially in the LCF regime, it is likely that both lines of the strain life model would be required to obtain a consistent fit. Fig. \ref{Fig:Homogenous_cases}(c) shows a significant decrease in the quality of the fit when the parameter $C$ is omitted, which highlights its importance for getting a sensible fatigue limit. The probabilistic fatigue curves show a decrease in the uncertainty with increasing applied stress amplitude, which is a widely observed phenomenon in fatigue, and enforced with $g^{-1}(\frac{\Delta \varepsilon}{2})$ in the numerator of the Weibull scale parameter, i.e. the scale reduces for higher loads and lower $N_R$ for a fixed shape parameter.

When optimisation is performed on only the porous data points (results shown in Fig. \ref{Fig:Heterogenous_cases}) both the two-line model and one line model seem to give good fits. However, as the porous data points contain information about both LCF and HCF behavior of the alloy, it is better to work with the two-line strain-life model. Material points around pores that are highly loaded should behave in the LCF regime and points in the non-loaded zones should behave in the HCF regime. 

\subsection{Enabling transferability of fatigue models between different types of pore populations}
Any fatigue lifetime model that needs to be accurate across different loading regimes (LCF and HCF) first needs to be properly identified with experimental data across the entire load spectra. The multi-scale method's capability of treating different alloys of porous material as made of the same base material is thus advantageous. The base material is considered to have a unique behavior, thus sparse data on different types of pore populations across various loading spectra can be exploited with ease. This is because the multi-scale method, by design, only requires a small representative tomography of the pore population for each grade of porosity. Additionally, the required FEA computations on statistically representative pore fields are accelerated by the plastic corrector \cite{Palchoudhary2024}.

The numerical results on the aluminium data-set demonstrate this: the model is able to treat the non-porous grade and the porous grade to be made of the same base material, and identification of a transferable model between these two grades is possible with lesser data than the simplified approach of treating the porous material as a different material than the non-porous one.

\subsection{Diversity of exploitable lifetime data for a model transferable between pore populations}
Two types of lifetime data are usable to identify a fatigue lifetime model with the proposed multi-scale approach. For example, to identify the model in the HCF regime, one may either load porous specimens at very low levels of stress such that the base material around the pores is loaded in the HCF regime, or by loading specimens made of the base material itself in the HCF regime. This can be particularly helpful if lifetime data on the base material (without porosity) is expensive or time consuming to obtain, or if a limited amount of lifetime data is present for each category of porosity.

For the example presented in this paper in section \ref{sec:results_twoscale_porousonly}, the two-line fatigue model (identified on the limited set of porous fatigue experiments, across a small range of loading levels) lacks information on distinguishing between the fatigue regimes. However, when the model is given some non-porous fatigue lifetime points at lower stress levels (in section \ref{sec:jointresults_reduceddata}), it learns to better differentiate between the two regimes of fatigue behaviour of the base material. The base material behavior in the HCF regime is identified using these non-porous points. In porous specimens, the base material near pores, being loaded at a higher level, informs the LCF fatigue behaviour. In other words, the model learns about different regimes of fatigue behavior of the base material from the lifetime data of the two grades of porosity. 

\section{Conclusions\label{sec:conclusion}}
In this paper, we propose a probabilistic fatigue lifetime model combined with a multi-scale method to model the lifetime of structures with pores whose exact distribution is unknown. The effect of the pores on the fatigue lifetime is modelled using statistical methods based on a small representative tomography of the porous material. The multi-scale aspect is facilitated by a FEA accelerator for lifetime criterion computations when non-linearity is present in the material model.

The identification procedure involves a maximum likelihood estimate, utilizing pre-computed statistically representative heterogeneous 3D elasto-plastic fields obtained via FEA during optimisation. Run-outs are used as valid experimental data by assigning them a finite probability of future failure.

The multi-scale approach developed in this paper allows materials with varying grades of porosity to be treated as if made of the same base material, thus enabling the use of lifetime data from materials with different pore populations for the identification of a singular fatigue lifetime model. This approach reduces the overall lifetime data required for identifying the fatigue lifetime model as compared to approaches that treat each porous grade as a different material. 

Additionally, the model naturally accounts for the fact that, in porous materials, fatigue resistance at the subsurface level is lower than that at the core. The model also considers the statistical size effect, where larger volumes fail earlier under iso-stress conditions.

The model allows for the approximation of Wöhler curves across arbitrary geometries, enabling lifetime estimation even without the need for full tomography of the structure. This is achieved using a small representative volume containing information about the pores, making the model both efficient and effective. The prediction time is comparable to the cost of an elastic-static FEA computation, as the plasticity in the stabilized cycle is computed through the use of a plastic corrector \cite{Palchoudhary2024}. However, the model does require meshes where pores are explicitly represented, as a simple homogenization of the porous material does not reproduce the multi-scale method's prediction for all geometries. Additionally, the optimisation process, particularly when solving fatigue lifetime computation via residual minimization, is resource-intensive.

\section{Acknowledgements}
The authors would like to thank the French National Research Agency (ANR-20-THIA-0022) for the partial funding of this project. The direct financial support of
Mines Paris - PSL is also acknowledged. The authors are particularly grateful to Pierre Osmond (CETIM) for tomographies of the porous specimens.

%
\section*{Conflict of interest}
The authors declare that they have no known competing financial interests or personal
relationships that could have appeared to influence the work reported in this paper.

\bibliography{mybibfile}

\begin{thebibliography}{46}
\expandafter\ifx\csname natexlab\endcsname\relax\def\natexlab#1{#1}\fi
\providecommand{\url}[1]{\texttt{#1}}
\providecommand{\href}[2]{#2}
\providecommand{\path}[1]{#1}
\providecommand{\DOIprefix}{doi:}
\providecommand{\ArXivprefix}{arXiv:}
\providecommand{\URLprefix}{URL: }
\providecommand{\Pubmedprefix}{pmid:}
\providecommand{\doi}[1]{\href{http://dx.doi.org/#1}{\path{#1}}}
\providecommand{\Pubmed}[1]{\href{pmid:#1}{\path{#1}}}
\providecommand{\bibinfo}[2]{#2}
\ifx\xfnm\relax \def\xfnm[#1]{\unskip,\space#1}\fi
\bibitem[{Ezanno et~al.(2010)Ezanno, Doudard, Calloch, Millot, and
  Heuzé}]{Ezanno2010}
\bibinfo{author}{A.~Ezanno}, \bibinfo{author}{C.~Doudard},
  \bibinfo{author}{S.~Calloch}, \bibinfo{author}{T.~Millot},
  \bibinfo{author}{J.-L. Heuzé},
\newblock \bibinfo{title}{Fast characterization of high-cycle fatigue
  properties of a cast copper–aluminum alloy by self-heating measurements
  under cyclic loadings},
\newblock \bibinfo{journal}{Procedia Engineering} \bibinfo{volume}{2}
  (\bibinfo{year}{2010}) \bibinfo{pages}{967--976}.
  \DOIprefix\doi{https://doi.org/10.1016/j.proeng.2010.03.105},
  \bibinfo{note}{fatigue 2010}.
\bibitem[{Le et~al.(2018)Le, Saintier, Morel, Bellett, and Osmond}]{Le2018}
\bibinfo{author}{V.-D. Le}, \bibinfo{author}{N.~Saintier},
  \bibinfo{author}{F.~Morel}, \bibinfo{author}{D.~Bellett},
  \bibinfo{author}{P.~Osmond},
\newblock \bibinfo{title}{Investigation of the effect of porosity on the high
  cycle fatigue behaviour of cast al-si alloy by x-ray micro-tomography},
\newblock \bibinfo{journal}{International Journal of Fatigue}
  \bibinfo{volume}{106} (\bibinfo{year}{2018}) \bibinfo{pages}{24--37}.
  \URLprefix
  \url{https://www.sciencedirect.com/science/article/pii/S0142112317303778}.
  \DOIprefix\doi{https://doi.org/10.1016/j.ijfatigue.2017.09.012}.
\bibitem[{{Matpadi Raghavendra} et~al.(2024){Matpadi Raghavendra}, Maurel,
  Marcin, and Proudhon}]{Matpadi2024}
\bibinfo{author}{A.~K. {Matpadi Raghavendra}}, \bibinfo{author}{V.~Maurel},
  \bibinfo{author}{L.~Marcin}, \bibinfo{author}{H.~Proudhon},
\newblock \bibinfo{title}{Fatigue life prediction at mesoscopic scale of
  samples containing casting defects: A novel energy based non-local model},
\newblock \bibinfo{journal}{International Journal of Fatigue}
  \bibinfo{volume}{188} (\bibinfo{year}{2024}) \bibinfo{pages}{108485}.
  \URLprefix
  \url{https://www.sciencedirect.com/science/article/pii/S0142112324003438}.
  \DOIprefix\doi{https://doi.org/10.1016/j.ijfatigue.2024.108485}.
\bibitem[{{El Khoukhi} et~al.(2019){El Khoukhi}, Morel, Saintier, Bellett,
  Osmond, Le, and Adrien}]{ElKhoukhi2019}
\bibinfo{author}{D.~{El Khoukhi}}, \bibinfo{author}{F.~Morel},
  \bibinfo{author}{N.~Saintier}, \bibinfo{author}{D.~Bellett},
  \bibinfo{author}{P.~Osmond}, \bibinfo{author}{V.-D. Le},
  \bibinfo{author}{J.~Adrien},
\newblock \bibinfo{title}{Experimental investigation of the size effect in high
  cycle fatigue: Role of the defect population in cast aluminium alloys},
\newblock \bibinfo{journal}{International Journal of Fatigue}
  \bibinfo{volume}{129} (\bibinfo{year}{2019}) \bibinfo{pages}{105222}.
  \DOIprefix\doi{https://doi.org/10.1016/j.ijfatigue.2019.105222}.
\bibitem[{Romano et~al.(2019)Romano, Miccoli, and Beretta}]{romano2019}
\bibinfo{author}{S.~Romano}, \bibinfo{author}{S.~Miccoli},
  \bibinfo{author}{S.~Beretta},
\newblock \bibinfo{title}{A new fe post-processor for probabilistic fatigue
  assessment in the presence of defects and its application to am parts},
\newblock \bibinfo{journal}{International Journal of Fatigue}
  \bibinfo{volume}{125} (\bibinfo{year}{2019}) \bibinfo{pages}{324--341}.
  \URLprefix
  \url{https://www.sciencedirect.com/science/article/pii/S014211231930132X}.
  \DOIprefix\doi{https://doi.org/10.1016/j.ijfatigue.2019.04.008}.
\bibitem[{Lacourt(2019)}]{Lacourt2019}
\bibinfo{author}{L.~Lacourt}, \bibinfo{title}{Étude numérique de la nocivité
  des défauts dans les soudures}, \bibinfo{type}{Thèse de doctorat},
  Université Paris Sciences et Lettres, \bibinfo{address}{Matériaux},
  \bibinfo{year}{2019}. \bibinfo{note}{NNT : 2019PSLEM050, tel-02512870}.
\bibitem[{Shirani and Härkegård(2012)}]{Shirani2012}
\bibinfo{author}{M.~Shirani}, \bibinfo{author}{G.~Härkegård},
\newblock \bibinfo{title}{Damage tolerant design of cast components based on
  defects detected by 3d x-ray computed tomography},
\newblock \bibinfo{journal}{International Journal of Fatigue}
  \bibinfo{volume}{41} (\bibinfo{year}{2012}) \bibinfo{pages}{188--198}.
  \URLprefix
  \url{https://www.sciencedirect.com/science/article/pii/S0142112311002660}.
  \DOIprefix\doi{https://doi.org/10.1016/j.ijfatigue.2011.09.011},
  \bibinfo{note}{fatigue Design \& Material Defects}.
\bibitem[{Bercelli et~al.(2021)Bercelli, Moyne, Dhondt, Doudard, Calloch, and
  Beaudet}]{Bercelli2021}
\bibinfo{author}{L.~Bercelli}, \bibinfo{author}{S.~Moyne},
  \bibinfo{author}{M.~Dhondt}, \bibinfo{author}{C.~Doudard},
  \bibinfo{author}{S.~Calloch}, \bibinfo{author}{J.~Beaudet},
\newblock \bibinfo{title}{A probabilistic approach for high cycle fatigue of
  wire and arc additive manufactured parts taking into account process-induced
  pores},
\newblock \bibinfo{journal}{Additive Manufacturing} \bibinfo{volume}{42}
  (\bibinfo{year}{2021}) \bibinfo{pages}{101989}. \URLprefix
  \url{https://www.sciencedirect.com/science/article/pii/S2214860421001548}.
  \DOIprefix\doi{https://doi.org/10.1016/j.addma.2021.101989}.
\bibitem[{Talemi(2020)}]{Talemi2020}
\bibinfo{author}{R.~Talemi},
\newblock \bibinfo{title}{A numerical study on effects of randomly distributed
  subsurface hydrogen pores on fretting fatigue behaviour of aluminium
  alsi10mg},
\newblock \bibinfo{journal}{Tribology International} \bibinfo{volume}{142}
  (\bibinfo{year}{2020}) \bibinfo{pages}{105997}. \URLprefix
  \url{https://www.sciencedirect.com/science/article/pii/S0301679X19305146}.
  \DOIprefix\doi{https://doi.org/10.1016/j.triboint.2019.105997}.
\bibitem[{Hou et~al.(2024)Hou, Hu, Wauters, and Talemi}]{Hou2024}
\bibinfo{author}{Y.~Hou}, \bibinfo{author}{Z.~Hu},
  \bibinfo{author}{T.~Wauters}, \bibinfo{author}{R.~Talemi},
\newblock \bibinfo{title}{Combined effect of random porosity and surface defect
  on fatigue lifetime of additively manufactured micro-sized ti6al4v
  components: An investigation based on numerical analysis and machine learning
  approach},
\newblock \bibinfo{journal}{Theoretical and Applied Fracture Mechanics}
  \bibinfo{volume}{131} (\bibinfo{year}{2024}) \bibinfo{pages}{104451}.
  \URLprefix
  \url{https://www.sciencedirect.com/science/article/pii/S0167844224002003}.
  \DOIprefix\doi{https://doi.org/10.1016/j.tafmec.2024.104451}.
\bibitem[{Manson(1953)}]{Manson1953}
\bibinfo{author}{S.~S. Manson}, \bibinfo{title}{Behavior of materials under
  conditions of thermal stress}, \bibinfo{type}{Technical Note}
  \bibinfo{number}{TN 2933}, National Advisory Committee for Aeronautics
  (NACA), \bibinfo{year}{1953}.
\bibitem[{Astm(2015)}]{astm2015}
\bibinfo{author}{E.~Astm},
\newblock \bibinfo{title}{739-91. standard practice for statistical analysis of
  linear or linearized stress-life (sn) and strain-life ($\varepsilon$-n)
  fatigue data},
\newblock \bibinfo{journal}{ASTM International}  (\bibinfo{year}{2015}).
\bibitem[{Brown and Miller(1973)}]{Brown1973}
\bibinfo{author}{M.~Brown}, \bibinfo{author}{K.~Miller},
\newblock \bibinfo{title}{A theory for fatigue under multiaxial stress-strain
  conditions},
\newblock \bibinfo{journal}{Proceedings of the Institute of Mechanical
  Engineers} \bibinfo{volume}{187} (\bibinfo{year}{1973})
  \bibinfo{pages}{745--756}.
\bibitem[{Fatemi and Socie(1988)}]{Fatemi1988}
\bibinfo{author}{A.~Fatemi}, \bibinfo{author}{D.~Socie},
\newblock \bibinfo{title}{A critical plane approach to multiaxial fatigue
  damage including out-of-phase loading},
\newblock \bibinfo{journal}{Fatigue and Fracture of Engineering Materials and
  Structures} \bibinfo{volume}{11} (\bibinfo{year}{1988})
  \bibinfo{pages}{149--166}.
\bibitem[{Karolczuk and Macha(2005)}]{karolczuk2005}
\bibinfo{author}{A.~Karolczuk}, \bibinfo{author}{E.~Macha},
\newblock \bibinfo{title}{A review of critical plane orientations in multiaxial
  fatigue failure criteria of metallic materials},
\newblock \bibinfo{journal}{International Journal of Fracture}
  \bibinfo{volume}{134} (\bibinfo{year}{2005}) \bibinfo{pages}{267--304}.
  \DOIprefix\doi{10.1007/s10704-005-1088-2}.
\bibitem[{Zok(2017)}]{Zok2017}
\bibinfo{author}{F.~W. Zok},
\newblock \bibinfo{title}{On weakest link theory and weibull statistics},
\newblock \bibinfo{journal}{Journal of the American Ceramic Society}
  \bibinfo{volume}{100} (\bibinfo{year}{2017}) \bibinfo{pages}{1265--1268}.
  \URLprefix
  \url{https://ceramics.onlinelibrary.wiley.com/doi/abs/10.1111/jace.14665}.
  \DOIprefix\doi{https://doi.org/10.1111/jace.14665}.
  \href{http://arxiv.org/abs/https://ceramics.onlinelibrary.wiley.com/doi/pdf/10.1111/jace.14665}{{\tt
  arXiv:https://ceramics.onlinelibrary.wiley.com/doi/pdf/10.1111/jace.14665}}.
\bibitem[{Liu et~al.(2020)Liu, Wang, Hu, and Mao}]{Liu2020}
\bibinfo{author}{X.~Liu}, \bibinfo{author}{R.~Wang}, \bibinfo{author}{D.~Hu},
  \bibinfo{author}{J.~Mao},
\newblock \bibinfo{title}{A calibrated weakest-link model for probabilistic
  assessment of lcf life considering notch size effects},
\newblock \bibinfo{journal}{International Journal of Fatigue}
  \bibinfo{volume}{137} (\bibinfo{year}{2020}) \bibinfo{pages}{105631}.
  \URLprefix
  \url{https://www.sciencedirect.com/science/article/pii/S0142112320301626}.
  \DOIprefix\doi{https://doi.org/10.1016/j.ijfatigue.2020.105631}.
\bibitem[{Li et~al.(2022)Li, Zhu, Liao, Correia, Berto, and Wang}]{Li2022}
\bibinfo{author}{X.-K. Li}, \bibinfo{author}{S.-P. Zhu},
  \bibinfo{author}{D.~Liao}, \bibinfo{author}{J.~A. Correia},
  \bibinfo{author}{F.~Berto}, \bibinfo{author}{Q.~Wang},
\newblock \bibinfo{title}{Probabilistic fatigue modelling of metallic materials
  under notch and size effect using the weakest link theory},
\newblock \bibinfo{journal}{International Journal of Fatigue}
  \bibinfo{volume}{159} (\bibinfo{year}{2022}) \bibinfo{pages}{106788}.
  \URLprefix
  \url{https://www.sciencedirect.com/science/article/pii/S0142112322000676}.
  \DOIprefix\doi{https://doi.org/10.1016/j.ijfatigue.2022.106788}.
\bibitem[{Taylor(1999)}]{Taylor1999}
\bibinfo{author}{D.~Taylor},
\newblock \bibinfo{title}{Geometrical effects in fatigue: a unifying
  theoretical model},
\newblock \bibinfo{journal}{International Journal of Fatigue}
  \bibinfo{volume}{21} (\bibinfo{year}{1999}) \bibinfo{pages}{413--420}.
  \URLprefix
  \url{https://www.sciencedirect.com/science/article/pii/S0142112399000079}.
  \DOIprefix\doi{https://doi.org/10.1016/S0142-1123(99)00007-9}.
\bibitem[{Kuguel(1961)}]{Kuguel1961}
\bibinfo{author}{R.~Kuguel},
\newblock \bibinfo{title}{A relation between theoretical stress concentration
  factor and fatigue notch factor deduced from the concept of highly stressed
  volume},
\newblock in: \bibinfo{booktitle}{ASTM proc}, volume~\bibinfo{volume}{61},
  \bibinfo{year}{1961}, pp. \bibinfo{pages}{732--748}.
\bibitem[{Sonsino et~al.(1997)Sonsino, Kaufmann, and Grubisic}]{Sonsino1997}
\bibinfo{author}{C.~M. Sonsino}, \bibinfo{author}{H.~Kaufmann},
  \bibinfo{author}{V.~Grubisic},
\newblock \bibinfo{title}{Transferability of material data for the example of a
  randomly loaded forged truck stub axle},
\newblock \bibinfo{journal}{SAE Technical Paper Series} \bibinfo{volume}{No.
  970708} (\bibinfo{year}{1997}) \bibinfo{pages}{1--22}.
\bibitem[{He et~al.(2022)He, Zhu, Taddesse, and Niu}]{He2022}
\bibinfo{author}{J.-C. He}, \bibinfo{author}{S.-P. Zhu}, \bibinfo{author}{A.~T.
  Taddesse}, \bibinfo{author}{X.~Niu},
\newblock \bibinfo{title}{Evaluation of critical distance, highly stressed
  volume, and weakest-link methods in notch fatigue analysis},
\newblock \bibinfo{journal}{International Journal of Fatigue}
  \bibinfo{volume}{162} (\bibinfo{year}{2022}) \bibinfo{pages}{106950}.
  \URLprefix
  \url{https://www.sciencedirect.com/science/article/pii/S0142112322002195}.
  \DOIprefix\doi{https://doi.org/10.1016/j.ijfatigue.2022.106950}.
\bibitem[{Mu{\~{n}}iz-Calvente et~al.(2015)Mu{\~{n}}iz-Calvente,
  Fern{\'{a}}ndez~Canteli, Shlyannikov, and Castillo}]{munizcalvente2015}
\bibinfo{author}{M.~Mu{\~{n}}iz-Calvente},
  \bibinfo{author}{A.~Fern{\'{a}}ndez~Canteli},
  \bibinfo{author}{V.~Shlyannikov}, \bibinfo{author}{E.~Castillo},
\newblock \bibinfo{title}{Probabilistic weibull methodology for fracture
  prediction of brittle and ductile materials},
\newblock in: \bibinfo{booktitle}{Damage Mechanics: Theory, Computation and
  Practice}, volume \bibinfo{volume}{784} of \textit{\bibinfo{series}{Applied
  Mechanics and Materials}}, \bibinfo{publisher}{Trans Tech Publications Ltd},
  \bibinfo{year}{2015}, pp. \bibinfo{pages}{443--451}.
  \DOIprefix\doi{10.4028/www.scientific.net/AMM.784.443}.
\bibitem[{Lanning et~al.(2003)Lanning, Nicholas, and Palazotto}]{Lanning2003}
\bibinfo{author}{D.~B. Lanning}, \bibinfo{author}{T.~Nicholas},
  \bibinfo{author}{A.~Palazotto},
\newblock \bibinfo{title}{Hcf notch predictions based on weakest-link failure
  models},
\newblock \bibinfo{journal}{International Journal of Fatigue}
  \bibinfo{volume}{25} (\bibinfo{year}{2003}) \bibinfo{pages}{835--841}.
  \URLprefix
  \url{https://www.sciencedirect.com/science/article/pii/S0142112303001567}.
  \DOIprefix\doi{https://doi.org/10.1016/S0142-1123(03)00156-7},
  \bibinfo{note}{international Conference on Fatigue Damage of Structural
  Materials IV}.
\bibitem[{Karolczuk and Palin-Luc(2013)}]{Karolczuk2013}
\bibinfo{author}{A.~Karolczuk}, \bibinfo{author}{T.~Palin-Luc},
\newblock \bibinfo{title}{Modelling of stress gradient effect on fatigue life
  using weibull based distribution function},
\newblock \bibinfo{journal}{Journal of Theoretical and Applied Mechanics}
  \bibinfo{volume}{51} (\bibinfo{year}{2013}) \bibinfo{pages}{297--311}.
\bibitem[{Palchoudhary et~al.(2024)Palchoudhary, Peter, Maurel, Ovalle, and
  Kerfriden}]{Palchoudhary2024}
\bibinfo{author}{A.~Palchoudhary}, \bibinfo{author}{S.~Peter},
  \bibinfo{author}{V.~Maurel}, \bibinfo{author}{C.~Ovalle},
  \bibinfo{author}{P.~Kerfriden}, \bibinfo{title}{A plastic correction
  algorithm for full-field elasto-plastic finite element simulations : critical
  assessment of predictive capabilities and improvement by machine learning},
  \bibinfo{year}{2024}. \href{http://arxiv.org/abs/2402.06313}{{\tt
  arXiv:2402.06313}}.
\bibitem[{Le(2016)}]{LePhD2016}
\bibinfo{author}{V.~D. Le}, \bibinfo{title}{Etude de l’influence des
  hétérogénéités microstructurales sur la tenue en fatigue à grand nombre
  de cycles des alliages d’aluminium de fonderie}, Ph.D. thesis, ENSAM,
  \bibinfo{year}{2016}. \URLprefix \url{http://www.theses.fr/2016ENAM0012},
  \bibinfo{note}{thèse de doctorat dirigée par Morel, Franck Saintier,
  Nicolas et Bellett, Daniel Mécanique-matériaux Paris}.
\bibitem[{Le et~al.(2016)Le, Morel, Bellett, Saintier, and
  Osmond}]{VietDucLe2016}
\bibinfo{author}{V.-D. Le}, \bibinfo{author}{F.~Morel},
  \bibinfo{author}{D.~Bellett}, \bibinfo{author}{N.~Saintier},
  \bibinfo{author}{P.~Osmond},
\newblock \bibinfo{title}{Multiaxial high cycle fatigue damage mechanisms
  associated with the different microstructural heterogeneities of cast
  aluminium alloys},
\newblock \bibinfo{journal}{Materials Science and Engineering: A}
  \bibinfo{volume}{649} (\bibinfo{year}{2016}) \bibinfo{pages}{426--440}.
  \DOIprefix\doi{https://doi.org/10.1016/j.msea.2015.10.026}.
\bibitem[{{El Khoukhi} et~al.(2022){El Khoukhi}, Morel, Saintier, Bellett,
  Osmond, Le, and Adrien}]{Elkhoukhi2022}
\bibinfo{author}{D.~{El Khoukhi}}, \bibinfo{author}{F.~Morel},
  \bibinfo{author}{N.~Saintier}, \bibinfo{author}{D.~Bellett},
  \bibinfo{author}{P.~Osmond}, \bibinfo{author}{V.-D. Le},
  \bibinfo{author}{J.~Adrien},
\newblock \bibinfo{title}{Scatter and size effect in high cycle fatigue of cast
  aluminum-silicon alloys: A comprehensive experimental investigation},
\newblock \bibinfo{journal}{Procedia Structural Integrity} \bibinfo{volume}{38}
  (\bibinfo{year}{2022}) \bibinfo{pages}{611--620}. \URLprefix
  \url{https://www.sciencedirect.com/science/article/pii/S2452321622002761}.
  \DOIprefix\doi{https://doi.org/10.1016/j.prostr.2022.03.063},
  \bibinfo{note}{fatigue Design 2021, International Conference Proceedings, 9th
  Edition}.
\bibitem[{LE et~al.(2015)LE, Morel, Bellett, Pessard, Saintier, and
  Osmond}]{VietDucLe2015}
\bibinfo{author}{V.-D. LE}, \bibinfo{author}{F.~Morel},
  \bibinfo{author}{D.~Bellett}, \bibinfo{author}{E.~Pessard},
  \bibinfo{author}{N.~Saintier}, \bibinfo{author}{P.~Osmond},
\newblock \bibinfo{title}{Microstructural-based analysis and modelling of the
  fatigue behaviour of cast al-si alloys},
\newblock \bibinfo{journal}{Procedia Engineering} \bibinfo{volume}{133}
  (\bibinfo{year}{2015}) \bibinfo{pages}{562--575}.
  \DOIprefix\doi{https://doi.org/10.1016/j.proeng.2015.12.630},
  \bibinfo{note}{fatigue Design 2015, International Conference Proceedings, 6th
  Edition}.
\bibitem[{Doudard et~al.(2004)Doudard, Calloch, Hild, Cugy, and
  Galtier}]{Doudard2004}
\bibinfo{author}{C.~Doudard}, \bibinfo{author}{S.~Calloch},
  \bibinfo{author}{F.~Hild}, \bibinfo{author}{P.~Cugy},
  \bibinfo{author}{A.~Galtier},
\newblock \bibinfo{title}{Identification of the scatter in high cycle fatigue
  from temperature measurements},
\newblock \bibinfo{journal}{Comptes Rendus Mécanique} \bibinfo{volume}{332}
  (\bibinfo{year}{2004}) \bibinfo{pages}{795--801}.
  \DOIprefix\doi{https://doi.org/10.1016/j.crme.2004.06.002}.
\bibitem[{Ni and Mahadevan(2004)}]{Ni2004}
\bibinfo{author}{K.~Ni}, \bibinfo{author}{S.~Mahadevan},
\newblock \bibinfo{title}{Strain-based probabilistic fatigue life prediction of
  spot-welded joints},
\newblock \bibinfo{journal}{International Journal of Fatigue}
  \bibinfo{volume}{26} (\bibinfo{year}{2004}) \bibinfo{pages}{763--772}.
  \URLprefix
  \url{https://www.sciencedirect.com/science/article/pii/S0142112303002834}.
  \DOIprefix\doi{https://doi.org/10.1016/j.ijfatigue.2003.10.021}.
\bibitem[{Pessard et~al.(2011)Pessard, Morel, Morel, and Bellett}]{Pessard2011}
\bibinfo{author}{E.~Pessard}, \bibinfo{author}{F.~Morel},
  \bibinfo{author}{A.~Morel}, \bibinfo{author}{D.~Bellett},
\newblock \bibinfo{title}{Modelling the role of non-metallic inclusions on the
  anisotropic fatigue behaviour of forged steel},
\newblock \bibinfo{journal}{International Journal of Fatigue}
  \bibinfo{volume}{33} (\bibinfo{year}{2011}) \bibinfo{pages}{568--577}.
  \URLprefix
  \url{https://www.sciencedirect.com/science/article/pii/S0142112310002471}.
  \DOIprefix\doi{https://doi.org/10.1016/j.ijfatigue.2010.10.012}.
\bibitem[{Koutiri et~al.(2013)Koutiri, Bellett, Morel, and
  Pessard}]{Koutiri2013}
\bibinfo{author}{I.~Koutiri}, \bibinfo{author}{D.~Bellett},
  \bibinfo{author}{F.~Morel}, \bibinfo{author}{E.~Pessard},
\newblock \bibinfo{title}{A probabilistic model for the high cycle fatigue
  behaviour of cast aluminium alloys subject to complex loads},
\newblock \bibinfo{journal}{International Journal of Fatigue}
  \bibinfo{volume}{47} (\bibinfo{year}{2013}) \bibinfo{pages}{137--147}.
  \URLprefix
  \url{https://www.sciencedirect.com/science/article/pii/S0142112312002472}.
  \DOIprefix\doi{https://doi.org/10.1016/j.ijfatigue.2012.08.004}.
\bibitem[{Li et~al.(2016)Li, Wen, Lu, Wang, and Deng}]{Li2016}
\bibinfo{author}{H.~Li}, \bibinfo{author}{D.~Wen}, \bibinfo{author}{Z.~Lu},
  \bibinfo{author}{Y.~Wang}, \bibinfo{author}{F.~Deng},
\newblock \bibinfo{title}{Identifying the probability distribution of fatigue
  life using the maximum entropy principle},
\newblock \bibinfo{journal}{Entropy} \bibinfo{volume}{18}
  (\bibinfo{year}{2016}). \URLprefix
  \url{https://www.mdpi.com/1099-4300/18/4/111}.
  \DOIprefix\doi{10.3390/e18040111}.
\bibitem[{Desmorat(2002)}]{Desmorat2002}
\bibinfo{author}{R.~Desmorat},
\newblock \bibinfo{title}{Fast estimation of localized plasticity and damage by
  energetic methods},
\newblock \bibinfo{journal}{International Journal of Solids and Structures}
  \bibinfo{volume}{39} (\bibinfo{year}{2002}) \bibinfo{pages}{3289--3310}.
  \URLprefix
  \url{https://www.sciencedirect.com/science/article/pii/S0020768302000021}.
  \DOIprefix\doi{https://doi.org/10.1016/S0020-7683(02)00002-1}.
\bibitem[{Kitagawa(1976)}]{Kitagawa1976}
\bibinfo{author}{H.~Kitagawa},
\newblock \bibinfo{title}{Applicability of fracture mechanics to very small
  cracks or the cracks in the early stage},
\newblock in: \bibinfo{booktitle}{Proceedings of 2nd ICM, Cleveland},
  \bibinfo{year}{1976}, pp. \bibinfo{pages}{627--631}.
\bibitem[{Geuzaine and Remacle(2020)}]{gmsh2020}
\bibinfo{author}{C.~Geuzaine}, \bibinfo{author}{J.-F. Remacle},
  \bibinfo{title}{Gmsh}, \bibinfo{year}{2020}. \URLprefix
  \url{http://http://gmsh.info/}.
\bibitem[{Chaboche(1989)}]{Chaboche1989}
\bibinfo{author}{J.~Chaboche},
\newblock \bibinfo{title}{Constitutive equations for cyclic plasticity and
  cyclic viscoplasticity},
\newblock \bibinfo{journal}{International Journal of Plasticity}
  \bibinfo{volume}{5} (\bibinfo{year}{1989}) \bibinfo{pages}{247--302}.
  \DOIprefix\doi{https://doi.org/10.1016/0749-6419(89)90015-6}.
\bibitem[{{M. S. Alnaes, J. Blechta, J. Hake, A. Johansson, B. Kehlet, A. Logg,
  C. Richardson, J. Ring, M. E. Rognes and G. N. Wells.}(2015)}]{fenics2015}
\bibinfo{author}{{M. S. Alnaes, J. Blechta, J. Hake, A. Johansson, B. Kehlet,
  A. Logg, C. Richardson, J. Ring, M. E. Rognes and G. N. Wells.}},
  \bibinfo{title}{The fenics project version 1.5, archive of numerical software
  3}, \bibinfo{year}{2015}. \URLprefix
  \url{doi.org/10.11588/ans.2015.100.20553}.
\bibitem[{Pollak and Palazotto(2009)}]{Pollak2009}
\bibinfo{author}{R.~D. Pollak}, \bibinfo{author}{A.~N. Palazotto},
\newblock \bibinfo{title}{A comparison of maximum likelihood models for fatigue
  strength characterization in materials exhibiting a fatigue limit},
\newblock \bibinfo{journal}{Probabilistic Engineering Mechanics}
  \bibinfo{volume}{24} (\bibinfo{year}{2009}) \bibinfo{pages}{236--241}.
  \URLprefix
  \url{https://www.sciencedirect.com/science/article/pii/S026689200800057X}.
  \DOIprefix\doi{https://doi.org/10.1016/j.probengmech.2008.06.006}.
\bibitem[{Lee et~al.(2023)Lee, Norovrinchen, and Sumiyadorj}]{Lee2023}
\bibinfo{author}{S.~Lee}, \bibinfo{author}{O.~Norovrinchen},
  \bibinfo{author}{C.~Sumiyadorj},
\newblock \bibinfo{title}{Fatigue strength estimation based on the maximum
  likelihood method},
\newblock in: \bibinfo{booktitle}{Proceedings of the second International
  Conference on Resources and Technology (RESAT 2023)},
  \bibinfo{publisher}{Atlantis Press}, \bibinfo{year}{2023}, pp.
  \bibinfo{pages}{98--106}. \URLprefix
  \url{https://doi.org/10.2991/978-94-6463-318-4_8}.
  \DOIprefix\doi{10.2991/978-94-6463-318-4_8}.
\bibitem[{Pegues et~al.(2017)Pegues, Roach, Williamson, and
  Shamsaei}]{Pegues2017}
\bibinfo{author}{J.~Pegues}, \bibinfo{author}{M.~Roach},
  \bibinfo{author}{R.~Williamson}, \bibinfo{author}{N.~Shamsaei},
\newblock \bibinfo{title}{Effect of specimen surface area size on fatigue
  strength of additively manufactured ti-6al-4v parts},
\newblock \bibinfo{year}{2017}.
\bibitem[{Makkonen(2001)}]{Makkonen2001}
\bibinfo{author}{M.~Makkonen},
\newblock \bibinfo{title}{Statistical size effect in the fatigue limit of
  steel},
\newblock \bibinfo{journal}{International Journal of Fatigue}
  \bibinfo{volume}{23} (\bibinfo{year}{2001}) \bibinfo{pages}{395--402}.
  \URLprefix
  \url{https://www.sciencedirect.com/science/article/pii/S0142112301000032}.
  \DOIprefix\doi{https://doi.org/10.1016/S0142-1123(01)00003-2}.
\bibitem[{{El Khoukhi} et~al.(2021){El Khoukhi}, Morel, Saintier, Bellett,
  Osmond, and Le}]{ElKhoukhi2021}
\bibinfo{author}{D.~{El Khoukhi}}, \bibinfo{author}{F.~Morel},
  \bibinfo{author}{N.~Saintier}, \bibinfo{author}{D.~Bellett},
  \bibinfo{author}{P.~Osmond}, \bibinfo{author}{V.-D. Le},
\newblock \bibinfo{title}{Probabilistic modeling of the size effect and scatter
  in high cycle fatigue using a monte-carlo approach: Role of the defect
  population in cast aluminum alloys},
\newblock \bibinfo{journal}{International Journal of Fatigue}
  \bibinfo{volume}{147} (\bibinfo{year}{2021}) \bibinfo{pages}{106177}.
  \URLprefix
  \url{https://www.sciencedirect.com/science/article/pii/S0142112321000372}.
  \DOIprefix\doi{https://doi.org/10.1016/j.ijfatigue.2021.106177}.
\bibitem[{Besson et~al.(2012)Besson, Leriche, Foerch, and
  Cailletaud}]{Besson1998}
\bibinfo{author}{J.~Besson}, \bibinfo{author}{R.~Leriche},
  \bibinfo{author}{R.~Foerch}, \bibinfo{author}{G.~Cailletaud},
\newblock \bibinfo{title}{Object-oriented programming applied to the finite
  element method part ii. application to material behaviors},
\newblock \bibinfo{journal}{Revue Européenne des Éléments}
  \bibinfo{volume}{7} (\bibinfo{year}{2012}) \bibinfo{pages}{567--588}.
  \DOIprefix\doi{10.1080/12506559.1998.10511322}.

\end{thebibliography}

\clearpage
\begin{appendices}

\section{Quality of strain-life fatigue lifetime criterion approximation}\label{appendixB}

\begin{figure}[htbp]
    \centering
    \includegraphics[width=0.9\textwidth]{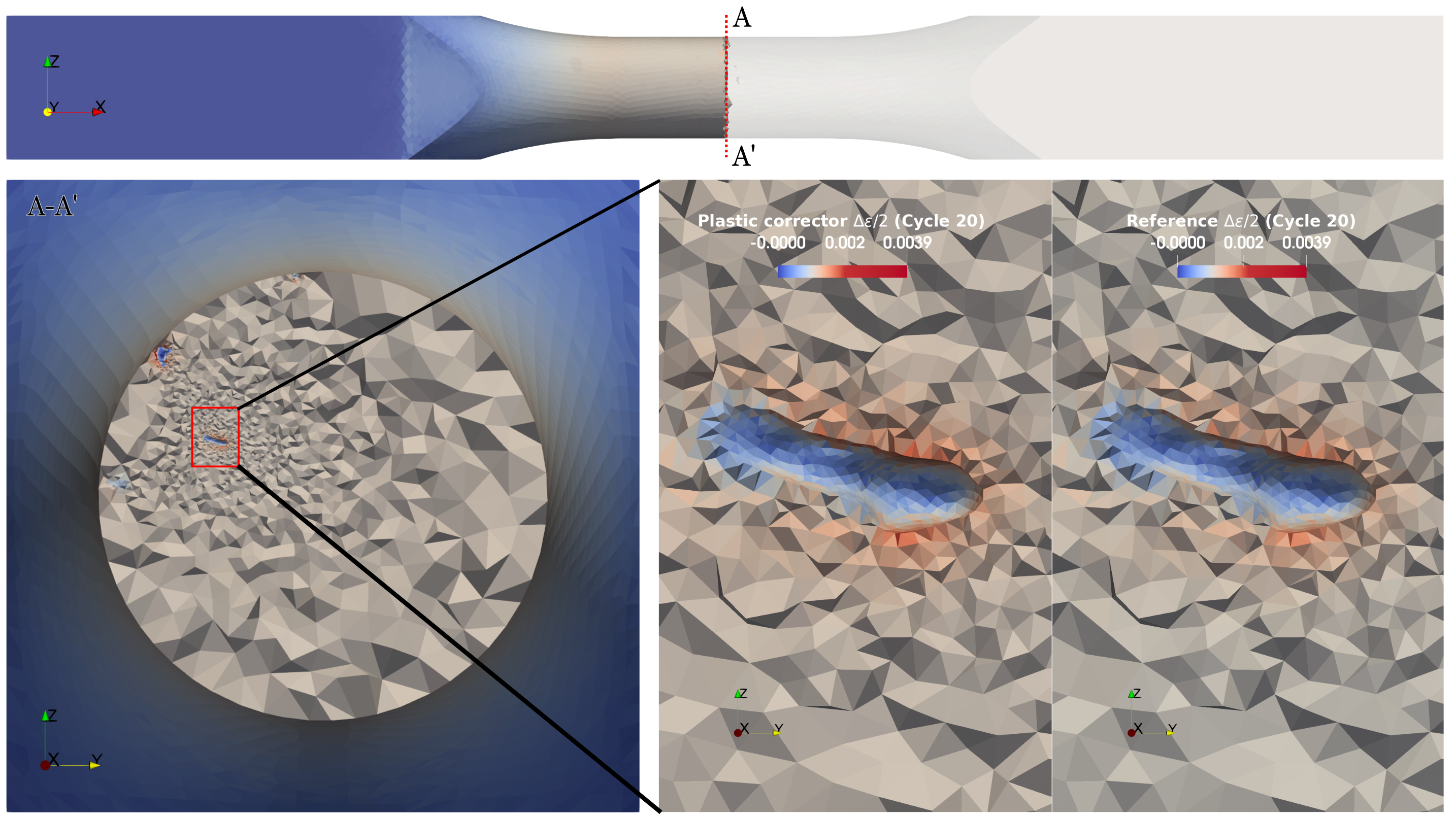}
    \caption{\label{Fig:RealPoresApproximationQuality} A comparison between $\Delta \varepsilon$ computed via the plastic corrector \cite{Palchoudhary2024} and a reference computation via Z-Set \cite{Besson1998} in a specimen containing a subvolume of pores. The loading corresponds to $\Sigma_a = 80$ MPa  (around 47\% of R0), which is the highest level of loading for which experimental high-cycle fatigue data is available.} 
\end{figure}

\begin{figure}[htbp]
    \centering
    \begin{subfigure}[b]{0.42\textwidth}
        \includegraphics[width=\textwidth]{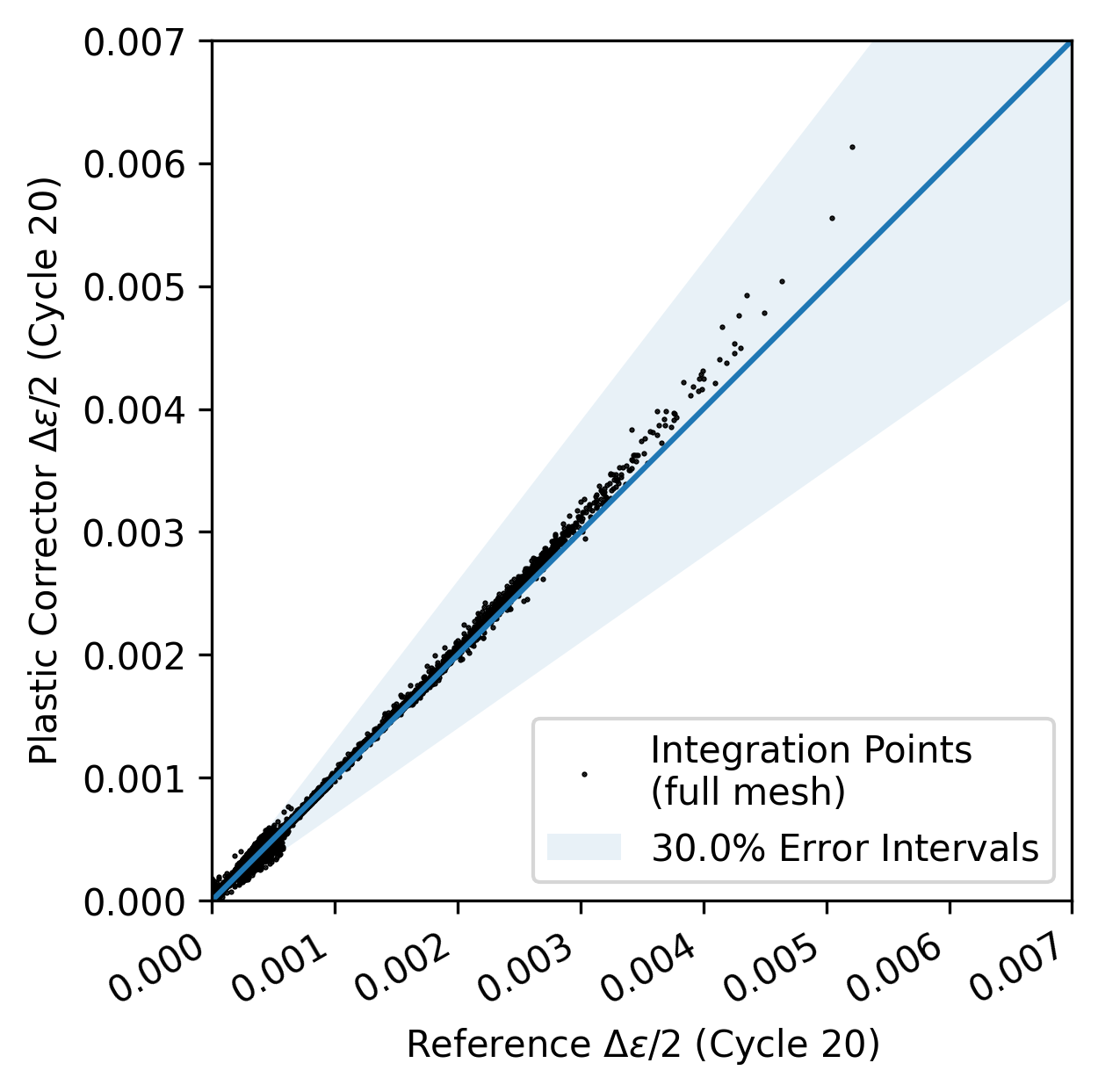}
        \caption{}
    \end{subfigure}
    \begin{subfigure}[b]{0.42\textwidth}
        \includegraphics[width=\textwidth]{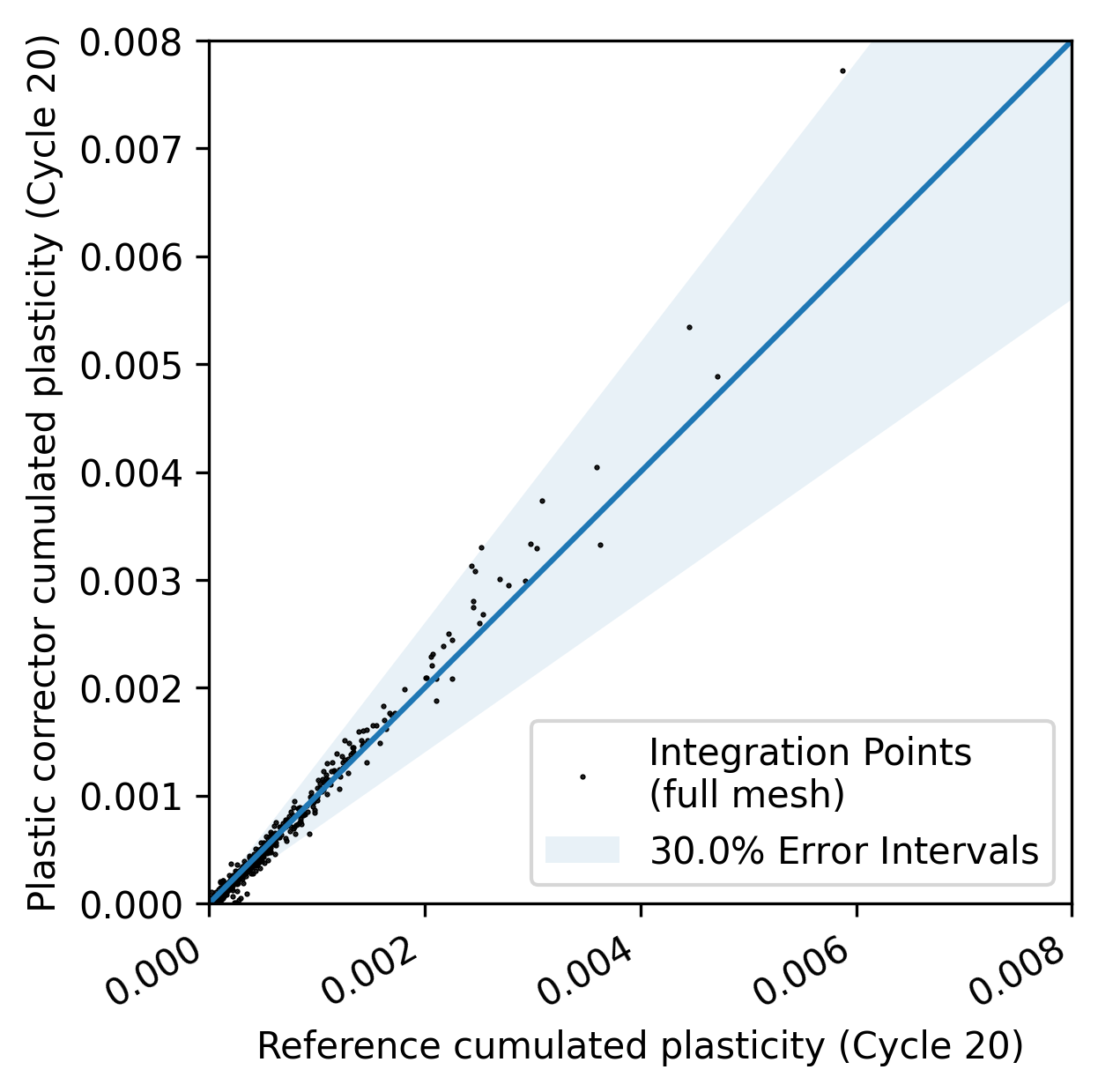}
        \caption{}
    \end{subfigure}
    \caption{\label{Fig:RealPoresApproximationFullComparison} A scatter plot comparing  $\Delta \varepsilon$ and $\Delta p$ computed via the plastic corrector \cite{Palchoudhary2024} and a reference computation via Z-Set \cite{Besson1998} in all the integration points of a specimen containing a subvolume of pores. The loading corresponds to $\Sigma_a = 80$ MPa  (around 47\% of R0), which is the highest level of loading for which experimental high-cycle fatigue data is available.} 
    \end{figure}

A plastic corrector is used for computation of the full-field $\Delta \varepsilon$ in the synthetically generated porous specimens \cite{Palchoudhary2024}. Figs. \ref{Fig:RealPoresApproximationQuality} and \ref{Fig:RealPoresApproximationFullComparison} show the quality of approximations as compared to a reference computation performed using the Z-Set suite \cite{Besson1998}. The plastic corrector approximates an elastoplastic solution pointwise using a plastic correction
heuristic. The heuristic is a modified Neuber rule coupled with a local proportional evolution rule, which, along with the constitutive relations, form a closed set of equations which are solved for the approximated elasto-plastic response.


\clearpage    
\section{Analytical form of the failure density of a specimen undergoing heterogeneous stresses}\label{appendixC}

The life of the porous specimen is dependent on the elements that constitute its volumes:

    \begin{equation}
        \textrm{Prob}(N_{\textrm{R}}^{\textrm{s}} \geq N) = \prod_{*\in \mathcal{E}} \textrm{Prob}(N_{\textrm{R}}^{\textrm{*}} \geq N)
    \end{equation}

       Therefore, the cumulative probability distribution (CDF) of the specimen failing is given by:

\begin{equation}
        F_{N_{\textrm{R}}^{\textrm{s}}}(N) = 1 - \prod_{*\in \mathcal{E}} \left(1 - F_{N_{\textrm{R}}^{\textrm{*}}}(N)\right)
\end{equation}

The CDF of failure of an element was defined as: 

\begin{equation}
    F_{N_{\textrm{R}}^{\textrm{*}}}(N) = 1 - \exp{\left( - \left\{ \frac{N}{\lambda} \right\}^m \right)}
\end{equation}

Substituting this expression in the expression for the CDF of the specimen, we get:

\begin{equation}
    F_{N_{\textrm{R}}^{\textrm{s}}}(N) = 1 - \prod_{*\in \mathcal{E}} \left( \exp\left(-\left\{\frac{N}{\lambda}\right\}^m\right)\right)
\end{equation}

\begin{equation}
    F_{N_{\textrm{R}}^{\textrm{s}}}(N) = 1 - \exp \left(\sum_{*\in \mathcal{E}} \left(-\left\{\frac{N}{\lambda}\right\}^m\right)\right)
\end{equation}

\begin{equation}
    F_{N_{\textrm{R}}^{\textrm{s}}}(N) = 1 - \exp \left(-N^m \sum_{*\in \mathcal{E}} \left\{\frac{1}{\lambda}\right\}^m\right)
\end{equation}

This is equivalent to a Weibull distribution with Weibull scale parameter denoted as $\lambda^s$:

\begin{equation}
    F_{N_{\textrm{R}}^{\textrm{s}}}(N) = 1 - \exp \left(-\left\{\frac{N}{\lambda^s}\right\}^m\right)
\end{equation}

\begin{equation}
    \lambda^s = \frac{1}{\left\{\sum\limits_{* \in \mathcal{E}}   \frac{1}{\lambda^{m}}\right\}^{1/m}}
\end{equation}

\begin{equation}
    N_{\textrm{R}}^{\textrm{s}} \sim \mathcal{W} \left( \lambda^s , m \right)
\end{equation}

The probability density function (PDF) and the cumulative distribution function (CDF) of the specimen's fatigue lifetime are denoted by $f_{N_{\textrm{R}}^{\textrm{s}}}(N; \Sigma_a,\mu)$ and $F_{N_{\textrm{R}}^{\textrm{s}}}(N; \Sigma_a,\mu)$ to show the dependence of these functions on the set of parameters $\mu$ and the applied stress amplitude $\Sigma_a$.

\end{appendices}
\end{document}